\documentclass[11pt]{report}
\usepackage[a4paper,width=150mm,top=25mm,bottom=25mm]{geometry}
\usepackage[utf8]{inputenc}
\usepackage{hyperref}
\hypersetup{
    colorlinks,
    citecolor=black,
    filecolor=black,
    linkcolor=black,
    urlcolor=black,
    breaklinks=true
}
\usepackage{xstring}
\usepackage{adjustbox}
\usepackage{algorithm}
\usepackage{algpseudocode}
\usepackage[titletoc]{appendix}
\usepackage{amssymb}
\usepackage{amsmath}
\usepackage{bbm}
\usepackage{bm}
\usepackage{booktabs}
\usepackage[style=numeric-comp, sorting=none]{biblatex}
\usepackage{braket}
\usepackage{catchfile}
\usepackage[capitalize]{cleveref}
\usepackage{color}
\usepackage{datatool}
\usepackage{easy-todo}
\usepackage{float}
\usepackage{sourcecodepro}
\usepackage[T1]{fontenc}
\usepackage{graphicx}
\usepackage{listings}
\usepackage{multirow}
\usepackage{mathtools}
\usepackage{notoccite}
\usepackage{siunitx}
\usepackage{tabularx}
\usepackage{pdfpages}
\usepackage{pythonhighlight}

\graphicspath{ {images/}{results/plots/} }
\newcommand*\samethanks[1][\value{footnote}]{\footnotemark[#1]}

\newcommand{\tr}{\text{tr}}

\newcommand{\R}{\mathbb{R}}
\newcommand{\N}{\mathbb{N}}
\newcommand{\mat}[1]{\mathbf{#1}}
\newcommand{\binset}{\{0,1\}}

\newcommand{\spause}{s_{\text{pause}}}
\newcommand{\tpause}{t_{\text{pause}}}
\newcommand{\Deltapause}{\Delta_{\text{pause}}}
\newcommand{\Deltaquench}{\Delta_{\text{quench}}}
\newcommand{\alphaquench}{\alpha_{\text{quench}}}
\newcommand{\squench}{s_{\text{quench}}}
\newcommand{\tquench}{t_{\text{quench}}}
\newcommand{\ptheory}{p_{\text{theory}}}
\newcommand{\pdata}{p_{\text{data}}}
\newcommand{\psamples}{p_{\text{samples}}}
\newcommand{\pmodel}{p_{\text{model}}}
\newcommand{\betahat}{\hat{\beta}}
\newcommand{\rcs}{\gamma_{\text{relative}}}
\newcommand{\trelative}{t_{\text{relative}}}
\DeclarePairedDelimiterX{\infdivx}[2]{(}{)}{#1\;\delimsize\|\;#2}
\newcommand{\DKL}{D_{\text{KL}}\infdivx}
\renewcommand{\vec}[1]{\mathbf{#1}}

\DeclarePairedDelimiter{\abs}{\lvert}{\rvert}

\makeatletter
\newcommand*{\transpose}{%
  {\mathpalette\@transpose{}}%
}
\newcommand*{\@transpose}[2]{%
  \raisebox{\depth}{$\m@th#1\intercal$}%
}
\makeatother
\newcommand{\T}{^{\transpose}}

\title{
    {Quantum Boltzmann Machines}\\
    {\large Applications in Quantitative Finance}
}
\author{
    {\LARGE Cameron Perot\thanks{\href{cameron.perot@pm.me}{cameron.perot@pm.me}}\vspace{1cm}}\\
    {Master's Thesis\vspace{0.1cm}}\\
    {\small submitted to\vspace{0.1cm}}\\
    {The Faculty of Mathematics, Computer Science, and Natural Sciences}\\
    {of RWTH Aachen University\vspace{0.1cm}}\\
    {\small written at\vspace{0.1cm}}\\
    {Jülich Supercomputing Centre}\\
    {Forschungszentrum Jülich\vspace{1cm}}\\
    {First Examiner: Prof. Dr. Kristel Michielsen\thanks{RWTH Aachen University, D-52056 Aachen, Germany}\textsuperscript{,}\thanks{Jülich Supercomputing Centre, Institute for Advanced Simulation, Forschungszentrum Jülich, D-52425 Jülich, Germany}}\\
    {Second Examiner: Prof. Dr. Holger Rauhut\samethanks[2]}\\
    {Adviser: Dr. Dennis Willsch\samethanks[3]\vspace{0.5cm}}
}
\date{July 4, 2022}

\begin{document}
\maketitle
\pagenumbering{roman}

\chapter*{Abstract}

In this thesis we explore using the D-Wave Advantage 4.1 quantum annealer to sample from quantum Boltzmann distributions and train quantum Boltzmann machines (QBMs).
We focus on the real-world problem of using QBMs as generative models to produce synthetic foreign exchange market data and analyze how the results stack up against classical models based on restricted Boltzmann machines (RBMs).
Additionally, we study a small 12-qubit problem which we use to compare samples obtained from the Advantage 4.1 with theory, and in the process gain vital insights into how well the Advantage 4.1 can sample quantum Boltzmann random variables and be used to train QBMs.
Through this, we are able to show that the Advantage 4.1 can sample classical Boltzmann random variables to some extent, but is limited in its ability to sample from quantum Boltzmann distributions.
Our findings indicate that QBMs trained using the Advantage 4.1 are much noisier than those trained using simulations and struggle to perform at the same level as classical RBMs.
However, there is the potential for QBMs to outperform classical RBMs if future generation annealers can generate samples closer to the desired theoretical distributions.

\newpage\thispagestyle{empty}\mbox{}

\tableofcontents

\chapter{Introduction}
\label{ch:introduction}
\pagenumbering{arabic}
In recent years we have seen the inception of cloud-based quantum computing, with a number of different providers offering various services.
In terms of maturity, the quantum computing industry as a whole is still in the early stages and there are a lot of obstacles left to overcome before mainstream adoption.
Quantum computing is not only trying to advance the theory and technology, but also yearning for practical applications in which quantum computing offers advantages over classical computing.

There are two main branches of quantum computing: universal quantum computing, i.e., gate-based quantum computing, and adiabatic quantum computing, i.e., quantum annealing.
In our work here we focus on the latter, as current generation devices are slightly more mature and have much higher numbers of qubits than the former.
We discuss the theory behind quantum annealing later in~\cref{sec:quantum_annealing}.
One such cloud-based quantum computing service is D-Wave's Leap platform~\cite{dwave_leap}, which allows users to access quantum annealers and other solvers across the world.

D-Wave is a pioneer in this field, having been researching and developing quantum annealers since 1999.
They revolutionized the field with the release of the world's first commercially available quantum annealer in 2011~\cite{zyga_2011}.
Since then, they have released a new version every 2-3 years, each having more qubits and couplers than the previous.
Their latest version, the D-Wave Advantage, has over 5000 qubits with 15 connections per qubit~\cite{dwave_advantage}.

In this thesis we take a journey into the field of quantum machine learning and explore the possibilities of using quantum Boltzmann machines (QBMs) as generative models for real-world financial data.
As we will see, there is a deep connection between the quantum Boltzmann machine and quantum annealing, allowing one to train QBMs using a quantum annealer.

Risk management is one of the most important components of the financial system, and in 2008 it failed, leading to the financial crisis which wreaked havoc on economies around the world.
The success of risk management hinges on how accurately the underlying risk models capture the true behavior of the market.
Therefore, it is essential that we continuously strive to find new and innovative ways of modeling that can help us understand the real risks involved and implement policies to effectively mitigate such risks.

In the globalized economy of today, foreign exchange (forex) fluctuations expose a number of firms to a lot of risk if not properly mitigated.
Forex markets had a daily volume of \$6.6T in 2019~\cite{bis_2019}, the majority of which was concentrated in a few major pairs.
In the 2019 paper \textit{The Market Generator}~\cite{kondratyev_2019}, Kondratyev and Schwarz detail how a classical restricted Boltzmann machine (RBM) can be used to generate synthetic forex data, and the advantages it offers over traditional parametric models.
We use their work as a basis to build our classical models upon, which we then use as a reference to compare our quantum models with.

In~\cref{ch:data_analysis}, we start by visualizing the data set in various ways to get an idea how it is distributed.
We further analyze quantitative metrics to get a better understanding of some of the intricacies of the data set.
Finally, we go through and detail how we preprocess the data set into a model-friendly format.

With the data set in hand, we move to explaining the theory behind the classical RBM in~\cref{ch:rbm} and describing some of the difficulties associated with training and using classical RBMs.
We then train several classical models on the data set discussed in~\cref{ch:data_analysis} using different preprocessing methods and compare them with each other using visualizations and a number of quantitative metrics.

In~\cref{ch:qbm}, we start from the theory of quantum Boltzmann machines, detailing how they work and their connection to quantum annealing.
We study a small 12-qubit problem which we can simulate, allowing us to compare annealer performance with that of theory, and gaining key insights into how to train and use QBMs.
With those insights, we move to the final stage of training a model using the data set from~\cref{ch:data_analysis}, then assessing the performance versus the classical models from~\cref{ch:rbm}.
Additionally, we cover some of the challenges of using D-Wave quantum annealers to train QBMs in~\cref{sec:challenges}.

Lastly, we summarize our findings in~\cref{ch:conclusion}, as well as discuss future directions in which this research can be expanded.

In addition to the research and results presented here, we also introduce the open source Python package \texttt{qbm}~\cite{qbm} to make it easier for the community to train and study quantum Boltzmann machines.
All work presented here is reproducible (except for that involving quantum measurements), and the code is available on GitHub~\footnote{\url{https://github.com/cameronperot/qbm-quant-finance}}.

\chapter{Data Analysis \& Preprocessing}
\label{ch:data_analysis}
\section{Data Analysis}
Our raw data set consists of the daily open, high, low, and close (OHLC) values for the time period 1999-01-01 through 2019-12-31 of the following major currency pairs
\begin{itemize}
    \item EURUSD - Euro € / U.S. Dollar \$
    \item GBPUSD - British Pound Sterling £ / U.S. Dollar \$
    \item USDCAD - U.S. Dollar \$ / Canadian Dollar \$
    \item USDJPY - U.S. Dollar \$ / Japanese Yen ¥
\end{itemize}
obtained from Dukascopy historical data feed~\cite{dukascopy}.
We filter the data set to remove days with zero volume, as well as NYSE and LSE holidays, resulting in 5165 training samples.
Here we use the notation \( x_\text{open} \), \( x_\text{high} \), \( x_\text{low} \), and \( x_\text{close} \) to denote the open, high, low, and close values of a currency pair on a particular day.

Given that the raw data values are on an absolute basis, we need to convert them to relative terms in order to be able to compare data from different time periods on a more equal footing.
The natural way to do so is to use the intraday returns
\begin{align}
    r = \frac{x_\text{close} - x_\text{open}}{x_\text{open}}.
\end{align}
However, this is not necessarily the best way to approach this.
Instead, we opt to use the log returns
\begin{align}
    \tilde{r}
        = \log(1+r)
        = \log\bigg( \frac{x_\text{close}}{x_\text{open}} \bigg)
\end{align}
due to several advantages, such as log-normality and small \( r \) approximation~\cite{quantivity_2012}.

\begin{figure}[!htb]
    \begin{center}
        \includegraphics[width=1\linewidth]{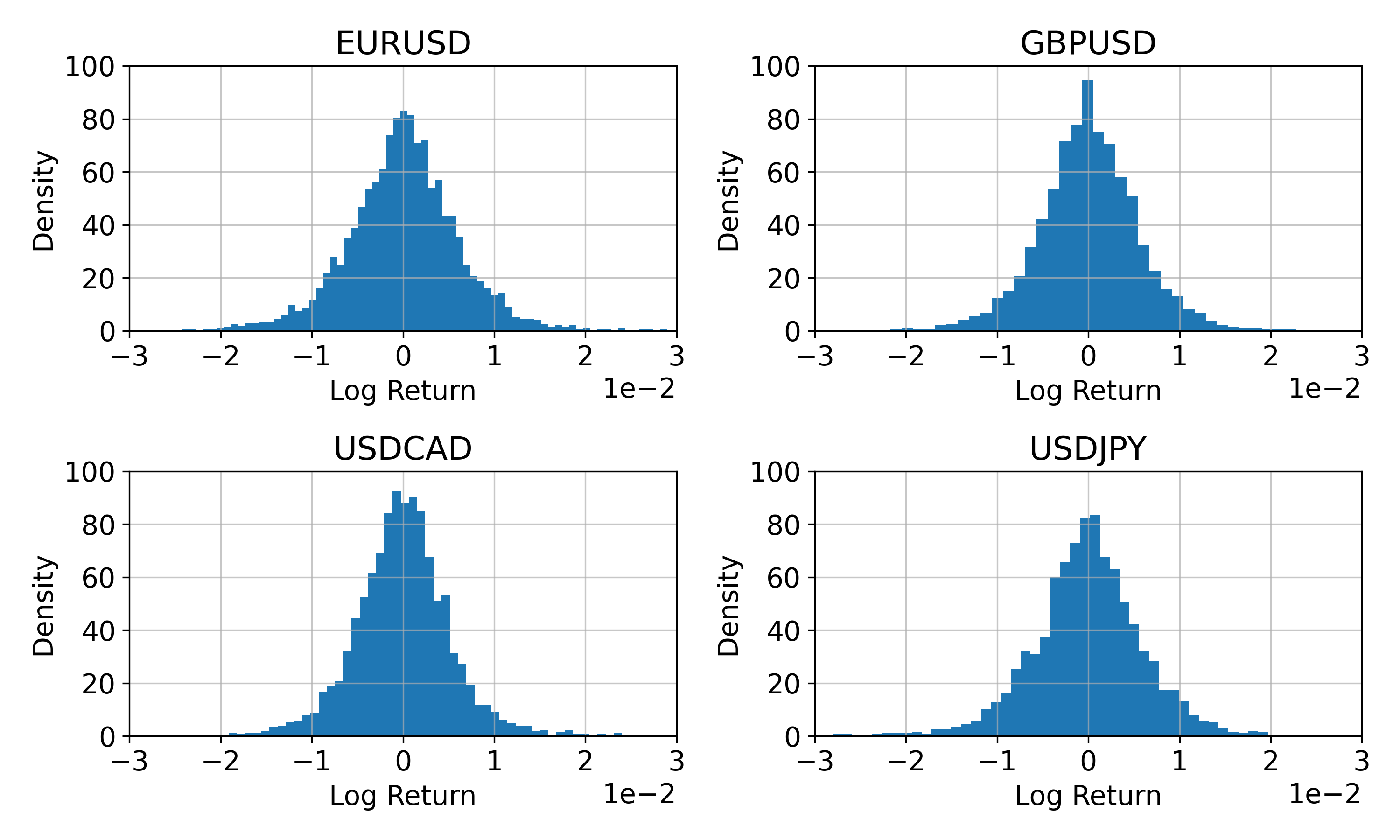}
    \end{center}
    \caption{Histograms of the log returns data set.}
    \label{fig:histograms_raw}
\end{figure}

We begin our analysis by taking a look at the histograms depicted in~\cref{fig:histograms_raw}.
From visual examination we see that the log returns are roughly normally distributed with the statistics given in~\cref{tbl:data_log_returns_raw_stats}.

\begin{table}[!htb]
    \centering
    \begin{adjustbox}{max width=\textwidth}
        \begin{tabular}{l r r}
\multicolumn{3}{c}{\textbf{Log Returns Data Set Statistics}} \\
\toprule
Currency Pair & Mean & Standard Deviation \\
\midrule
EURUSD & \( 5.15 \cdot 10^{-5} \) & \( 6.17 \cdot 10^{-3} \) \\
GBPUSD & \( -8.49 \cdot 10^{-6} \) & \( 5.73 \cdot 10^{-3} \) \\
USDCAD & \( -5.04 \cdot 10^{-5} \) & \( 5.40 \cdot 10^{-3} \) \\
USDJPY & \( -6.31 \cdot 10^{-5} \) & \( 6.32 \cdot 10^{-3} \) \\
\bottomrule
\end{tabular}

    \end{adjustbox}
    \caption{Statistics of the log returns data set.}
    \label{tbl:data_log_returns_raw_stats}
\end{table}

\begin{figure}[!htb]
    \begin{center}
        \includegraphics[width=1\linewidth]{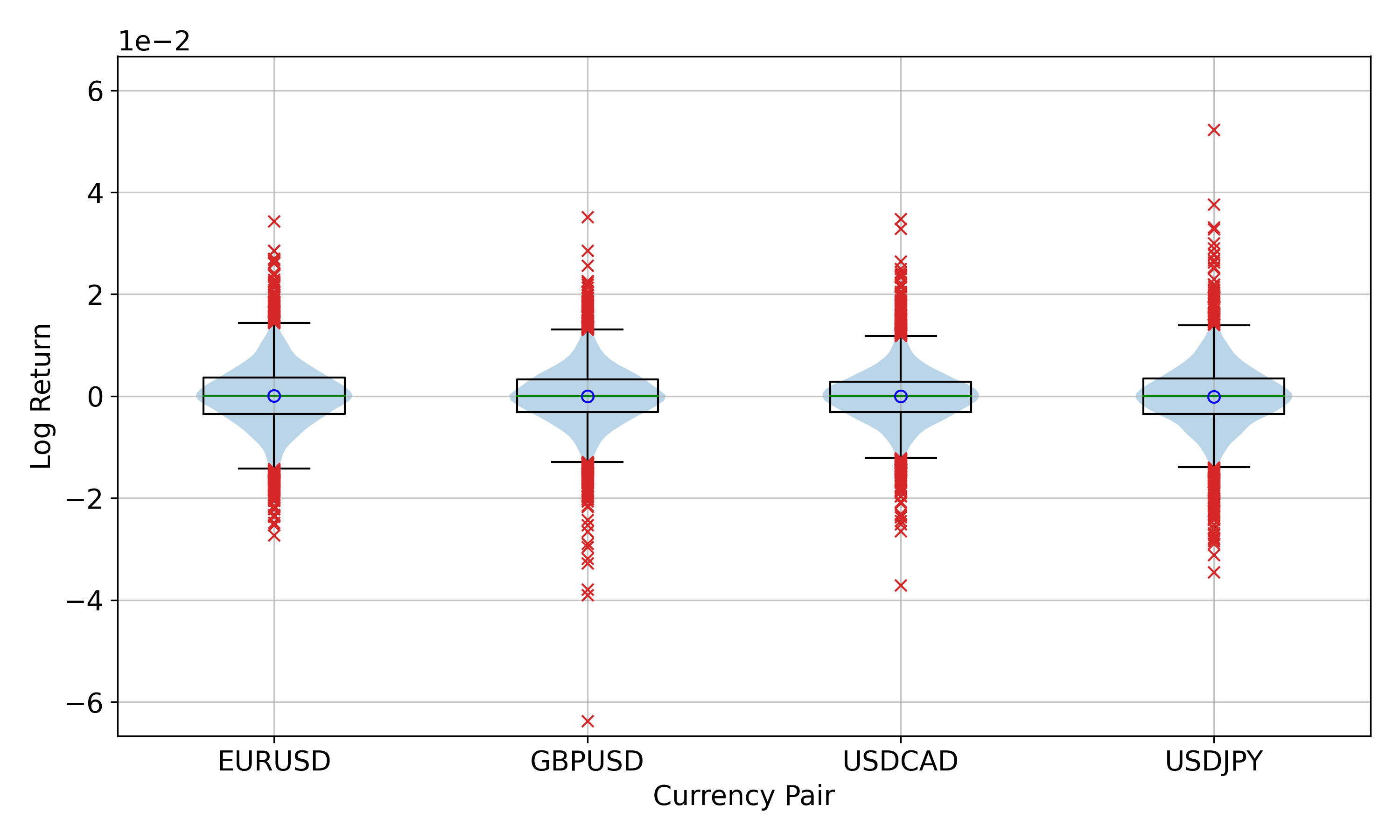}
    \end{center}
    \caption{Violin and box plot of the log returns data set illustrating the distribution of the outliers.}
    \label{fig:violin_raw}
\end{figure}

We also visualize the log returns in a violin and box plot in~\cref{fig:violin_raw} to identify outliers and see how they are distributed.
Two major outliers clearly stand out from the rest: one to the downside for the GBPUSD pair, and another to the upside for the USDJPY pair.
The former occurred on 2016-06-24, the day the Brexit referendum result was announced~\cite{brexit_gov_uk}.
The latter occurred on 2008-10-28, right in the midst of the financial crisis when people were talking about the end of the Yen carry trade~\cite{jpy_carry_trade_nyt}.
In the final training data set, we remove outliers greater than \( 10\sigma \) from the mean, resulting in only removing the day corresponding to the Brexit referendum result, which lies \( 11.1\sigma \) below the mean.

\begin{figure}[!htb]
    \begin{center}
        \includegraphics[width=1\linewidth]{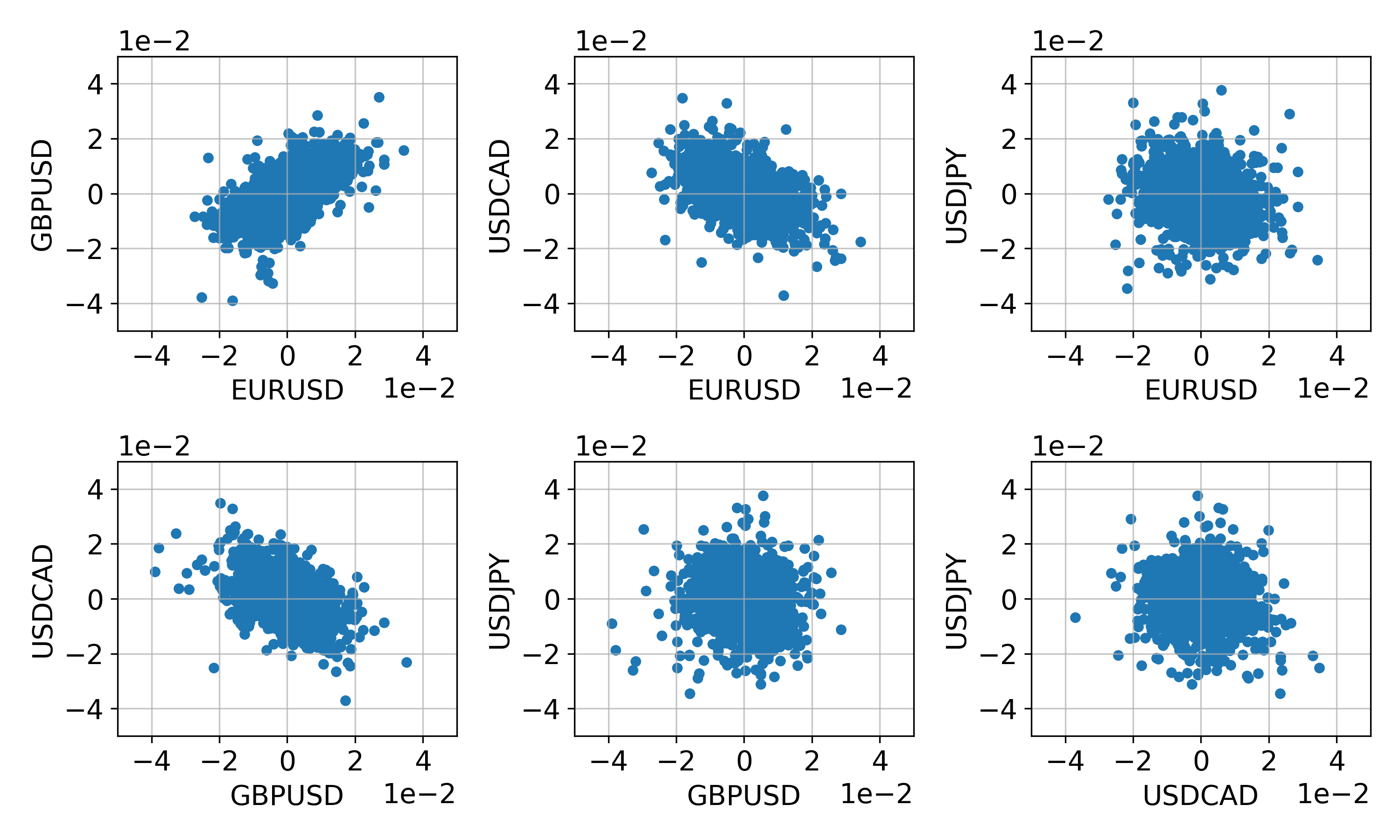}
    \end{center}
    \caption{Scatter plots of the log returns data set.}
    \label{fig:scatters}
\end{figure}

Next we examine the correlations between the currency pairs to get an idea of the interdependencies between them.
We visualize this with scatter plots shown in~\cref{fig:scatters} where we observe a clear positive correlation between EURUSD/GBPUSD, and clear negative correlations between EURUSD/USDCAD and GBPUSD/USDCAD, where the / is used to denote the pairs being compared against each other.
This is further verified by the Pearson \( r \), Spearman \( \rho \), and Kendall \( \tau \) correlation coefficients laid out in~\cref{tbl:data_correlation_coefficients}.
Furthermore, we find the correlation coefficients to be positive for pairs of the form \( X \)USD/\( Y \)USD, and negative for pairs of the form \( X \)USD/USD\( Y \), for \( X,Y \in \) \{EUR, GBP, CAD, JPY\}, as expected.
Details on how the correlation coefficients are computed and how to interpret them can be found in~\cref{app:correlation_coefficients}.

\begin{table}[!htb]
    \centering
    \begin{adjustbox}{max width=\textwidth}
        \begin{tabular}{l r r r}
\multicolumn{4}{c}{\textbf{Correlation Coefficients}} \\
\toprule
Currency Pairs & Pearson & Spearman & Kendall \\
\midrule
EURUSD/GBPUSD & 0.62 & 0.62 & 0.44 \\
EURUSD/USDCAD & -0.44 & -0.41 & -0.29 \\
EURUSD/USDJPY & -0.26 & -0.30 & -0.21 \\
GBPUSD/USDCAD & -0.42 & -0.37 & -0.26 \\
GBPUSD/USDJPY & -0.14 & -0.21 & -0.15 \\
USDCAD/USDJPY & 0.00 & 0.06 & 0.04 \\
\bottomrule
\end{tabular}

    \end{adjustbox}
    \caption{Correlation coefficients of the log returns data set.}
    \label{tbl:data_correlation_coefficients}
\end{table}

\section{Data Preprocessing}
The models in the following chapters require the training data to be in the form of bit vectors, so we must first convert our data set to such a form.
Let \( \mat{X} \in \R^{4 \times N} \) represent the training data set of log returns with \( N \) samples, where training samples are vectors in the column space, thus element \( x_{ij} \) represents the \( i \)th currency pair log return for the \( j \)th training sample.

To discretize the data, we rescale and round the entries of \( \mat{X} \) to integer values in \( \{0, 1, \dots, 2^{n_\text{bits}} - 1\} \), represented by the matrix \( \mat{X}' \in \N^{4 \times N} \) with entries
\begin{align}
    x_{ij}' = \bigg\lfloor \frac{x_{ij} - \min_k \{x_{ik}\}}{\max_k \{x_{ik}\} - \min_k \{x_{ik}\}} \cdot (2^{n_\text{bits}} - 1) \bigg\rceil,
\end{align}
where \( \lfloor \ \cdot \ \rceil \) denotes rounding to the nearest integer.

A new matrix \( \mat{V} \in \binset^{4\cdot n_\text{bits} \times N} \) is then created with the columns being the \( n_\text{bits} \)-length bit vectors corresponding to the binary representation of the entries of the columns of \( \mat{X}' \) concatenated together.
For example, if \( \vec{x}' = (x_1',x_2',x_3',x_4') \) is a column of \( \mat{X}' \) and the function \( \text{bitvector}(x') \) takes in an integer \( x' \) and returns an \( n_\text{bits} \)-bit binary representation bit vector, then the corresponding column in \( \mat{V} \) is
\begin{align}
    \vec{v} = \begin{bmatrix}
        \text{bitvector}(x_1') \\
        \text{bitvector}(x_2') \\
        \text{bitvector}(x_3') \\
        \text{bitvector}(x_4') \\
    \end{bmatrix}
    \in \binset^{4\cdot n_\text{bits}}.
\end{align}

For this research we take \( n_\text{bits} = 16 \), giving us a training set \( \mat{V} \in \binset^{64 \times N} \), thus our training samples are bit vectors of length 64.
The discretization errors associated with this conversion and data set are on the order of \( 10^{-7} \), well within the desired tolerance for this purpose.

\subsection{Data Transformation}\label{sec:outlier_transform}
Due to how the data is linearly converted to a discrete form before rounding, it opens up the possibility of the discretized data being clustered in the mid-range values if large outliers are present.
To mitigate this, we use a transformation to reduce the gap between outliers by scaling outliers beyond a certain threshold \( \tau \) using the procedure detailed in~\cref{alg:transformation}.
We call this the \textit{outlier power transformation}.

\begin{figure}[!htb]
    \begin{center}
        \includegraphics[width=1\linewidth]{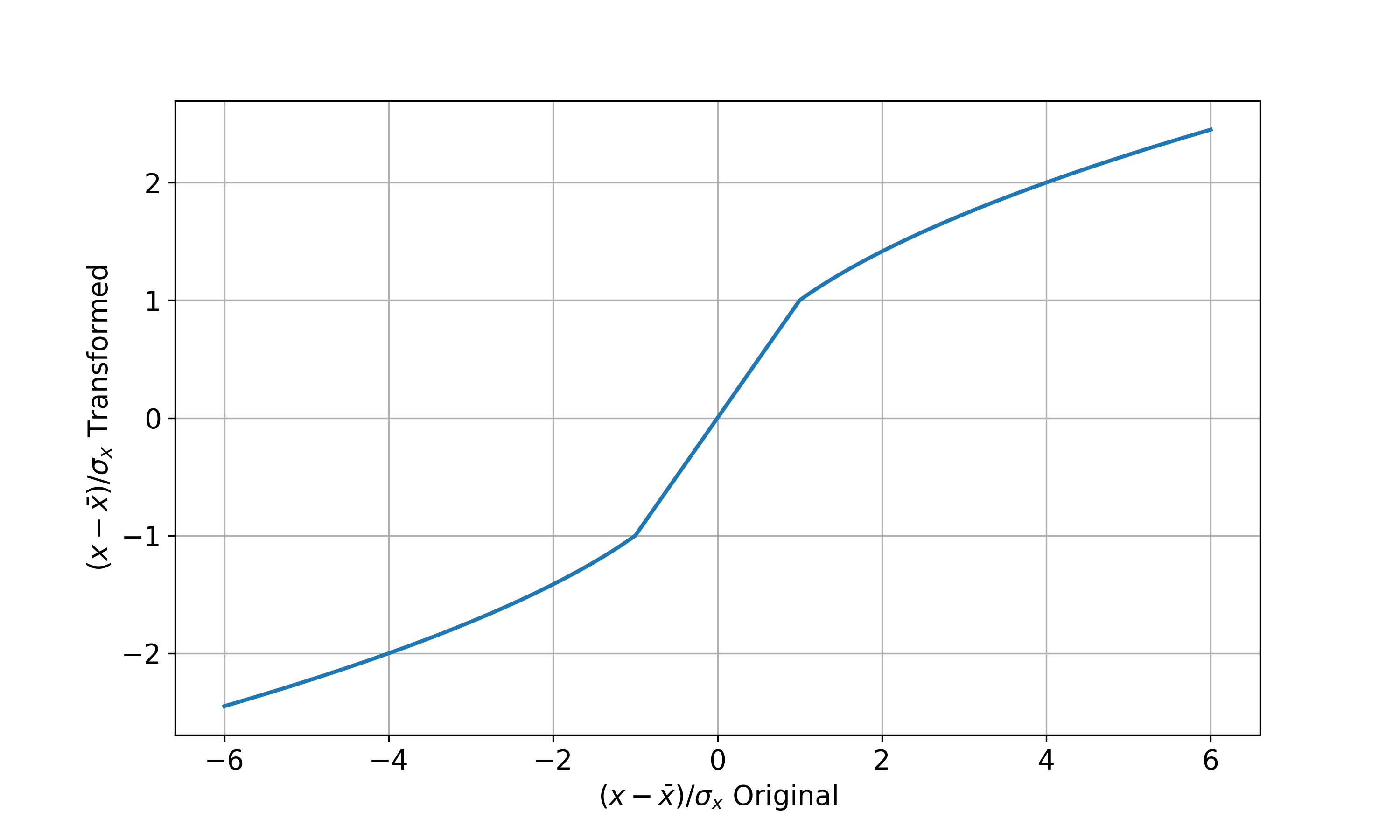}
    \end{center}
    \caption{Transformation defined in~\cref{alg:transformation} using \( \tau = 1 \) and \( \alpha = 0.5 \), for the purpose of reducing large gaps in the discretized data set by scaling outliers above \( \tau \) standard deviations.}
    \label{fig:data_transformation}
\end{figure}

In practice, we take \( \tau = 1 \) and \( \alpha = 0.5 \), thus the standardized data points above one standard deviation are mapped to their square roots, as illustrated in~\cref{fig:data_transformation}.
We tested a few other combinations of \( \tau \) and \( \alpha \), but found these values to produce the best model results out of those we tried; of course this could likely be further optimized.
The effect this transformation has on the model results versus the base dataset can be seen in~\cref{sec:rbm_results}.
This transformation is invertible when \( \bar{x} \), \( \sigma_x \), and \( \delta \) are saved.

\begin{algorithm}
\caption{Outlier Power Transformation}
\begin{algorithmic}[1]
    \Procedure{Transform}{$\vec{x}, \alpha, \tau$}
            \Comment $\alpha$ is the power, $\tau$ is the threshold
        \State $N \gets \text{length}(\vec{x})$
        \State $\bar{x} \gets \frac{1}{N} \sum_{i=1}^{N} x_i$
        \State $\sigma_{x} \gets \sqrt{\frac{1}{N} \sum_{i=1}^{N} (x_i - \bar{x})^2}$
        \State $\delta \gets \tau - \tau^\alpha$
            \Comment ensures the transformation is bijective
        \For {$i$ in 1 to $N$}
            \State $x_i \gets (x_i - \bar{x}) / \sigma_x$
                \Comment standardize
            \If {$x_i > \tau$}
                \State $x_i \gets (\abs{x_i}^\alpha + \delta) \cdot \text{sign}(x_i)$
                    \Comment scale standardized values beyond $\tau$
            \EndIf
            \State $x_i \gets x_i \cdot \sigma_x + \bar{x}$
                \Comment undo standardization
        \EndFor
    \EndProcedure
\end{algorithmic}
\label{alg:transformation}
\end{algorithm}

\begin{figure}[!htb]
    \begin{center}
        \includegraphics[width=1\linewidth]{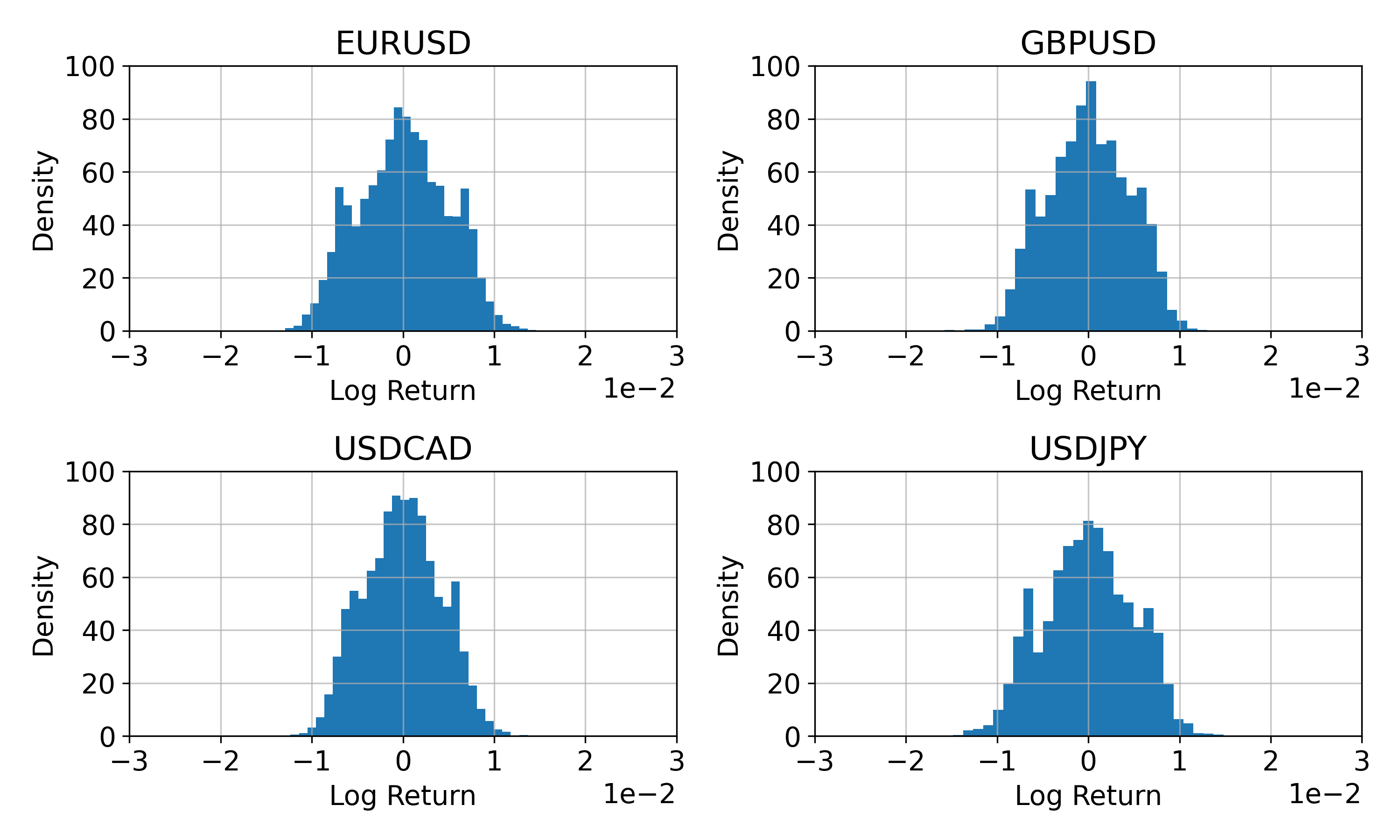}
    \end{center}
    \caption{Histograms of the outlier power-transformed log returns data set.}
    \label{fig:histograms_transformed}
\end{figure}

\begin{figure}[!htb]
    \begin{center}
        \includegraphics[width=1\linewidth]{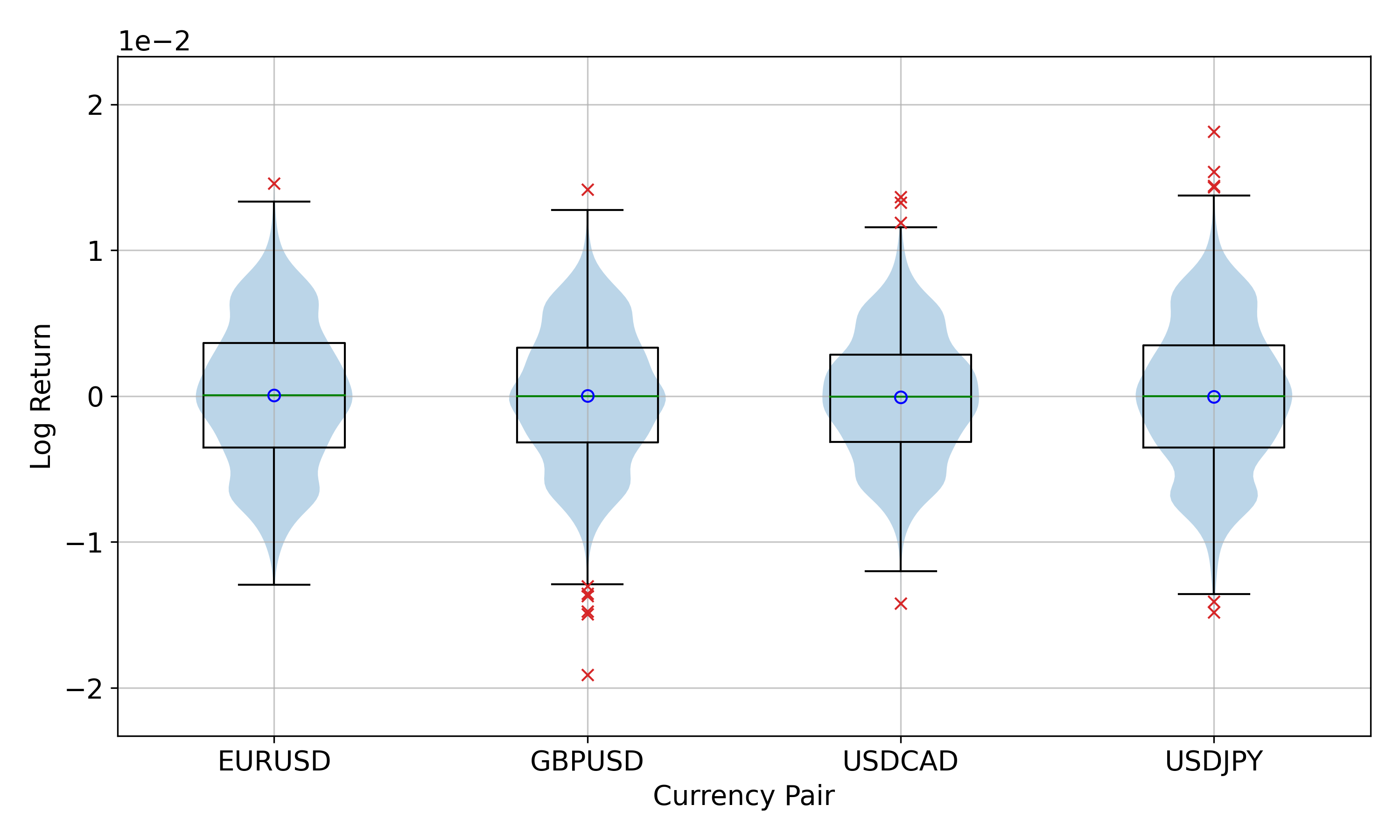}
    \end{center}
    \caption{Violin and box plot of the outlier power-transformed log returns data set illustrating the distribution of the rescaled outliers.}
    \label{fig:violin_transformed}
\end{figure}

Histograms of the transformed data set are shown in~\cref{fig:histograms_transformed}, and a violin and box plot is shown in~\cref{fig:violin_transformed}.
In these, we observe the appearance of "shoulders" around the threshold \( \tau = 1 \) standard deviation, and that the transformed outliers appear much less extreme, allowing us to better utilize the full range of discrete values.
\cref{tbl:data_log_returns_transformed_stats} shows that the transformation reduces the standard deviations to roughly \( 78\% \) of their originals values given in~\cref{tbl:data_log_returns_raw_stats}.

\begin{table}[!htb]
    \centering
    \begin{adjustbox}{max width=\textwidth}
        \begin{tabular}{l r r}
\multicolumn{3}{c}{\textbf{Transformed Log Returns Data Set Statistics}} \\
\toprule
Currency Pair & Mean & Standard Deviation \\
\midrule
EURUSD & \( 5.54 \cdot 10^{-5} \) & \( 4.88 \cdot 10^{-3} \) \\
GBPUSD & \( 1.66 \cdot 10^{-5} \) & \( 4.48 \cdot 10^{-3} \) \\
USDCAD & \( -6.42 \cdot 10^{-5} \) & \( 4.21 \cdot 10^{-3} \) \\
USDJPY & \( -4.68 \cdot 10^{-5} \) & \( 4.93 \cdot 10^{-3} \) \\
\bottomrule
\end{tabular}

    \end{adjustbox}
    \caption{Statistics of the outlier power-transformed log returns data set.}
    \label{tbl:data_log_returns_transformed_stats}
\end{table}

\subsection{Additional Information}
\begin{figure}[!htb]
    \begin{center}
        \includegraphics[width=1\linewidth]{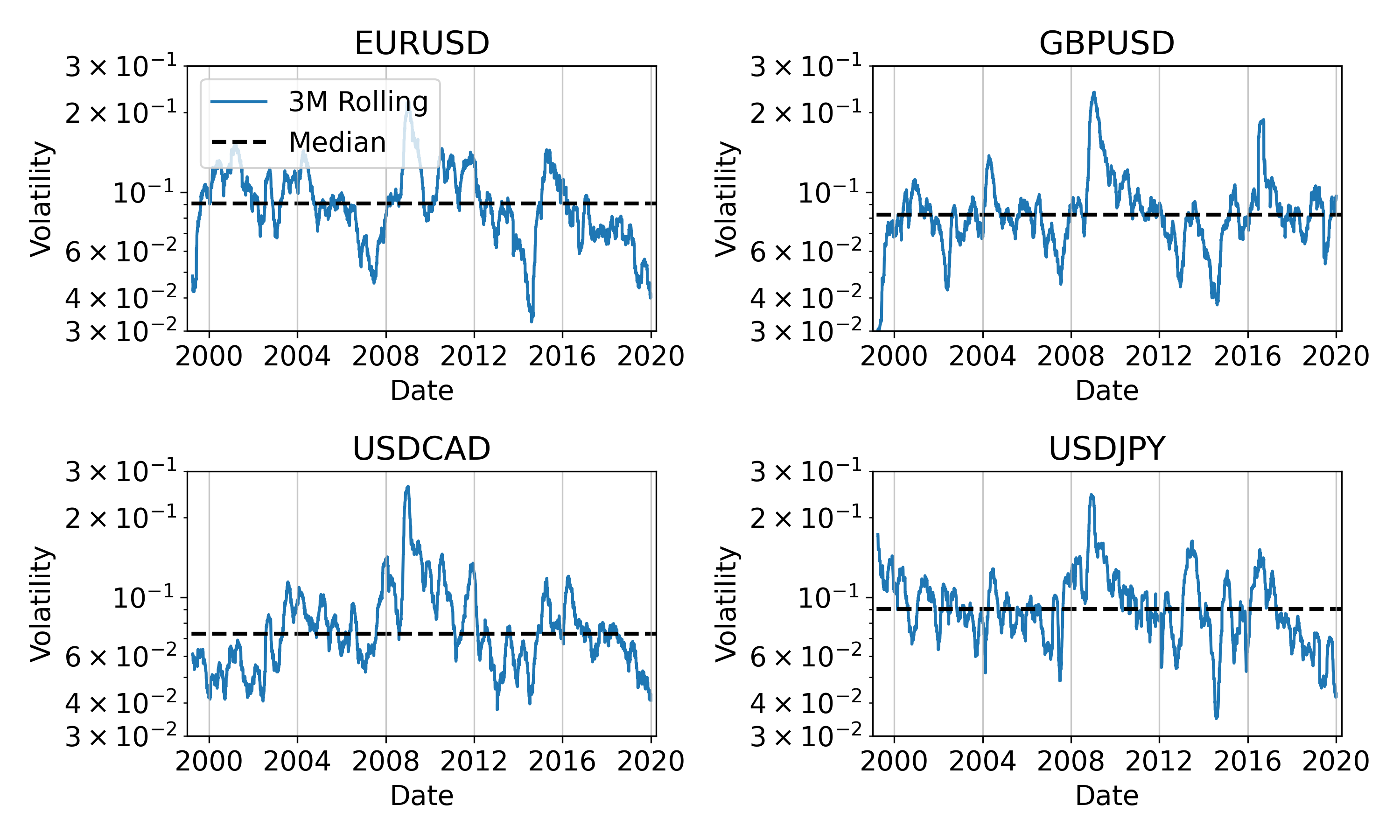}
    \end{center}
    \caption{3-month rolling volatilities of the log returns data set compared with their historical medians.}
    \label{fig:rolling_volatility}
\end{figure}

As mentioned in~\cite{kondratyev_2019}, one can use additional binary indicator variables to enrich the training data set.
One such bit of information is the rolling volatility relative to the historical median (see~\cref{app:annualized_volatility} for definition of annualized volatility).
If the 3-month rolling volatility is below (above) the historical median it is assigned a value of 0 (1) to indicate the low (high) volatility regime.
The 3-month rolling volatilities versus their historical medians are plotted in~\cref{fig:rolling_volatility}.

These additional binary indicator variables are then concatenated onto the training data set and fed to the model to make it more flexible by allowing for the model outputs to be conditioned on a specific volatility regime.
Adding one indicator for each of the four currency pairs increases the number of rows in our training data set by four, thus the volatility-concatenated data set is in the space \( \binset^{68 \times N} \).

\chapter{The Classical Restricted Boltzmann Machine}
\label{ch:rbm}
\section{Theory}
The restricted Boltzmann machine (RBM) is an energy-based model defined by the energy function~\cite{goodfellow_deep_learning}
\begin{align}
\begin{split}
    \label{eq:rbm_energy}
    E(\vec{v}, \vec{h})
        &= -\sum_{i=1}^{n_v} a_i v_i - \sum_{j=1}^{n_h} b_j h_j - \sum_{i=1}^{n_v} \sum_{j=1}^{n_h} v_i w_{ij} h_j \\
        &= -\vec{a}\T\vec{v} - \vec{b}\T\vec{h} - \vec{v}\T\mat{W}\vec{h},
\end{split}
\end{align}
where
\begin{itemize}
    \item \( \vec{v} \in \binset^{n_v} \) represents the visible units, with associated bias vector \( \vec{a} \in \R^{n_v} \).
    \item \( \vec{h} \in \binset^{n_h} \) represents the hidden units, with associated bias vector \( \vec{b} \in \R^{n_h} \).
    \item \( \mat{W} \in \R^{n_v \times n_h} \) represents the weights corresponding to the interaction strengths between visible and hidden units.
\end{itemize}

\begin{figure}[!htb]
    \begin{center}
        \includegraphics[width=1\linewidth]{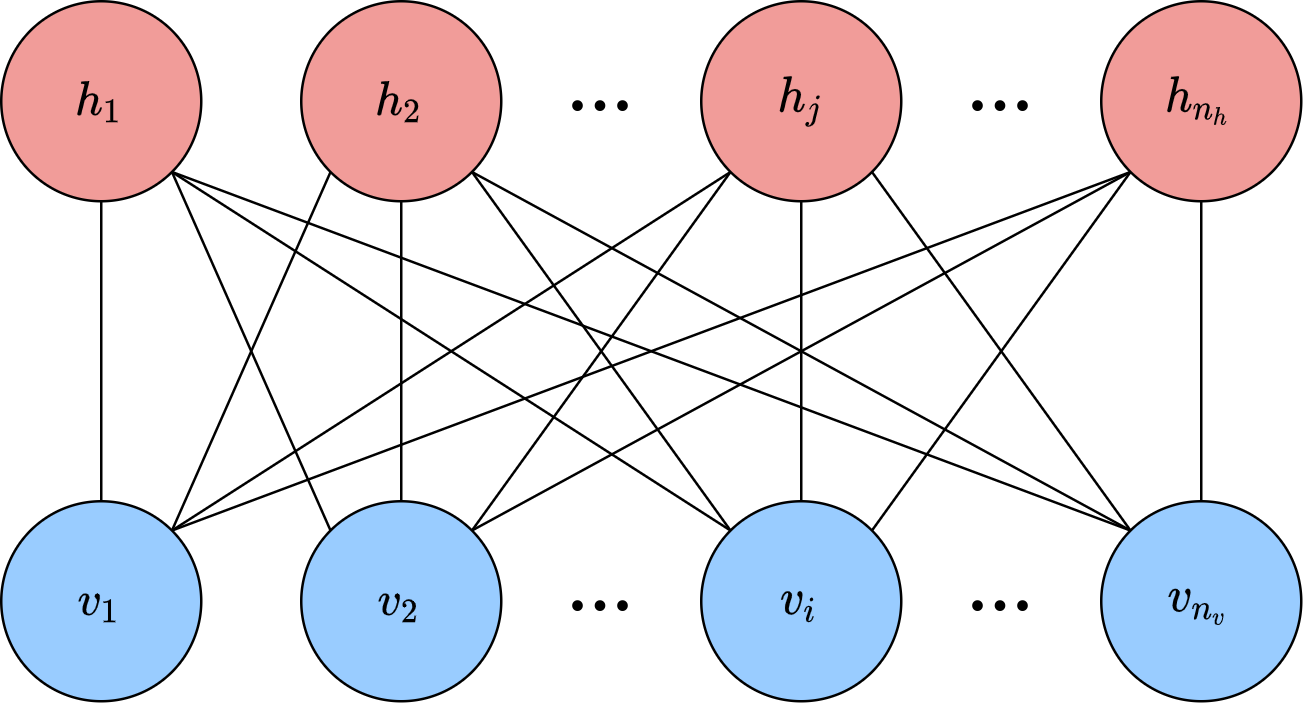}
    \end{center}
    \caption{Diagram of a restricted Boltzmann machine with \( n_v \) visible units and \( n_h \) hidden units.}
    \label{fig:rbm_diagram}
\end{figure}

It is termed \textit{restricted} due to the fact that there are no intralayer connections, i.e., visible units are only connected to hidden units, and vice versa.
An example diagram is depicted in~\cref{fig:rbm_diagram}.

The probability to find the system in the configuration \( (\vec{v},\vec{h}) \) is given by the Boltzmann distribution (with \( \beta = 1/kT = 1 \))
\begin{align}
    \label{eq:rbm_joint_probability}
    p(\vec{v}, \vec{h}) = \frac{1}{Z} e^{-E(\vec{v},\vec{h})},
\end{align}
with intractable~\cite{long_servedio_2010} partition function
\begin{align}
    \label{eq:rbm_partition_function}
    Z = \sum_{\vec{v},\vec{h}} e^{-E(\vec{v},\vec{h})},
\end{align}
where \( \sum_{\vec{v},\vec{h}} \) denotes the sum over all possible configurations of \( \vec{v} \) and \( \vec{h} \).

The imposed restrictions on intralayer connections enable us to write the conditional probabilities of the layers as the product of the individual units' probabilities~\footnote{Here \( \sigma(x) \) is the element-wise logistic sigmoid function and \( \odot \) denotes element-wise multiplication.} (see \cref{app:conditional_probabilities_derivation} for derivation)
\begin{align}
\begin{split}
    p(\vec{h} | \vec{v})
        &= \prod_{j=1}^{n_h} \sigma\big( (2\vec{h} - 1) \odot (\vec{b} + \mat{W}\T\vec{v}) \big)_j, \\
    p(\vec{v} | \vec{h})
        &= \prod_{i=1}^{n_v} \sigma\big( (2\vec{v} - 1) \odot (\vec{a} + \mat{W}\vec{h}) \big)_i.
\end{split}
\end{align}

\subsection{Optimizing an RBM}
Due to the intractability of the partition function, the model cannot be solved exactly in general, thus we resort to other methods to optimize it such as likelihood maximization via gradient descent.
For data set distribution \( p_\text{data} \) and parameters \( \theta = (\mat{W}, \vec{a}, \vec{b}) \), the log-likelihood is given by
\begin{align}
\begin{split}
    \ell(\theta)
        &= \sum_{\vec{v}} p_{\text{data}}(\vec{v}) \log p(\vec{v}) \\
        &= \sum_{\vec{v}} p_{\text{data}}(\vec{v}) \log \bigg(\frac{1}{Z} \sum_\vec{h} e^{-E(\vec{v},\vec{h})}\bigg),
\end{split}
\end{align}
with gradients (see \cref{app:rbm_log_likelihood_derivation} for derivation)
\begin{align}
\begin{split}
    \partial_{w_{ij}} \ell(\theta)
        &= \langle v_i h_j \rangle_{\text{data}} - \langle v_i h_j \rangle_{\text{model}}, \\
    \partial_{a_i} \ell(\theta)
        &= \langle v_i \rangle_{\text{data}} - \langle v_i \rangle_{\text{model}}, \\
    \partial_{b_j} \ell(\theta)
        &= \langle h_j \rangle_{\text{data}} - \langle h_j \rangle_{\text{model}}.
\end{split}
\end{align}

The part of the gradient under the data set distribution is referred to as the \textit{positive} phase, and the part under the model distribution is referred to as the \textit{negative} phase.
It is trivial to compute the expectation values in the positive phase, but not so much in the negative phase because \( p(\vec{v}) \) cannot be sampled directly.

\begin{figure}
    \begin{center}
        \includegraphics[width=1\linewidth]{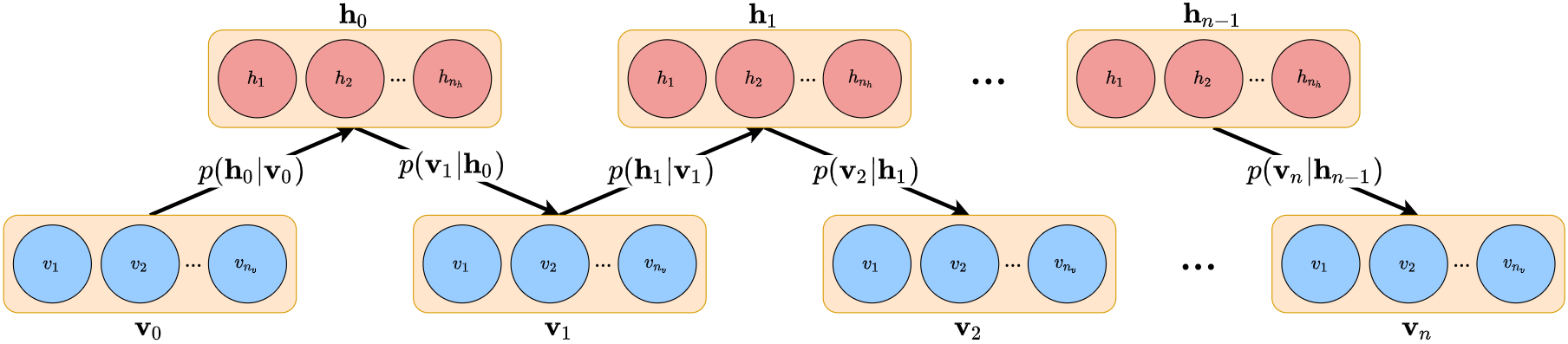}
    \end{center}
    \caption{Illustration of the \( n \)-step Gibbs sampling procedure.}
    \label{fig:gibbs_sampling_diagram}
\end{figure}

In practice the negative phase expectation values are sampled using a Markov chain Monte Carlo (MCMC) method.
This is done via Gibbs sampling~\cite{hinton_rbm_training}, which uses the conditional probabilities \( p(\vec{h}|\vec{v}) \) and \( p(\vec{v}|\vec{h}) \).
One starts with a visible vector and then samples the hidden units conditioned on the visible units, followed by sampling the visible units conditioned on the hidden units, and so forth until the desired thermalization threshold is reached.
The number of steps required to reach thermalization is model dependent and can be estimated by analyzing the autocorrelations of a sample chain generated by the model.
The algorithm for Gibbs sampling is given in \cref{alg:Gibbs} and illustrated in \cref{fig:gibbs_sampling_diagram}.
The algorithm is presented in a vectorized format for brevity.

\begin{algorithm}
\caption{Gibbs Sampling}
\begin{algorithmic}[1]
    \Procedure{Gibbs}{$\vec{v},n,\mat{W},\vec{a},\vec{b}$}
        \State $n_v \gets$ length$(\vec{a})$
        \State $n_h \gets$ length$(\vec{b})$
        \For{$k$ in 1 to $n$}
            \State $\vec{r} \sim$ Uniform$(0, 1, n_h)$
            \State $\vec{h} \gets \vec{r} < \sigma(\vec{b} + \mat{W}\T\vec{v})$
                \Comment $\sigma, <$ applied element-wise
            \State $\vec{r} \sim$ Uniform$(0, 1, n_v)$
            \State $\vec{v} \gets \vec{r} < \sigma(\vec{a} + \mat{W}\vec{h})$
                \Comment $\sigma, <$ applied element-wise
        \EndFor
        \State \Return $\vec{v}$
    \EndProcedure
\end{algorithmic}
\label{alg:Gibbs}
\end{algorithm}
The Uniform$(a, b, n)$ function in \cref{alg:Gibbs} produces a length \( n \) vector of uniform i.i.d. random variables on the interval $[a, b)$, and the \( < \) operator acts element-wise with \( (\text{true}, \text{false}) \mapsto (1, 0) \).

The standard procedure for training an RBM is called \( n \)-step contrastive divergence (CD-\( n \)), with \( n \) often taken to be one in practice~\cite{hinton_rbm_training}.
The algorithm is detailed in \cref{alg:CDn}, where one can see that \( n \) corresponds to how many Gibbs sampling steps are between the positive and negative phase gradients.
Applying the algorithm to a mini-batch is essentially the same except that one divides the learning rate by the size of the mini-batch to get a mini-batch averaged gradient.

\begin{algorithm}
\caption{$n$-Step Contrastive Divergence (CD-$n$)}
\begin{algorithmic}[1]
    \Procedure{CD}{$\vec{v}_+,n,\mat{W},\vec{a},\vec{b},\eta$}
        \Comment $\vec{v}_+$ is a training sample
        \State $\vec{h}_+ \gets \sigma(\vec{b} + \mat{W}\T\vec{v}_+)$
            \Comment $\sigma$ applied element-wise
        \State $\vec{v}_- \gets$ Gibbs$(\vec{v}_+,n,\mat{W},\vec{a},\vec{b})$
        \State $\vec{h}_- \gets \sigma(\vec{b} + \mat{W}\T\vec{v}_-)$
            \Comment $\sigma$ applied element-wise
        \State $\mat{W} \gets \mat{W} + \eta(\vec{v}_+ \vec{h}_+\T - \vec{v}_- \vec{h}_-\T)$
        \State $\vec{a} \gets \vec{a} + \eta(\vec{v}_+ - \vec{v}_-)$
        \State $\vec{b} \gets \vec{b} + \eta(\vec{h}_+ - \vec{h}_-)$
        \State \Return $\mat{W}, \vec{a}, \vec{b}$
    \EndProcedure
\end{algorithmic}
\label{alg:CDn}
\end{algorithm}

\section{The Classical Market Generator}\label{sec:classical_market_generator}
In \textit{The Market Generator}~\cite{kondratyev_2019} by Kondratyev and Schwarz, they show how an RBM can be used as a generative model to produce synthetic market data.
Specifically, they study how it performs on the log returns of forex data for the same currency pairs we use here for the time period 1999-2019.
In this section we use some of the same metrics, as well as a couple additional ones, so that we can verify our models achieve similar performance to theirs, as well as give us a good reference point to compare our quantum models within~\cref{ch:qbm}.

\subsection{Models}
We train and analyze four RBM models using variations of the filtered data set from~\cref{ch:data_analysis}, each with slightly different preprocessing procedures denoted by:
\begin{itemize}
    \item (B): base data set.
    \item (X): base data set transformed using~\cref{alg:transformation}.
    \item (V): base data set with additional volatility indicators.
    \item (XV): base data set transformed using~\cref{alg:transformation} with additional volatility indicators.
\end{itemize}
The models here have 64 (68 for ones with volatility indicators) visible units and 30 hidden units (the same as in~\cite{kondratyev_2019}) to act as regularized autoencoders.
We use a mini-batch size of 10, and an initial learning rate of \( 10^{-3} \) that decays by a factor of half every 1000 epochs after epoch 5000 as defined in~\cref{app:lr_exp_decay}, for a total of \( 10^4 \) epochs.
We base the models on a modified version of scikit-learn's~\cite{python_sklearn} BernoulliRBM class, which we forked~\footnote{https://github.com/cameronperot/scikit-learn/} to implement the ability to use a learning rate schedule with the BernoulliRBM class.

One of the drawbacks of the RBM is that it is not easy to track the training progress for our use case, as the pseudolikelihood metric implemented by the scikit-learn package is not necessarily a good proxy for our models' performances.
The Kullback-Leibler (KL) divergence of \( \pmodel \) from \( \pdata \), denoted \( \DKL{\pdata}{\pmodel} \), is a suitable quantity to track model performance as it measures the information loss associated with using the model distribution \( \pmodel \) to approximate the data set distribution \( \pdata \) (more information in~\cref{app:kl_divergence}).
However, due to the high number of epochs and the thermalization requirements of samples generated by the RBM, this is not very feasible because generating samples to compute the KL divergence every epoch significantly increases model training times.
Therefore, we only present the final results of the models.

\subsection{Results}\label{sec:rbm_results}
\subsubsection{Autocorrelations}
As mentioned before, the classical RBM sampling method is based on an MCMC algorithm, and thus samples produced via this method are autocorrelated.
Therefore, we first examine the autocorrelations to see how dependent samples are on the previous, so that we can get an idea of how many Gibbs steps are needed between samples to consider them statistically independent.
We use Gibbs sample chains of length \( 10^8 \) for this analysis.
More information about the autocorrelation function and time can be found in~\cref{app:autocorrelation_analysis}.

\begin{figure}[!htb]
    \begin{center}
        \includegraphics[width=1\linewidth]{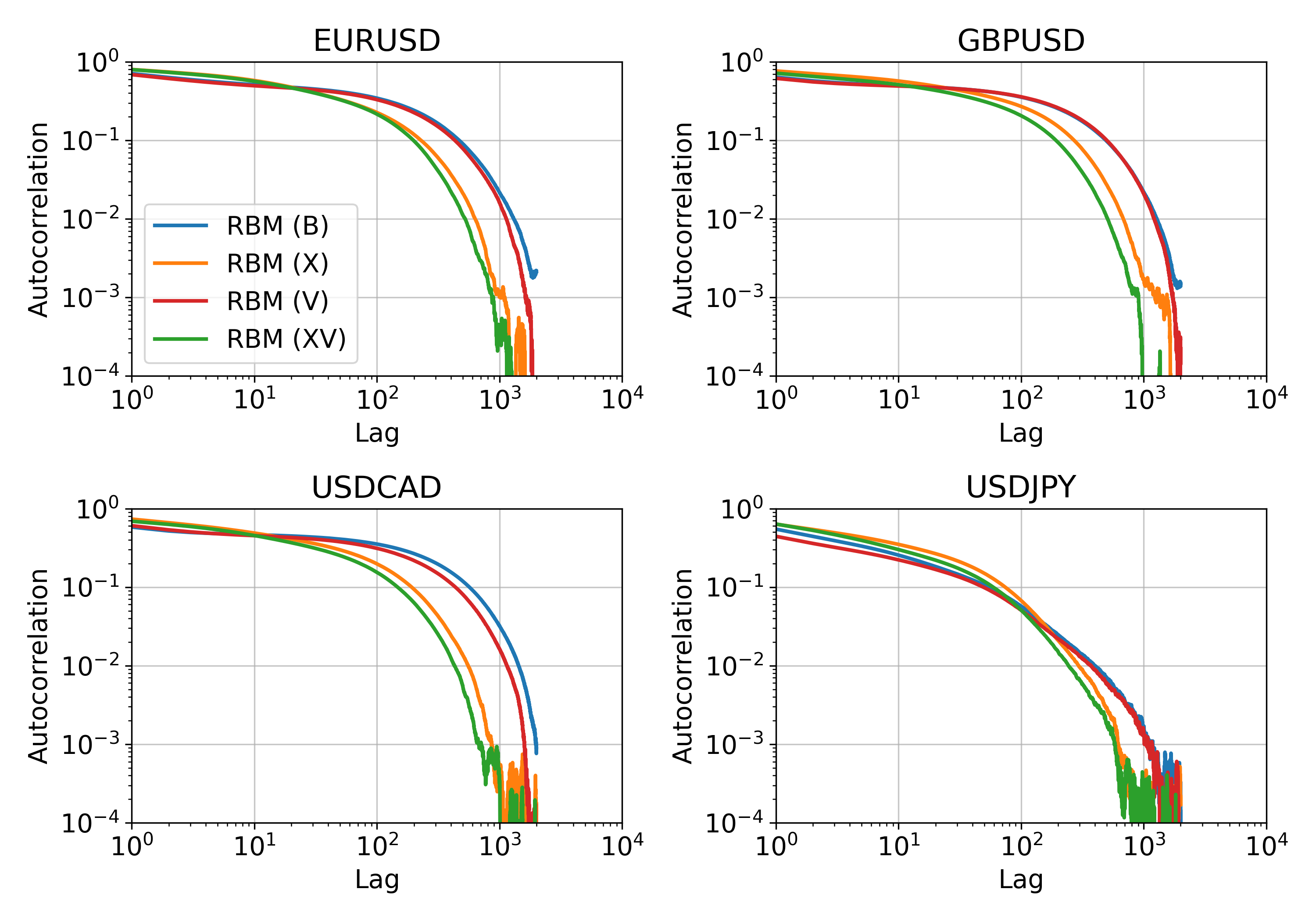}
    \end{center}
    \caption{
        Autocorrelation functions of the RBM models.
    }
    \label{fig:rbm_autocorrelation_functions}
\end{figure}

\cref{fig:rbm_autocorrelation_functions} shows the autocorrelation functions for the various models and currency pairs.
It is immediately clear that the autocorrelations fall off much sooner for the models trained on the transformed data sets for all currency pairs.
This observation is confirmed by examining the integrated autocorrelation times in~\cref{tbl:rbm_ac_times}.

It is not immediately clear why the transformed data sets lead to such shorter integrated autocorrelation times, but this is a welcome trend as it means that less sampling steps are required to reach thermalization.
\begin{table}[!htb]
    \centering
    \begin{adjustbox}{max width=\textwidth}
        \begin{tabular}{l r r r r}
\multicolumn{5}{c}{\textbf{Integrated Autocorrelation Times}} \\
\toprule
Currency Pair & \textbf{RBM (B)} & \textbf{RBM (X)} & \textbf{RBM (V)} & \textbf{RBM (XV)} \\
\midrule
EURUSD & 295.7 & 147.5 & 267.4 & 129.2 \\
GBPUSD & 307.0 & 173.2 & 308.9 & 121.6 \\
USDCAD & 340.6 & 120.0 & 258.8 & 91.3 \\
USDJPY & 33.9 & 46.7 & 28.8 & 36.7 \\
\bottomrule
\end{tabular}

    \end{adjustbox}
    \caption{
        Integrated autocorrelation times of the RBM models.
    }
    \label{tbl:rbm_ac_times}
\end{table}

The results in the rest of this section are derived from an ensemble of 100 sample sets consisting of \( 10^4 \) samples each, and \( 10^4 \) Gibbs sampling steps between samples to ensure thermalization.

\subsubsection{Marginal Distributions}
\begin{figure}[!htb]
    \begin{center}
        \includegraphics[width=1\linewidth]{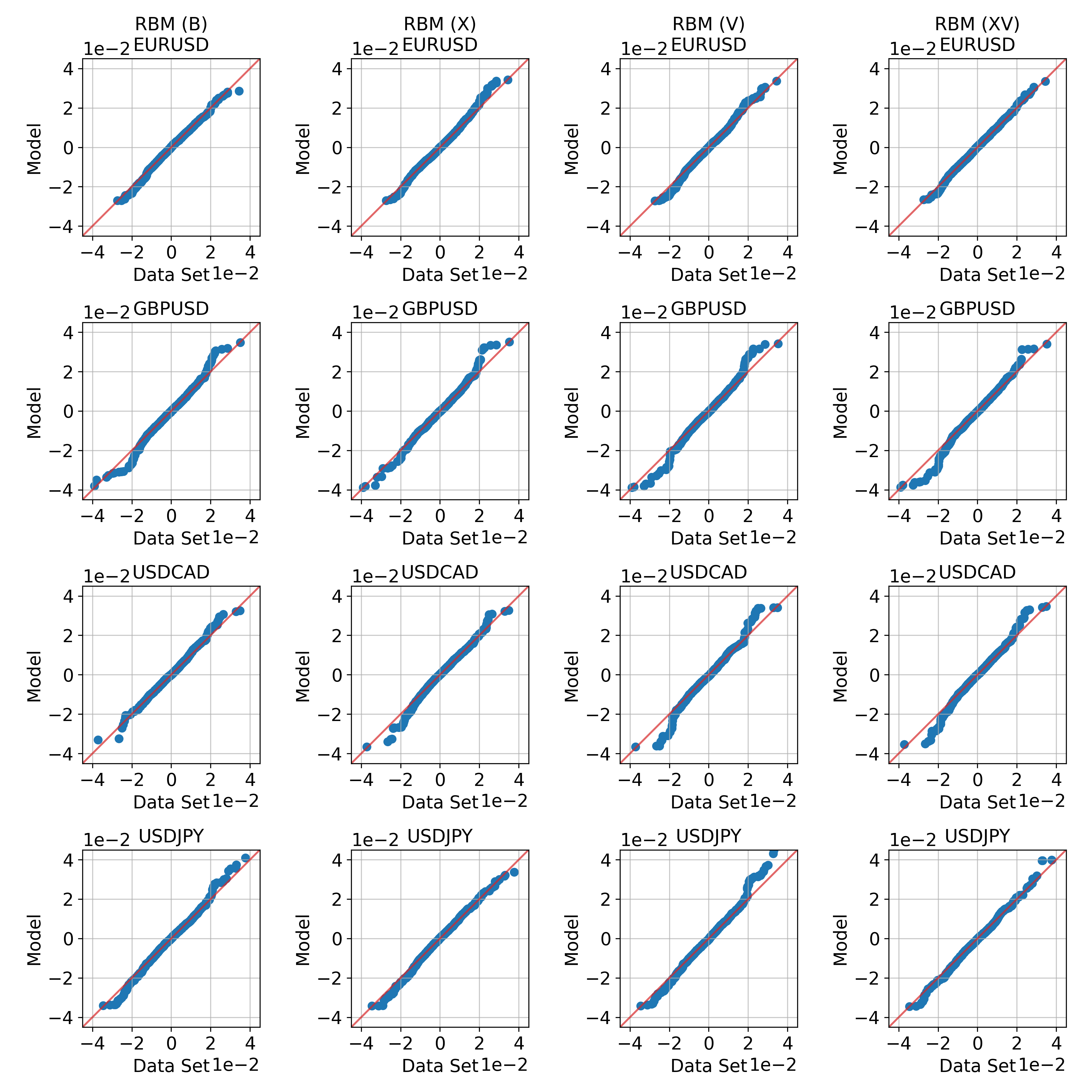}
    \end{center}
    \caption{Log return Q-Q plots of the RBM models for each currency pair. Note that these plots only use the same number of samples as the size of the training data set (5165), and thus are not entirely representative of the models' performances.}
    \label{fig:rbm_qq_plots}
\end{figure}

To get an idea of how well the models perform, we examine the KL divergences of the marginal distributions of each currency pair in~\cref{tbl:rbm_KL_divergences}.
Here we observe that all models reproduce the marginal distributions quite well, but the models trained on the transformed data sets perform slightly better, particularly on the USDCAD marginal.
The performance of the models on the marginal distributions is also visualized with Q-Q plots in~\cref{fig:rbm_qq_plots}.
More information on how the KL divergences are computed can be found in~\cref{app:kl_divergence_in_practice}.
\begin{table}[!htb]
    \centering
    \begin{adjustbox}{max width=\textwidth}
        \begin{tabular}{l r r r r}
\multicolumn{5}{c}{\(D_{\text{KL}}(p_\text{data} \ || \ p_\text{model})\)} \\
\toprule
Currency Pair & \textbf{RBM (B)} & \textbf{RBM (X)} & \textbf{RBM (V)} & \textbf{RBM (XV)} \\
\midrule
EURUSD & 0.010 $\pm$ 0.001 & 0.007 $\pm$ 0.001 & 0.011 $\pm$ 0.002 & 0.009 $\pm$ 0.001 \\
GBPUSD & 0.007 $\pm$ 0.001 & 0.006 $\pm$ 0.001 & 0.011 $\pm$ 0.001 & 0.007 $\pm$ 0.001 \\
USDCAD & 0.017 $\pm$ 0.002 & 0.007 $\pm$ 0.001 & 0.015 $\pm$ 0.002 & 0.008 $\pm$ 0.001 \\
USDJPY & 0.008 $\pm$ 0.001 & 0.007 $\pm$ 0.001 & 0.010 $\pm$ 0.001 & 0.009 $\pm$ 0.001 \\
\midrule
Mean & 0.010 $\pm$ 0.001 & 0.007 $\pm$ 0.001 & 0.011 $\pm$ 0.002 & 0.008 $\pm$ 0.001 \\
\bottomrule
\end{tabular}

    \end{adjustbox}
    \caption{
        KL divergences of the RBM models.
        The values are shown in the format mean \(\pm\) one standard deviation from an ensemble of 100 sample sets consisting of \( 10^4 \) samples each.
    }
    \label{tbl:rbm_KL_divergences}
\end{table}

\subsubsection{Correlations}
The distribution is in a sense more than just the sum of its parts.
Beyond learning the marginal distributions, the models should also capture the correlations between the currency pairs.
To verify this, we turn to the correlation coefficients in~\cref{tbl:rbm_correlation_coefficients} to see how well the models capture the correlations.
We find that the models reproduce the structure of the correlation coefficients reasonably well, with the models trained on the transformed data sets encoding more of the behavior.
\begin{table}[!htb]
    \centering
    \begin{adjustbox}{max width=\textwidth}
        \begin{tabular}{l r r r r r r}\multicolumn{7}{c}{\textbf{Correlation Coefficients}} \\
\toprule
& \multicolumn{3}{c}{\textbf{Data Set}} & \multicolumn{3}{c}{\textbf{RBM (B)}} \\
\cmidrule(lr){2-4}
\cmidrule(lr){5-7}
Currency Pairs & Pearson & Spearman & Kendall & Pearson & Spearman & Kendall \\
\midrule
EURUSD/GBPUSD & 0.62 & 0.62 & 0.44 & 0.48 $\pm$ 0.01 & 0.53 $\pm$ 0.01 & 0.38 $\pm$ 0.01 \\
EURUSD/USDCAD & -0.44 & -0.41 & -0.29 & -0.33 $\pm$ 0.01 & -0.34 $\pm$ 0.01 & -0.24 $\pm$ 0.01 \\
EURUSD/USDJPY & -0.26 & -0.30 & -0.21 & -0.21 $\pm$ 0.01 & -0.25 $\pm$ 0.01 & -0.17 $\pm$ 0.01 \\
GBPUSD/USDCAD & -0.42 & -0.37 & -0.26 & -0.31 $\pm$ 0.01 & -0.33 $\pm$ 0.01 & -0.22 $\pm$ 0.01 \\
GBPUSD/USDJPY & -0.14 & -0.21 & -0.15 & -0.15 $\pm$ 0.01 & -0.18 $\pm$ 0.01 & -0.13 $\pm$ 0.01 \\
USDCAD/USDJPY & 0.00 & 0.06 & 0.04 & 0.06 $\pm$ 0.01 & 0.07 $\pm$ 0.01 & 0.05 $\pm$ 0.01 \\
\midrule
& \multicolumn{3}{c}{\textbf{RBM (X)}} & \multicolumn{3}{c}{\textbf{RBM (V)}} \\
\cmidrule(lr){2-4}
\cmidrule(lr){5-7}
Currency Pairs & Pearson & Spearman & Kendall & Pearson & Spearman & Kendall \\
\midrule
EURUSD/GBPUSD & 0.56 $\pm$ 0.01 & 0.59 $\pm$ 0.01 & 0.42 $\pm$ 0.01 & 0.48 $\pm$ 0.01 & 0.54 $\pm$ 0.01 & 0.38 $\pm$ 0.01 \\
EURUSD/USDCAD & -0.39 $\pm$ 0.01 & -0.39 $\pm$ 0.01 & -0.27 $\pm$ 0.01 & -0.34 $\pm$ 0.01 & -0.36 $\pm$ 0.01 & -0.25 $\pm$ 0.01 \\
EURUSD/USDJPY & -0.24 $\pm$ 0.01 & -0.29 $\pm$ 0.01 & -0.19 $\pm$ 0.01 & -0.20 $\pm$ 0.01 & -0.23 $\pm$ 0.01 & -0.16 $\pm$ 0.01 \\
GBPUSD/USDCAD & -0.36 $\pm$ 0.01 & -0.35 $\pm$ 0.01 & -0.24 $\pm$ 0.01 & -0.30 $\pm$ 0.01 & -0.33 $\pm$ 0.01 & -0.22 $\pm$ 0.01 \\
GBPUSD/USDJPY & -0.16 $\pm$ 0.01 & -0.20 $\pm$ 0.01 & -0.13 $\pm$ 0.01 & -0.14 $\pm$ 0.01 & -0.17 $\pm$ 0.01 & -0.12 $\pm$ 0.01 \\
USDCAD/USDJPY & 0.05 $\pm$ 0.01 & 0.07 $\pm$ 0.01 & 0.05 $\pm$ 0.01 & 0.05 $\pm$ 0.01 & 0.07 $\pm$ 0.01 & 0.05 $\pm$ 0.01 \\
\midrule
& \multicolumn{3}{c}{\textbf{RBM (XV)}} & \\
\cmidrule(lr){2-4}
Currency Pairs & Pearson & Spearman & Kendall & & & \\
\midrule
EURUSD/GBPUSD & 0.54 $\pm$ 0.01 & 0.59 $\pm$ 0.01 & 0.42 $\pm$ 0.01 &  &  &  \\
EURUSD/USDCAD & -0.39 $\pm$ 0.01 & -0.38 $\pm$ 0.01 & -0.26 $\pm$ 0.01 &  &  &  \\
EURUSD/USDJPY & -0.22 $\pm$ 0.01 & -0.27 $\pm$ 0.01 & -0.19 $\pm$ 0.01 &  &  &  \\
GBPUSD/USDCAD & -0.36 $\pm$ 0.01 & -0.36 $\pm$ 0.01 & -0.24 $\pm$ 0.01 &  &  &  \\
GBPUSD/USDJPY & -0.16 $\pm$ 0.01 & -0.20 $\pm$ 0.01 & -0.13 $\pm$ 0.01 &  &  &  \\
USDCAD/USDJPY & 0.05 $\pm$ 0.01 & 0.07 $\pm$ 0.01 & 0.05 $\pm$ 0.01 &  &  &  \\
\bottomrule
\end{tabular}

    \end{adjustbox}
    \caption{Correlation coefficients of the data set vs.~samples generated by the RBM models. The RBM values are shown in the format mean \(\pm\) one standard deviation from an ensemble of 100 sample sets consisting of \( 10^4 \) samples each.}
    \label{tbl:rbm_correlation_coefficients}
\end{table}

\subsubsection{Volatilities}
Examining the historical volatilities in~\cref{tbl:rbm_volatilities} confirms the models can produce synthetic data with similar volatilities to the training data set, albeit marginally higher in all cases.
\begin{table}[!htb]
    \centering
    \begin{adjustbox}{max width=\textwidth}
        \begin{tabular}{l r r r r r}
\multicolumn{6}{c}{\textbf{Historical Volatilities}} \\
\toprule
Currency Pair & \textbf{Data Set} & \textbf{RBM (B)} & \textbf{RBM (X)} & \textbf{RBM (V)} & \textbf{RBM (XV)} \\
\midrule
EURUSD & 9.78\% & 9.98\% $\pm$ 0.11\% & 10.10\% $\pm$ 0.10\% & 10.18\% $\pm$ 0.11\% & 10.27\% $\pm$ 0.10\% \\
GBPUSD & 8.98\% & 9.34\% $\pm$ 0.11\% & 9.38\% $\pm$ 0.12\% & 9.55\% $\pm$ 0.12\% & 9.53\% $\pm$ 0.11\% \\
USDCAD & 8.56\% & 8.98\% $\pm$ 0.13\% & 9.01\% $\pm$ 0.11\% & 9.29\% $\pm$ 0.13\% & 9.12\% $\pm$ 0.12\% \\
USDJPY & 10.02\% & 10.26\% $\pm$ 0.13\% & 10.42\% $\pm$ 0.14\% & 10.82\% $\pm$ 0.16\% & 10.46\% $\pm$ 0.12\% \\
\bottomrule
\end{tabular}

    \end{adjustbox}
    \caption{Historical volatilities of the data set vs.~samples generated by the RBM models. The RBM values are shown in the format mean \(\pm\) one standard deviation from an ensemble of 100 sample sets consisting of \( 10^4 \) samples each.}
    \label{tbl:rbm_volatilities}
\end{table}

\subsubsection{Tails}
It is extremely important for the models to learn the tail events because these play a crucial role in financial risk management.
The models trained on the transformed data sets reproduce the lower tails a little better for most currency pairs, but overestimate some of the upper tails.
It is difficult to say overall if one model performs better than another here, as it really depends on what one wants to do with the generated data.
\begin{table}[!htb]
    \centering
    \begin{adjustbox}{max width=\textwidth}
        \begin{tabular}{l r r r r r}
\multicolumn{6}{c}{\textbf{Lower Tails (1st Percentile)}} \\
\toprule
Currency Pair & \textbf{Data Set} & \textbf{RBM (B)} & \textbf{RBM (X)} & \textbf{RBM (V)} & \textbf{RBM (XV)} \\
\midrule
EURUSD & -1.64\% & -1.80\% $\pm$ 0.04\% & -1.68\% $\pm$ 0.05\% & -1.90\% $\pm$ 0.05\% & -1.76\% $\pm$ 0.06\% \\
GBPUSD & -1.47\% & -1.59\% $\pm$ 0.04\% & -1.57\% $\pm$ 0.05\% & -1.63\% $\pm$ 0.05\% & -1.66\% $\pm$ 0.06\% \\
USDCAD & -1.40\% & -1.54\% $\pm$ 0.05\% & -1.57\% $\pm$ 0.06\% & -1.59\% $\pm$ 0.05\% & -1.58\% $\pm$ 0.06\% \\
USDJPY & -1.70\% & -2.03\% $\pm$ 0.07\% & -1.92\% $\pm$ 0.06\% & -2.17\% $\pm$ 0.09\% & -1.96\% $\pm$ 0.06\% \\
\bottomrule \\
\multicolumn{6}{c}{\textbf{Upper Tails (99th Percentile)}} \\
\toprule
Currency Pair & \textbf{Data Set} & \textbf{RBM (B)} & \textbf{RBM (X)} & \textbf{RBM (V)} & \textbf{RBM (XV)} \\
\midrule
EURUSD & 1.62\% & 1.59\% $\pm$ 0.04\% & 1.84\% $\pm$ 0.06\% & 1.70\% $\pm$ 0.04\% & 1.81\% $\pm$ 0.05\% \\
GBPUSD & 1.42\% & 1.45\% $\pm$ 0.04\% & 1.54\% $\pm$ 0.04\% & 1.53\% $\pm$ 0.03\% & 1.57\% $\pm$ 0.04\% \\
USDCAD & 1.51\% & 1.61\% $\pm$ 0.04\% & 1.53\% $\pm$ 0.05\% & 1.60\% $\pm$ 0.05\% & 1.56\% $\pm$ 0.05\% \\
USDJPY & 1.59\% & 1.56\% $\pm$ 0.04\% & 1.60\% $\pm$ 0.05\% & 1.61\% $\pm$ 0.04\% & 1.61\% $\pm$ 0.05\% \\
\bottomrule
\end{tabular}

    \end{adjustbox}
    \caption{Lower and upper tails, i.e., 1st and 99th percentiles, of the data set vs.~samples generated by the RBM models. The RBM values are shown in the format mean \(\pm\) one standard deviation from an ensemble of 100 sample sets consisting of \( 10^4 \) samples each.}
    \label{tbl:rbm_tails}
\end{table}

\begin{figure}[!htb]
    \begin{center}
        \includegraphics[width=1\linewidth]{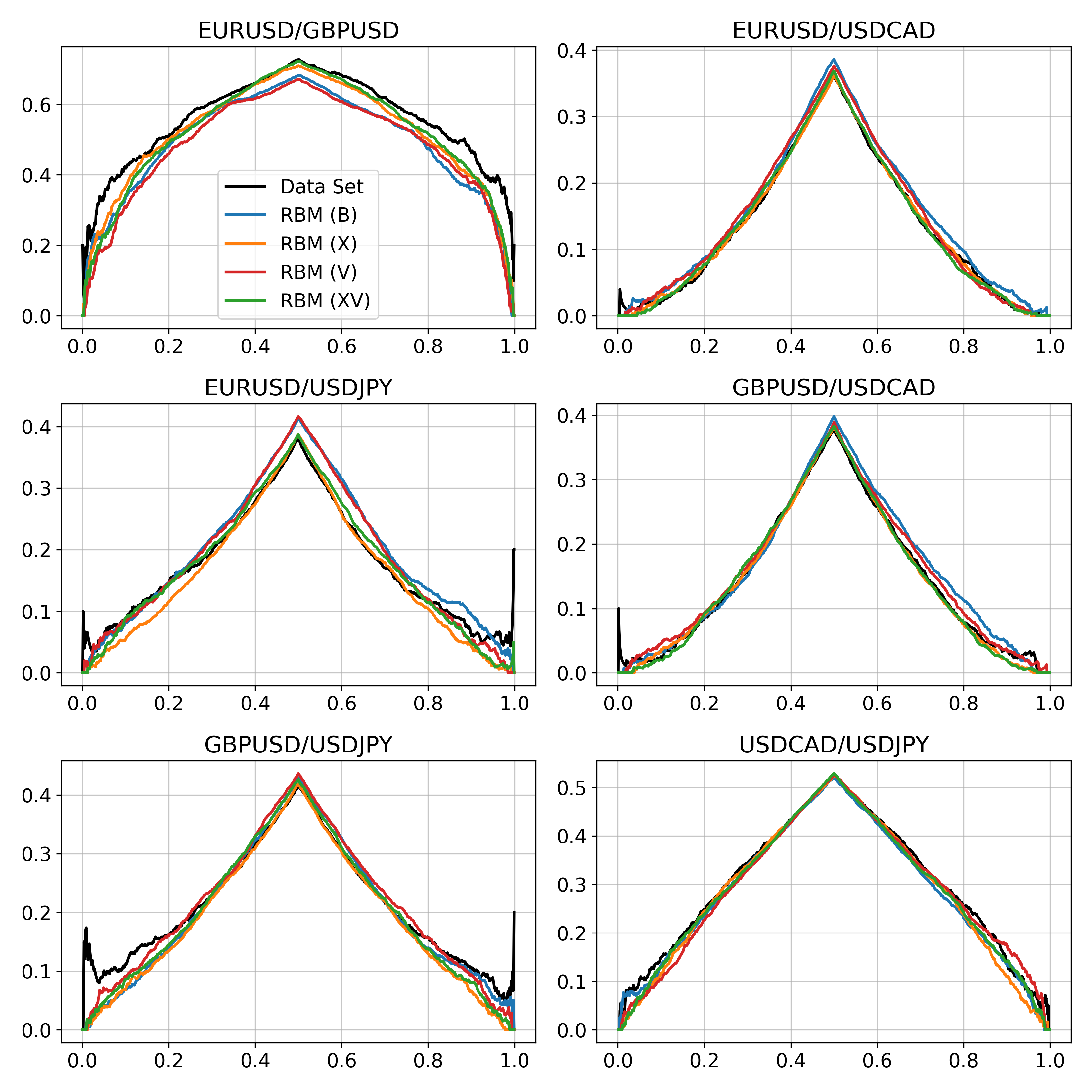}
    \end{center}
    \caption{Tail concentration functions of the data set vs.~samples generated by the RBM models.}
    \label{fig:rbm_tail_concentrations}
\end{figure}

We also study the tail concentration functions (see~\cref{app:tail_concentration_functions} for definitions and interpretations) between currency pairs in~\cref{fig:rbm_tail_concentrations}.
Here we see that all models perform quite well for the most part except for a few of the extreme regions in the EURUSD/GBPUSD, EURUSD/USDJPY, and GBPUSD/USDJPY plots.

\subsubsection{Conditional Sampling}
For the data sets with additional volatility indicators, we have the ability to condition on these indicators to sample from a specific volatility regime.
This is useful, for example, if we are trying to generate real-world data that fits the current volatility landscape.

This leads us to look at the conditional volatilities, i.e., seeing how well the models reproduce the volatilities from the two volatility regimes.
Laid out in~\cref{tbl:rbm_conditional_volatilities}, we observe that the samples produced by the RBMs have slightly lower (higher) volatilities in the high (low) regime, but are overall in good agreement with the data set.
\begin{table}[!htb]
    \centering
    \begin{adjustbox}{max width=\textwidth}
        \begin{tabular}{l r r r r r r}
\multicolumn{7}{c}{\textbf{Conditional Volatilities}} \\
\toprule
& \multicolumn{3}{c}{\textbf{Low Regime}} & \multicolumn{3}{c}{\textbf{High Regime}} \\
\cmidrule(lr){2-4}
\cmidrule(lr){5-7}
Currency Pair & \textbf{Data Set} & \textbf{RBM (V)} & \textbf{RBM (XV)} & \textbf{Data Set} & \textbf{RBM (V)} & \textbf{RBM (XV)} \\
\midrule
EURUSD & 6.72\% & 7.67\% $\pm$ 0.23\% & 7.60\% $\pm$ 0.24\% & 13.04\% & 13.19\% $\pm$ 0.31\% & 12.92\% $\pm$ 0.27\% \\
GBPUSD & 6.67\% & 7.45\% $\pm$ 0.21\% & 7.50\% $\pm$ 0.22\% & 12.69\% & 12.13\% $\pm$ 0.31\% & 11.61\% $\pm$ 0.28\% \\
USDCAD & 6.05\% & 6.72\% $\pm$ 0.22\% & 6.40\% $\pm$ 0.21\% & 12.86\% & 12.53\% $\pm$ 0.37\% & 12.14\% $\pm$ 0.30\% \\
USDJPY & 7.36\% & 9.01\% $\pm$ 0.32\% & 8.76\% $\pm$ 0.27\% & 13.15\% & 12.63\% $\pm$ 0.38\% & 12.41\% $\pm$ 0.31\% \\
\bottomrule
\end{tabular}

    \end{adjustbox}
    \caption{Conditional historical volatilities of the data set vs.~samples generated by the RBM models. The RBM values are shown in the format mean \(\pm\) one standard deviation from an ensemble of 100 sample sets consisting of \( 10^4 \) samples each.}
    \label{tbl:rbm_conditional_volatilities}
\end{table}

\subsection{Summary}
The classical RBM results presented in this section are in line with those obtained by Kondratyev and Schwarz in~\cite{kondratyev_2019}, and the differences can likely be accounted for by the different data sets used in training (e.g., different sources, different filtering, etc.), model hyperparameters, and the stochastic nature of the models.
This further confirms that the RBM is performant and can be used to generate synthetic data from distributions with intricate structures, such as the correlations and volatilities seen here.

Overall it is difficult to say if one of the models performs better than the others, as it depends on the desired use case, but the models trained on the transformed data sets do yield lower KL divergence values and capture more of the correlations between currency pairs.
This offers evidence that the results might be able to be further improved through the use of more advanced data preprocessing methods.
We do not investigate these possibilities any further though, given that this is not the main scope of this thesis.
The results in this section act mainly as a point of reference to compare the quantum models within the next chapter.

\chapter{The Quantum Boltzmann Machine}
\label{ch:qbm}
\section{Theory}
The Quantum Boltzmann Machine detailed here is based on the work in \textit{Quantum Boltzmann Machine} by Amin et al.~\cite{amin_2018}.
In this section we use spin eigenvalues \( +1 \) and \( -1 \) rather than binary values \( 0 \) and \( 1 \), respectively, in order to maintain consistency with the language of quantum mechanics.
We start with the \( n \)-qubit Hamiltonian
\begin{align}
    H = -\sum_{i=1}^{n} \Gamma_i \sigma_i^x -\sum_{i=1}^{n} b_i \sigma_i^z - \sum_{i=1}^{n}\sum_{j=i+1}^{n} w_{ij} \sigma_i^z \sigma_j^z,
\end{align}
where
\begin{align}
\begin{split}
    \sigma_i^x
        &= I^{\otimes i-1} \otimes \sigma_x \otimes I^{\otimes n-i}, \\
    \sigma_i^z
        &= I^{\otimes i-1} \otimes \sigma_z \otimes I^{\otimes n-i},
\end{split}
\end{align}
with \( \sigma_x \) and \( \sigma_z \) being the Pauli \( x \) and \( z \) matrices, and \( I \) being the \( 2 \times 2 \) identity matrix.
We denote the first \( n_v \) qubits as the visible units and the last \( n_h \) qubits as the hidden units, thus we have a total of \( n_v + n_h = n \) qubits.

The system's distribution is modeled by the density matrix
\begin{align}
    \rho = \frac{1}{Z} e^{-H},
\label{eq:density_operator}
\end{align}
where \( e^{-H} = \sum_{n=0}^{\infty} \frac{1}{n!} (-H)^n \) is the matrix exponential, and \( Z = \tr(e^{-H}) \) is the partition function.
The probability to observe the system in state \( \ket{\vec{v},\vec{h}} \) is given by
\begin{align}
    p(\vec{v},\vec{h})
        &= \tr(\ket{\vec{v},\vec{h}}\bra{\vec{v},\vec{h}} \rho),
\end{align}
and if we define the projection operator
\begin{align}
    \Lambda_\vec{v} = \ket{\vec{v}}\bra{\vec{v}} \otimes I^{\otimes n_h},
\end{align}
then the marginal probability to measure the visible units in state \( \ket{\vec{v}} \) is given by
\begin{align}
    p(\vec{v}) = \tr(\Lambda_{\vec{v}}\rho).
\end{align}

Using the probabilities above we can obtain the log-likelihood, which for data set distribution \( p_\text{data} \) and parameters \( \theta = (\mat{W}, \vec{a}, \vec{b}) \) is
\begin{align}
    \ell(\theta) = \sum_{\vec{v}} p_{\text{data}}(\vec{v}) \log \tr(\Lambda_\vec{v}\rho),
\end{align}
where \( \sum_{\vec{v}} \) denotes the sum over all possible configurations of \( \vec{v} \).

\subsection{Optimizing a QBM}
When optimizing a QBM, it is preferable to maximize the lower bound of the log-likelihood rather than maximizing the log-likelihood itself.
The reason for this is that the partial derivative of the log-likelihood with respect to the parameters has a term which is computationally expensive to compute, as discussed in \cref{app:qbm_log_likelihood_derivation}.
The lower bound of the log-likelihood is given by (see \cref{app:qbm_log_likelihood_lower_bound} for derivation)
\begin{align}
    \tilde{\ell}(\theta) = \sum_{\vec{v}} \pdata(\vec{v}) \log \tr(\rho_\vec{v}),
\end{align}
where we have what is referred to as the \textit{clamped} Hamiltonian, which for a given visible vector \( \vec{v} \) is
\begin{align}
    H_\vec{v}
        &= \braket{\vec{v}|H|\vec{v}},
\end{align}
with corresponding clamped density matrix
\begin{align}
    \rho_\vec{v}
        &= \frac{1}{Z_\vec{v}} e^{-H_\vec{v}},
\end{align}
and \( Z_\vec{v} = \tr(e^{-H_\vec{v}}) \).
This is called clamped because the visible qubits are held to the classical state of the visible vector \( \vec{v} \).

The associated derivatives with respect to the parameters of the lower bound are given by (see \cref{app:qbm_log_likelihood_lower_bound_derivative} for derivation)
\begin{align}
\begin{split}
    \partial_{w_{ij}} \tilde{\ell}(\theta)
        &= \langle \sigma_i^z \sigma_j^z \rangle_\text{data} - \langle \sigma_i^z \sigma_j^z \rangle_\text{model}, \\
    \partial_{b_i} \tilde{\ell}(\theta)
        &= \langle \sigma_i^z \rangle_\text{data} - \langle \sigma_i^z \rangle_\text{model},
\end{split}
\end{align}
where \( \langle \ \cdot \ \rangle_\text{data} \) is the expectation value with respect to the data set, and \( \langle \ \cdot \ \rangle_\text{model} \) is the expectation value with respect to the original density matrix.

If connections are restricted within the hidden layer, then the hidden unit probabilities are independent in the positive phase and can be computed easily, as shown in \cref{app:qbm_log_likelihood_lower_bound_derivative}.
This leads to positive phase expectation values of
\begin{align}
\begin{split}
    \langle \sigma_i^z \rangle_\text{data}
        &= \sum_\vec{v} \pdata(\vec{v}) v_i,
        \ i \in \mathcal{I}_v, \\
    \langle \sigma_i^z \rangle_\text{data}
        &= \sum_\vec{v} \pdata(\vec{v}) \frac{b_i'(\vec{v})}{D_i(\vec{v})} \tanh\big(D_i(\vec{v})\big), \ i \in \mathcal{I}_h, \\
    \langle \sigma_i^z \sigma_j^z \rangle_\text{data}
        &= \sum_\vec{v} \pdata(\vec{v}) v_i v_j,
        \ i, j \in \mathcal{I}_v, \\
    \langle \sigma_i^z \sigma_j^z \rangle_\text{data}
        &= \sum_\vec{v} \pdata(\vec{v}) v_i \frac{b_j'(\vec{v})}{D_j(\vec{v})} \tanh\big(D_j(\vec{v})\big), \ i \in \mathcal{I}_v, \ j \in \mathcal{I}_h,
\end{split}
\end{align}
where \( b_i'(\vec{v}) = b_i + (\mat{W}\T\vec{v})_i \), \( D_i(\vec{v}) = \sqrt{\Gamma_i^2 + b_i'(\vec{v})^2} \), \( \mathcal{I}_v = \{1, \dots, n_v\} \) represents the visible qubit indices, and \( \mathcal{I}_h = \{n_v + 1, \dots, n\} \) represents the hidden qubit indices.

\subsection{Quantum Annealing}\label{sec:quantum_annealing}
Quantum annealing, also known as adiabatic quantum computing, is a branch of quantum computing that is based on the adiabatic theorem, which in the (translated) words of Born and Fock~\cite{born_fock_1928}:
"A physical system remains in its instantaneous eigenstate if a given perturbation is acting on it slowly enough and if there is a gap between the eigenvalue and the rest of the Hamiltonian's spectrum."
This can be achieved by implementing a Hamiltonian of the form~\cite{qc_lecture_notes}
\begin{align}
    H(s) = A(s) H_{\text{initial}} + B(s) H_{\text{final}},
\end{align}
where \( s \in [0, 1] \).
For a linear anneal schedule \( s(t) = t / t_a \), where \( t_a \) is the annealing time.
\( H_{\text{initial}} \) is the initial Hamiltonian which describes the system at \( s = 0 \) and is responsible for introducing quantum fluctuations.
\( H_{\text{final}} \) is the final Hamiltonian which describes the system at \( s = 1 \) and is responsible for encoding the problem defined by the user.

The functions \( A(s) \) and \( B(s) \) must be such that they satisfy the relations
\begin{align}
\begin{split}
    A(0) &\gg B(0), \\
    A(1) &\ll B(1).
\end{split}
\end{align}

In essence, a quantum annealer starts in the ground state of the initial Hamiltonian, then slowly evolves the system over time so that it remains in the instantaneous ground state.
By the time the annealing process is completed, the Hamiltonian is just that of the problem, and if the system evolved adiabatically, then it should have remained in the instantaneous ground state.
Therefore, when the qubits are measured at the end, they should correspond to a low energy solution of the final Hamiltonian.

\subsubsection{D-Wave Quantum Annealer}
D-Wave quantum annealers implement a time-dependent Hamiltonian of the form~\cite{dwave_qa}
\begin{align}
    H(s) = A(s) \bigg( -\sum_{i=1}^{n} \sigma_i^x \bigg) + B(s) \bigg( \sum_{i=1}^{n} h_i \sigma_i^z + \sum_{i=1}^{n}\sum_{j=i+1}^{n} J_{ij} \sigma_i^z \sigma_j^z \bigg).
\end{align}
From this we see the initial Hamiltonian has the ground state where all qubits are aligned in the \( x \)-direction, i.e., \( \ket{+}^{\otimes n} \), which corresponds to an equal superposition of all possible states in the computational basis.
The final Hamiltonian corresponds to the Ising model described by the \( h_i \) and \( J_{ij} \) values.

\begin{figure}[!htb]
    \begin{center}
        \includegraphics[width=0.7\linewidth]{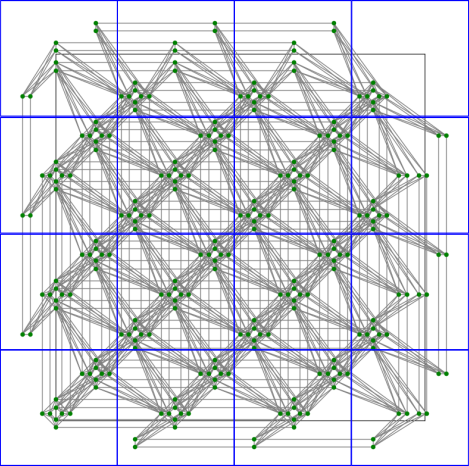}
    \end{center}
    \caption{
        A lattice with \( 4 \times 4 \) Pegasus unit cells (\( P_4 \)).
        The D-Wave Advantage QPU is based on a lattice with \( 16 \times 16 \) Pegasus unit cells (\( P_{16} \))~\cite{dwave_topologies}.
    }
    \label{fig:p4_unitcells}
\end{figure}

\begin{figure}[!htb]
    \begin{center}
        \includegraphics[width=1\linewidth]{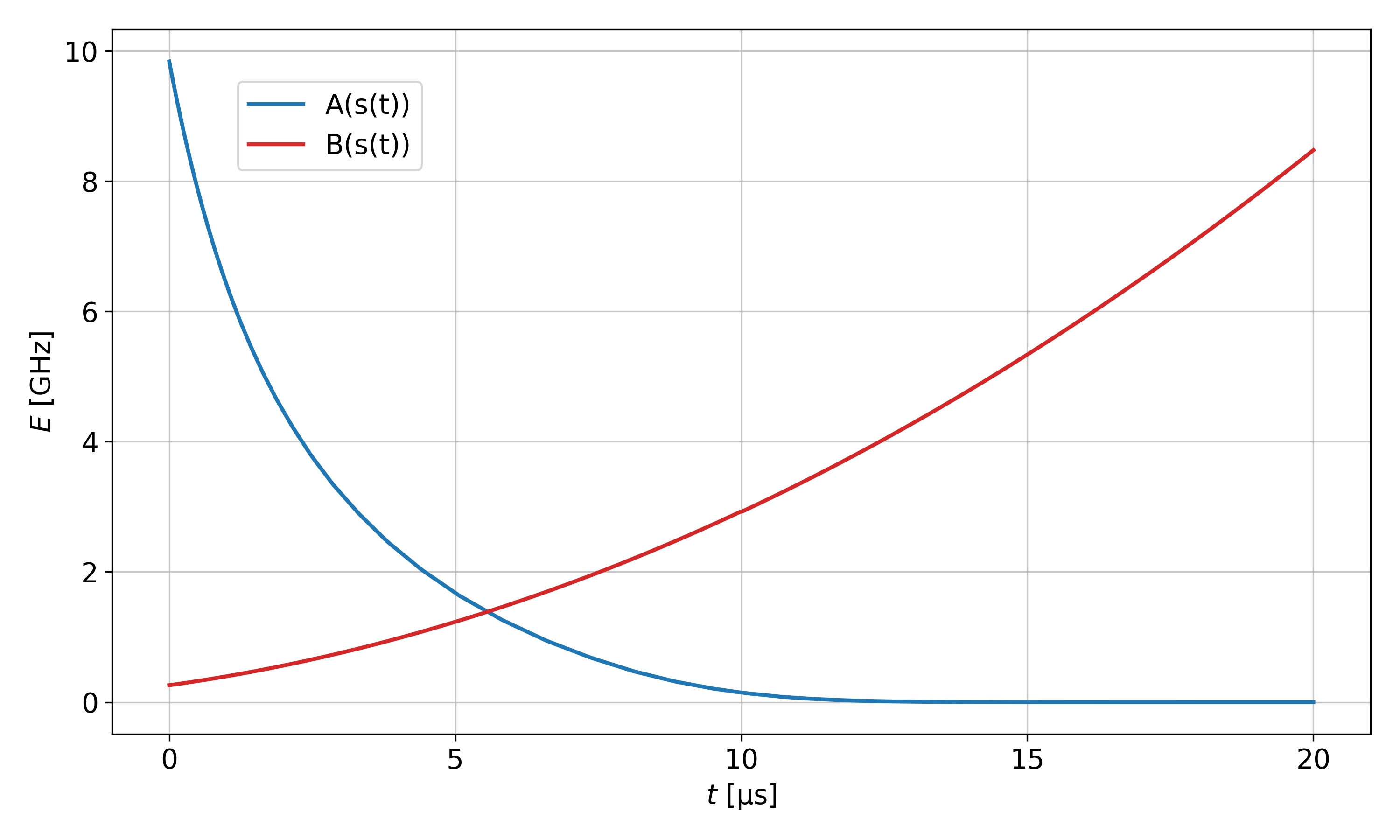}
    \end{center}
    \caption{
        Default anneal schedule of the D-Wave Advantage 4.1 with linear \( s(t) = t / t_a \) and \( t_a = 20 \ \si{\micro\second} \)~\cite{dwave_anneal_schedules}.
    }
    \label{fig:anneal_schedule_default}
\end{figure}

The quantum processing unit (QPU) is made up of superconducting qubits under the influence of external magnetic fluxes~\cite{qc_lecture_notes} that change the Hamiltonian from the initial to the final over the duration of the annealing process.
These qubits are arranged in a graph structure similar to that seen in~\cref{fig:p4_unitcells}.
The default anneal schedule for the D-Wave Advantage 4.1 is shown in~\cref{fig:anneal_schedule_default}

\subsubsection{Mapping the QBM to the D-Wave Quantum Annealer}
As stated in~\cite{amin_2018}, in order get a quantum annealer to sample from a quantum Boltzmann distribution, one would need to freeze the evolution at some point \( s^* \) during the annealing process and then perform the measurements.
The authors go on to say that this can be done in practice using a nonuniform \( s(t) \) that anneals slowly in the beginning, then quenches the system (completes the annealing as fast as possible) at the freeze-out point \( s^* \), if \( s^* \) is in the quasistatic regime.
In an earlier paper~\cite{amin_2015}, Amin showed that the quasistatic regime begins around 1 \si{\micro\second} for the D-Wave 2000Q, so it should not be an issue to reach the quasistatic regime for annealing times longer than 5 \si{\micro\second}.

Because a quantum annealer is a real-world physical device, samples generated with it have an associated temperature called the effective temperature.
To be more specific, the corresponding density operator is of the form
\begin{align}
    \rho(s, T) = \frac{1}{Z} e^{-\beta H(s)},
\label{eq:effective_density_operator}
\end{align}
where \( \beta = 1/kT \) is the effective inverse temperature.
In principle, \( \beta \) is an unknown quantity and must be determined in order to effectively use the annealer to generate samples from a quantum Boltzmann distribution.

Comparing the density operator of the QBM in~\cref{eq:density_operator} to the one in~\cref{eq:effective_density_operator} at the freeze-out point \( s^* \), we find
\begin{align}
\begin{split}
    \Gamma_i
        &= \beta A(s^*), \\
    b_i
        &= -\beta B(s^*) h_i, \\
    w_{ij}
        &= -\beta B(s^*) J_{ij}.
    \label{eq:qbm_scaling}
\end{split}
\end{align}
This enables us to map the QBM to the annealer if \( \beta \) can be determined to some reasonable degree of accuracy.

\subsubsection{Learning the Effective Inverse Temperature}\label{sec:learning_beta}
There is the possibility to treat \( \beta \) as a learnable parameter rather than having to choose a value empirically, as detailed by Xu and Oates in~\cite{xu_2021}.
The method is based on a log-likelihood maximization approach leading to parameter updates of the form (see \cref{app:learning_beta} for how one arrives at this result)
\begin{align}
    \Delta\betahat
        &= \eta_{\betahat}\big(\langle E \rangle_\text{data} - \langle E \rangle_\text{model}\big),
\end{align}
where \( \betahat = 1/k\hat{T} \) is the estimator of the effective inverse temperature, and \( \eta_{\betahat} \) is the associated learning rate.
We must note though, that this approach is only valid for classical Boltzmann distributions, but this fits our current use case as we will see in~\cref{sec:qbm_verifying_distribution}.

\subsubsection{D-Wave Ocean SDK}
D-Wave offers an easy-to-use Python package called Ocean SDK~\cite{dwave_ocean} to interact with their Leap~\cite{dwave_leap} cloud-based quantum annealing platform, which allows users to access various quantum annealers and other solvers around the globe.

One of the most important steps in solving a problem using a D-Wave annealer is finding an embedding, i.e., a mapping of the logical qubits to the physical qubits, and the SDK offers a heuristic method to do so.
If the problem cannot be directly embedded (1:1 logical:physical qubits), then a cluster of physical qubits called a chain is created to represent one logical qubit.
Chains introduce added complexity into the problem, because one then needs to tune the chain strength, i.e., the coupling constant between the qubits in the chains.
If the measured values of the qubits in a chain differ, this is called a chain break, and the system will report back the majority vote of the measured values in the chain.
Therefore, it is best to avoid chains if possible, but they are often a necessary evil for larger problems due to connectivity limitations.

Samples can be easily generated by the annealer using the \texttt{sample\_ising(h, J)} function which takes in the user-defined \( h_i \) and \( J_{ij} \) values and returns a sample set of specified size (maximum \( 10^4 \)).
The returned sample set contains the sampled state vectors (an array of shape \( (n_\text{samples}, n) \) with values \( \pm 1 \) corresponding to the qubit measurements), their energies, and other information about the run.

It must be noted that for the purposes of using a D-Wave annealer for quantum Boltzmann sampling, one must disable autoscaling to properly estimate the effective temperature, as per~\cref{eq:qbm_scaling}.
The \texttt{sample\_ising(h, J)} function has the keyword argument \texttt{autoscale=True}, which rescales the \( h_i \) and \( J_{ij} \) values by the factor~\cite{dwave_solver_parameters}
\begin{align}
\begin{split}
    r_\text{autoscale}
        = \max\Bigg\{
            &\max\bigg\{\frac{\max\{h_i\}}{\max\{h_\text{range}\}},0\bigg\},
            \max\bigg\{\frac{\min\{h_i\}}{\min\{h_\text{range}\}},0\bigg\}, \\
            &\max\bigg\{\frac{\max\{J_{ij}\}}{\max\{J_\text{range}\}},0\bigg\},
            \max\bigg\{\frac{\min\{J_{ij}\}}{\min\{J_\text{range}\}},0\bigg\}
        \Bigg\}.
\end{split}
\end{align}
This is because the main use case of D-Wave annealers is to maximize the probability of measuring the ground state, thus the problem is rescaled so that the \( h_i \) and \( J_{ij} \) values fully utilize the allowed range of values, essentially decreasing the effective temperature; therefore, we set \texttt{autoscale=False} to avoid this.
For the Advantage 4.1 system the allowed value ranges are \( h_\text{range} = [-4, 4] \) and \( J_\text{range} = [-1, 1] \)~\cite{dwave_solver_properties}.

\subsubsection{Generating More Robust Statistics}\label{sec:gauge}
QPUs are not perfect, and sometimes specific qubits or parts of the chip might have readout biases.
To mitigate such issues, one can perform a gauge transformation on the problem.
If we have an \( n \)-qubit problem, then we can generate a random vector \( \vec{r} \in \{+1, -1\}^n \) which allows us to change the submission to the solver without actually changing the underlying problem.
This is done by taking
\begin{align}
\begin{split}
    h_i
        &\rightarrow r_i h_i, \\
    J_{ij}
        &\rightarrow r_i r_j J_{ij}, \\
    (s_1, \dots, s_n)
        &\rightarrow (r_1 s_1, \dots, r_n s_n),
\end{split}
\end{align}
and then transforming the results back using the third relation above, where \( s_i \) is the measured value of qubit \( i \).

\subsubsection{Previous Work in This Field}
In recent years, a number of researchers have studied using D-Wave quantum annealers to train Boltzmann machines~\cite{adachi_2015,benedetti_2016,anschuetz_2019,wiebe_2019,rocutto_2020,dixit_2021,ilmo_2021,wilson_2021,xu_2021}.
The most common approach is to train a classical RBM with quantum assistance, i.e., using the annealer to generate the samples in the negative phase rather than using Gibbs sampling.
Classical RBMs trained with quantum assistance are a special case of the QBM, i.e., when \( s^* = 1 \) the problem reduces to a classical RBM because \( \lim_{s\rightarrow 1} \Gamma_i = 0 \) in~\cref{eq:qbm_scaling}.

One thing that stands out the most about some of the previous research is that very few discuss embeddings and anneal schedules, which as we will see in the next section are important for getting the best possible performance out of the annealer.
Therefore, we aim to create a basic framework with which one can use to approach the problem of using a D-Wave annealer to sample from a (quantum) Boltzmann distribution.

\section{12-Qubit Problem}\label{sec:qbm_12_qubit_problem}
In order to get a better understanding of how the QBM works, we study a small 12-qubit problem that can be solved exactly.
For this purpose we take a QBM with restrictions in both the visible and hidden layers and train it using the log-likelihood lower bound maximization approach; we call this a bound-based quantum restricted Boltzmann machine, or BQRBM for short.
We configure the model with 8 visible and 4 hidden units to act as a regularized autoencoder.

\subsection{Sampling From a Quantum Boltzmann Distribution}
Before training the model, we first need to assess the Advantage 4.1's ability to sample from quantum Boltzmann distributions.
To this end, we randomly generate the values of \( h_i \) and \( J_{ij} \) from a normal distribution with \( \mu = 0 \) and \( \sigma = 0.1 \), then use the KL divergence to compare samples generated by the Advantage 4.1 with theoretical distributions.

\subsubsection{Anneal Schedule Format}
\begin{figure}[!htb]
    \begin{center}
        \includegraphics[width=1\linewidth]{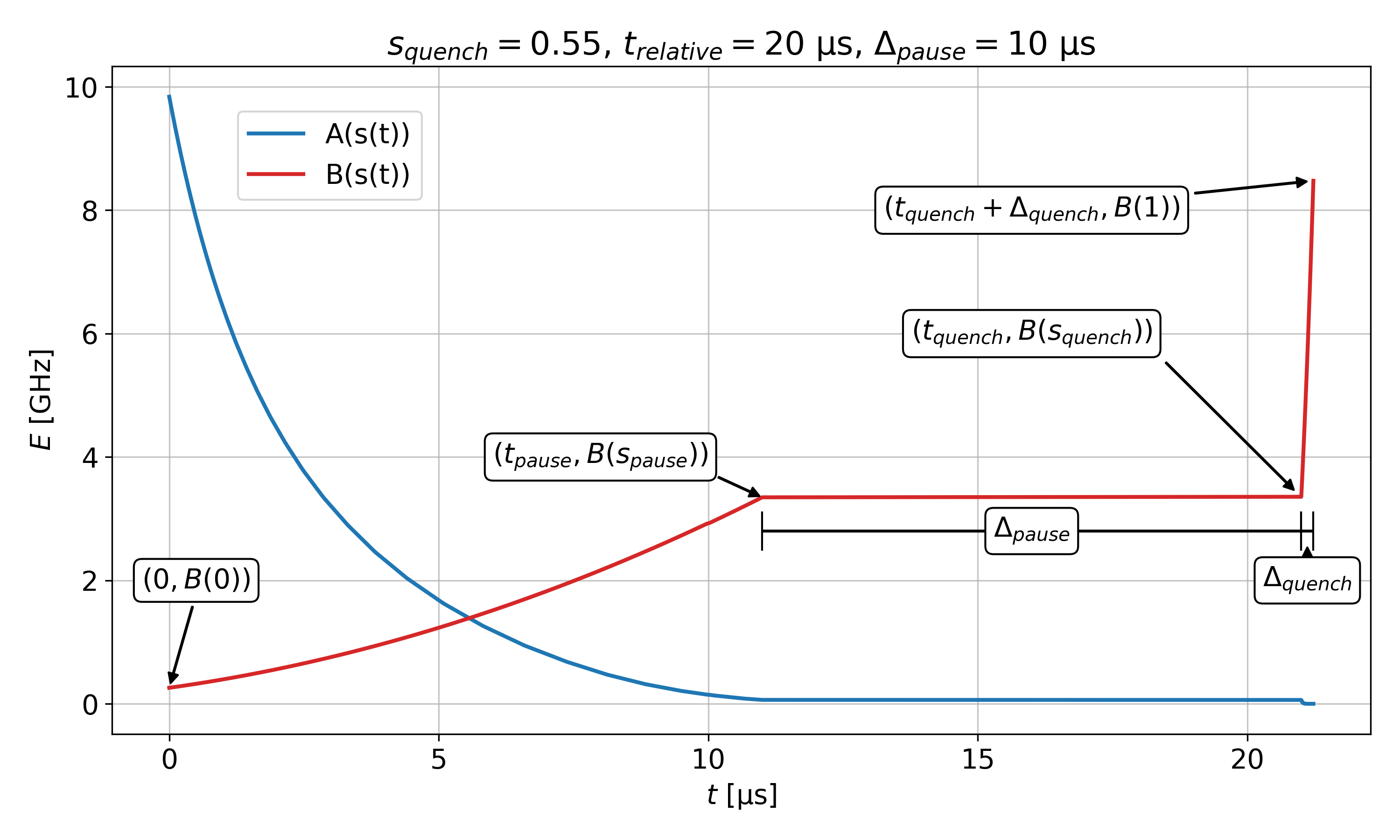}
    \end{center}
    \caption{
        Example of a custom pause-and-quench anneal schedule for the D-Wave Advantage 4.1~\cite{dwave_anneal_schedules}.
        Annotations indicate the points \( (t, B(s(t))) \), as well as the periods over which the annealing is paused and quenched.
    }
    \label{fig:anneal_schedule_annotated}
\end{figure}

The \( A(s) \) and \( B(s) \) values for a D-Wave annealer are fixed and depend on the specific system~\cite{dwave_anneal_schedules}, but the Ocean SDK allows us to define a nonuniform \( s(t) \) using a list of \( (t, s) \) tuples, which then determine the \( A(s(t)) \) and \( B(s(t)) \) curves.
In this section we use what we call pause-and-quench anneal schedules that
\begin{enumerate}
    \item start at \( (t = 0, s = 0) \),
    \item pause the system at \( (\tpause, \spause) \) for a duration of \( \Deltapause \),
    \item quench the system at \( (\tquench, \squench) \) over a duration of \( \Deltaquench \).
\end{enumerate}
Thus, the anneal schedules provided to the solver are of the form
\begin{align}
    [
        (0, 0),
        (\tpause, \spause),
        (\tquench, \squench),
        (\tquench + \Deltaquench, 1)
    ],
\end{align}
where
\begin{align}
\begin{split}
    \squench &\equiv \spause, \\
    \tpause &= \spause \cdot \trelative, \\
    \tquench &= \tpause + \Deltapause.
\end{split}
\end{align}
An annotated example of a custom pause-and-quench anneal schedule with \( \squench = 0.55 \), \( \trelative = 20 \ \si{\micro\second} \), and \( \Deltapause = 10 \ \si{\micro\second} \) is given in~\cref{fig:anneal_schedule_annotated}.

The minimum quench duration \( \Deltaquench \) is a function of \( \squench \) and is limited by the system's fastest anneal rate \( \alphaquench \)
\begin{align}
    \Deltaquench(\squench) = \frac{1 - \squench}{\alphaquench}.
\end{align}
The Advantage 4.1 system allows a maximum of \( \alphaquench = 2 \ \si{\micro\second}^{-1} \)~\cite{dwave_solver_parameters}.

\subsubsection{Verifying the Distribution}\label{sec:qbm_verifying_distribution}
\begin{figure}[!htb]
    \begin{center}
        \includegraphics[width=1\linewidth]{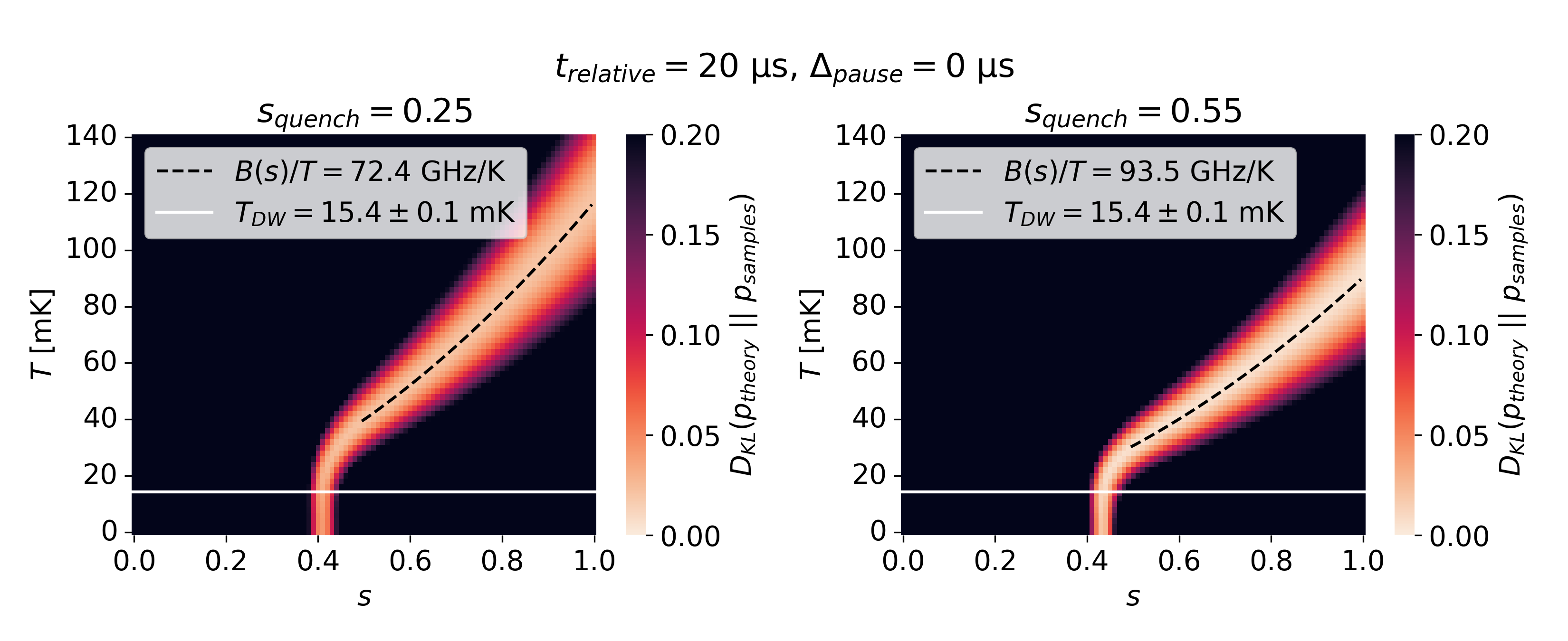}
    \end{center}
    \caption{
        Heatmaps of \( \DKL{\ptheory}{\psamples} \) comparing the distribution produced by samples from the D-Wave Advantage 4.1 to a set of theoretical QBM distributions for two different quench points using embedding 10.
        The dashed lines represent the optimal values of \( B(s) / T = \text{constant} \), computed by taking the value of \( T \) which produces the lowest KL divergence for each \( s \ge 0.5 \).
        Data represents an ensemble average over 10 random gauge sample sets consisting of \( 10^4 \) samples each.
    }
    \label{fig:dkl_min_heatmap}
\end{figure}

We use the KL divergence \( \DKL{\ptheory}{\psamples} \) to compare the probabilities of the energies computed from the samples returned by the Advantage 4.1 with the theoretical energy distributions for \( s = 0.01, 0.02, \dots, 1 \) and \( T = 10^{-3}, 2, 4, \dots, 200 \ \si{\milli\kelvin} \), which we visualize as heatmaps in~\cref{fig:dkl_min_heatmap}.
More information on how the KL divergences are computed can be found in~\cref{app:kl_divergence_in_practice}, how the density matrix (from which the theoretical distributions are obtained) is computed in~\cref{app:exact_rho_computation}, and the required constants in~\cref{app:constants}.

In the right heatmap, where \( \squench = 0.55 \), we observe a narrow band in which the Advantage 4.1-generated samples closely resemble a quantum Boltzmann distribution, and in fact the samples approximate multiple distributions depending on the effective temperature.
Marshall et al. present similar results using a D-Wave 2000Q in~\cite{marshall_2019}, in which they discuss if the distribution returned by the annealer fits that of a quantum Boltzmann distribution late in the anneal process when \( A(s^*) / B(s^*) \ll 1 \), then the distribution at \( s^* \) should be close to a classical Boltzmann distribution, i.e.,
\begin{align}
    e^{-\beta H(s^*)} \approx e^{-\beta B(s^*) H_\text{final}}.
\end{align}
This in turn means that not only is there one optimal \( s^* \) and effective temperature which models the distribution, but rather a set of them corresponding to a family of distributions for which \( \beta B(s^*) \) is constant.
Therefore, this explains the streak pattern in the heatmaps.

Furthermore, we observe that the left heatmap, where \( \squench = 0.25 \), is quite similar to the right one where \( \squench = 0.55 \), but with higher KL divergence values and temperatures.
This indicates that quenching at \( \squench = 0.25 \) produces samples that are distributed more as a classical Boltzmann distribution, and that we cannot generate samples from quantum Boltzmann distributions with \( s^* \lessapprox 0.45 \), at least not with the anneal schedules we use here.

It must also be noted that the effective temperature corresponding to the classical Boltzmann distribution \( (s^* = 1) \) is significantly higher than that of the D-Wave temperature of \( T_\text{DW} = 15.4 \pm 0.1 \ \si{\milli\kelvin} \)~\cite{dwave_leap}\footnote{Temperature obtained from the system properties in the Leap interface.}.
It is not entirely clear exactly why the effective temperature of the distribution is so much higher than the device temperature, but in~\cite{marshall_2019} they give several possible reasons, including the discrepancy between the temperature of the device and the qubits, fluctuations in the temperature while annealing, and control errors masquerading as higher temperatures.
In principle, higher effective temperatures are unwanted because they shrink the range of allowed values for the weights and biases as per~\cref{eq:qbm_scaling}, but there is not much one can do about this.

From analysis of the heatmaps and the fact that we cannot produce distributions with \( s^* \lessapprox 0.45 \), we conclude that nontrivial dynamics occur while the system is quenching, i.e., the system cannot quench fast enough.
It is difficult to compare directly since the 2000Q is a different system than the Advantage 4.1 we study here, but in~\cite{marshall_2019} they also allude to the possibility of nontrivial dynamics occurring.
The 2000Q allows for quenching with \( \alphaquench = 1 \), which is only a factor of two smaller than that of the Advantage 4.1.
Therefore, if as supposed in~\cite{marshall_2019} that the quench is not fast enough, then likely such a small difference in how fast the system can be quenched would not drastically change the results.

If we take a second to think about it, the qubits are oscillating at a frequency in terms of gigahertz.
This means that a quench duration of a few hundred nanoseconds still allows for a number of oscillations in the qubits, which is likely enough time for nontrivial dynamics to take place.
It would be interesting to verify via simulation how fast a quench must be in order to freeze out the distribution at the desired point \( s^* \).

We conclude that we are unable to reliably generate arbitrary quantum Boltzmann distributed samples using the Advantage 4.1 system.
Therefore, for the remainder of this thesis we focus on training models with \( s^* = 1 \) using classical Boltzmann distributed samples generated by the Advantage 4.1, also enabling us to use the aforementioned method of learning the effective temperature.

\subsubsection{Choosing an Embedding}
We compare 10 different heuristically generated embeddings based on how well they approximate the desired distribution.
In this embedding comparison, we use only direct embeddings (no chains), so the embeddings only differ by the location of the qubits on the chip, and pause-and-quench anneal schedules with \( \trelative = 20 \ \si{\micro\second} \) and \( \Deltapause = 0 \ \si{\micro\second} \).

\begin{figure}[!htb]
    \begin{center}
        \includegraphics[width=1\linewidth]{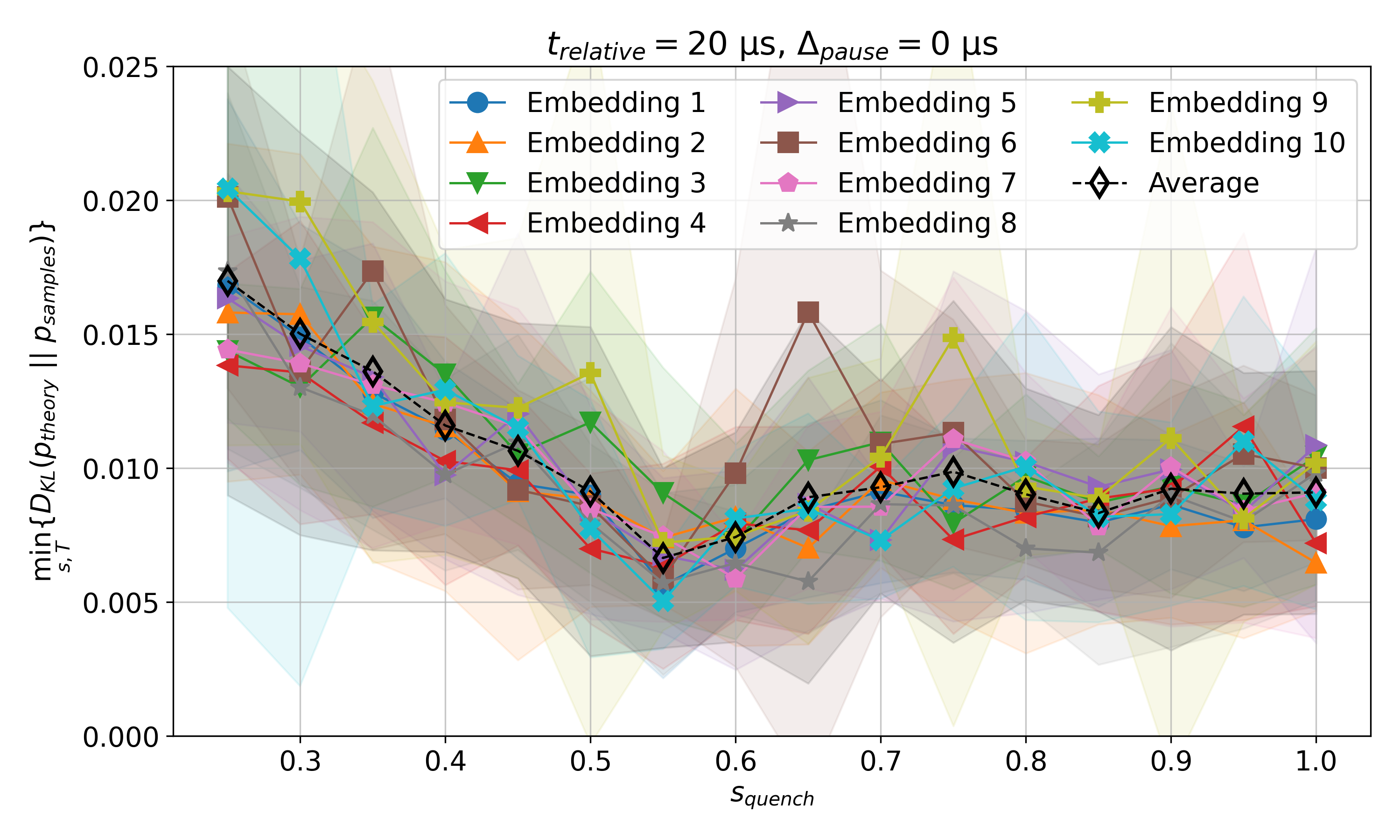}
    \end{center}
    \caption{
        Comparison of \( \min_{s,T}\big\{\DKL{\ptheory}{\psamples}\big\} \) for different embeddings and \( \squench \) values.
        Data represents an ensemble average over 10 random gauge sample sets consisting of \( 10^4 \) samples each.
        Shaded regions represent one standard deviation.
    }
    \label{fig:dkl_mins_embeddings}
\end{figure}

It is difficult to compare the heatmaps of all embeddings and quench points due to the higher dimensionality of the data, so we take the minimum KL divergence over \( s \) and \( T \), and plot it as a function of \( \squench \) in~\cref{fig:dkl_mins_embeddings}.
We immediately see how varied the results are depending on the embedding and quench point, highlighting the importance of choosing a good embedding and anneal schedule.

Our findings indicate that embedding 10 is likely a good choice because it produces the best results at \( \squench = 0.55 \).
The rest of the results in this subsection use embedding 10.

\subsubsection{Choosing an Anneal Schedule}\label{sec:choosing_an_anneal_schedule}
With the chosen embedding we want to see if there is a way in which we can alter the anneal schedule to further reduce the KL divergence.
We start with the same anneal schedule formula as before, except we introduce pausing before initiating the quench for durations \( \Deltapause = 0, 10, 100 \ \si{\micro\second} \), as well as the addition of \( \trelative = 100 \ \si{\micro\second} \).

\begin{figure}[!htb]
    \begin{center}
        \includegraphics[width=1\linewidth]{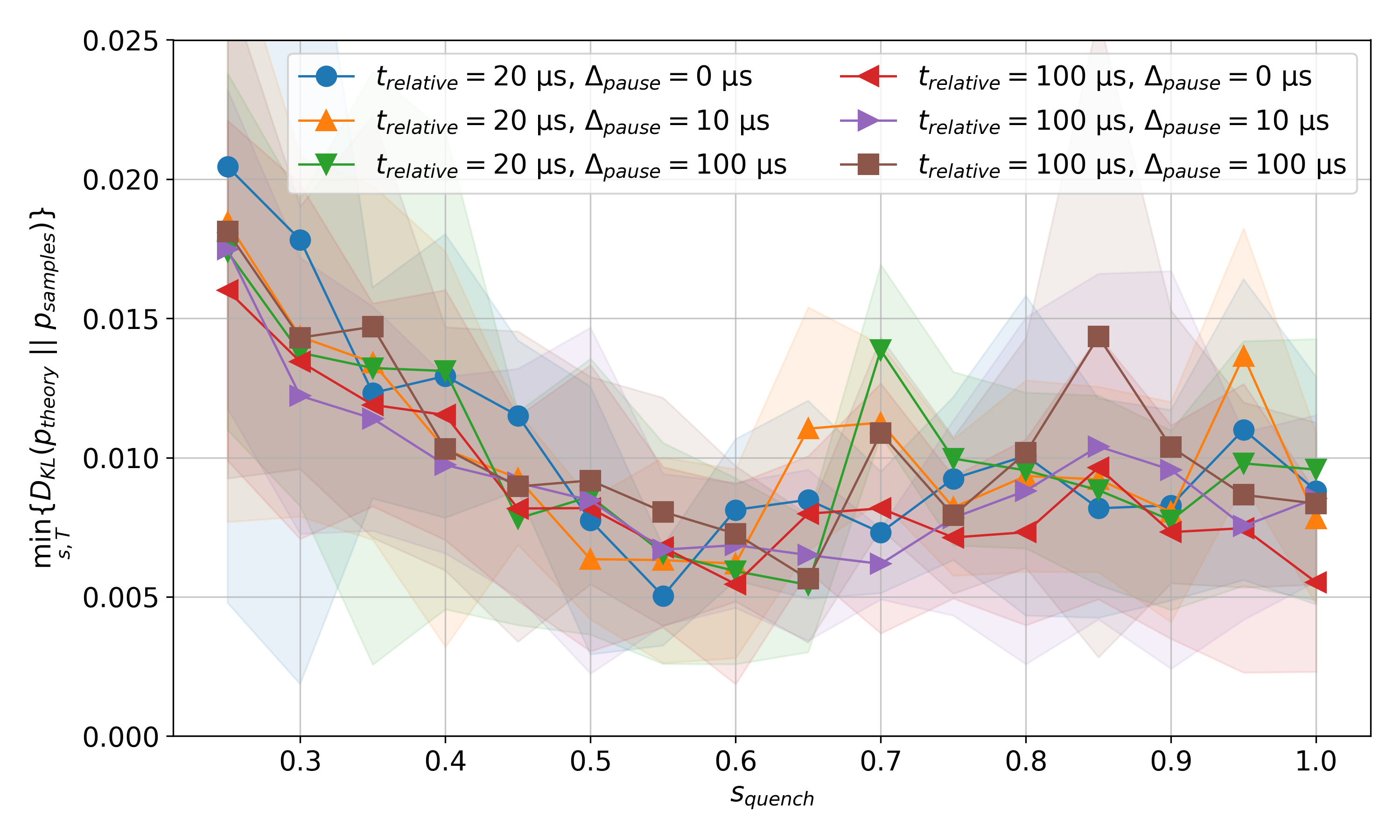}
    \end{center}
    \caption{
        Comparison of \( \min_{s,T}\big\{\DKL{\ptheory}{\psamples}\big\} \) for various pause-and-quench anneal schedules using embedding 10.
        Data represents an ensemble average over 10 random gauge sample sets consisting of \( 10^4 \) samples each.
        Shaded regions represent one standard deviation.
        Some of the sample sets with longer annealing times and pause durations contain less than \( 10^4 \) samples as to satisfy the maximum allowed run time of the D-Wave Advantage 4.1.
    }
    \label{fig:dkl_mins_embedding_05}
\end{figure}

\cref{fig:dkl_mins_embedding_05} illustrates that pausing and longer annealing times have little effect, and that quenching in the range of \( \squench \in [0.55, 0.6] \) produces the best results.
With this information, we opt to use an anneal schedule with \( \squench = 0.55 \), \( \trelative = 20 \ \si{\micro\second} \), and \( \Deltapause = 0 \ \si{\micro\second} \), as it offers a good balance between performance and QPU usage time.

\subsection{Training Data}
\begin{figure}[!htb]
    \begin{center}
        \includegraphics[width=1\linewidth]{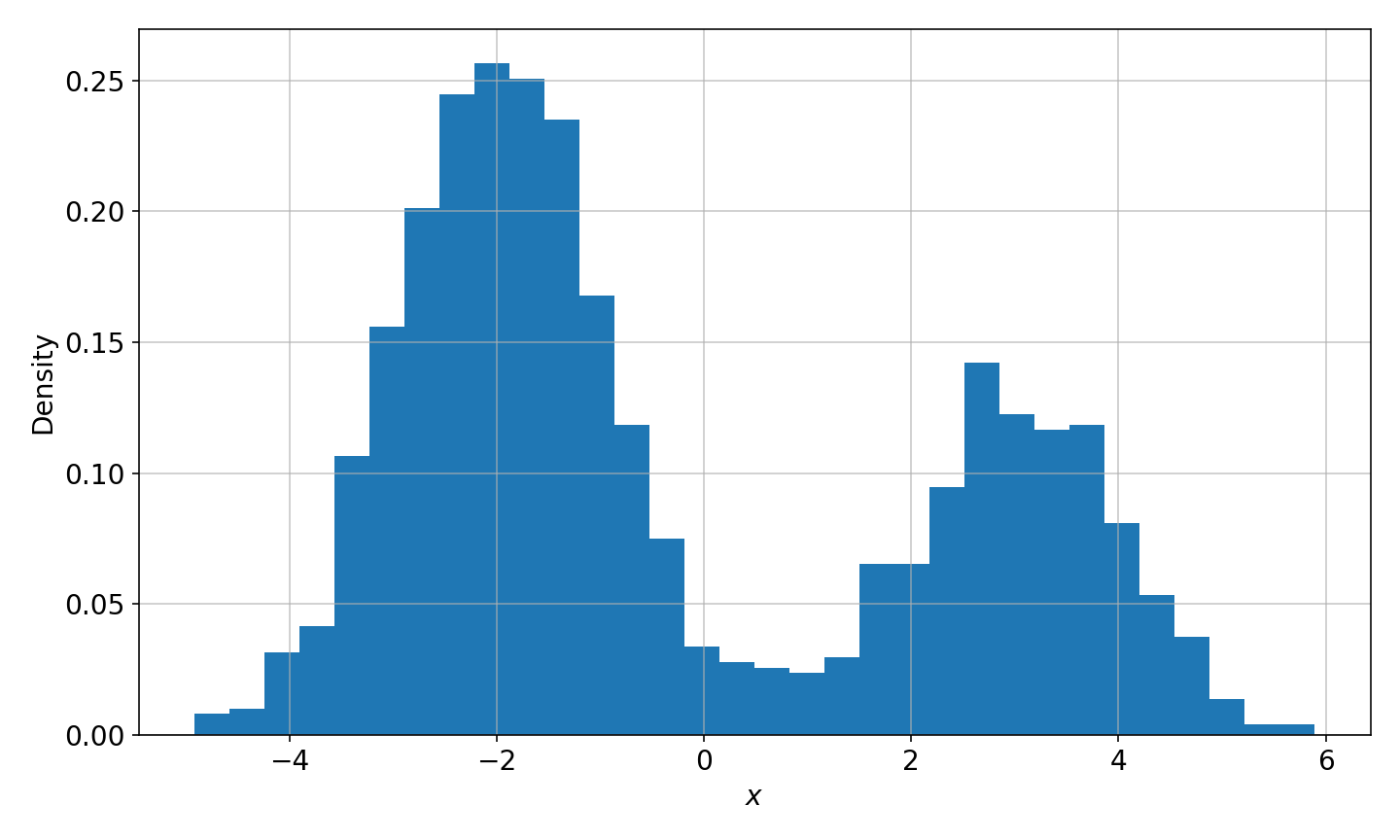}
    \end{center}
    \caption{
        Histogram of the training data set used in the 12-qubit problem.
    }
    \label{fig:hist_data}
\end{figure}

Having verified that the Advantage 4.1 can indeed produce Boltzmann distributed samples to some degree of accuracy, we proceed with training models using both a simulation and the Advantage 4.1.
We randomly generate a training data set consisting of 1500 samples, 1000 from a \( \mathcal{N}(-2, 1) \) distribution and 500 from a \( \mathcal{N}(3, 1) \) distribution, visualized in~\cref{fig:hist_data}.

\subsection{Simulation-based Model}
The first step is training a model using a simulation in which the samples are generated using the probabilities obtained from computing \( \rho \) exactly.
Here we use a mini-batch size of 10, \( s^* = 1 \), and an initial learning rate of \( \eta = 0.1 \) with a schedule that exponentially decays the learning rate every 10 epochs by a factor of 2 beginning at epoch 50 as defined in~\cref{app:lr_exp_decay}.
The learning rate \( \eta_{\hat{\beta}} \) for the parameter \( \betahat \) follows a similar schedule, except it has a decay period of 20 as opposed to 10, to allow for more range of motion in the \( \betahat \) parameter later in the training process if the estimate needs to adapt more quickly to a new effective \( \beta \).

\subsubsection{Results}\label{sec:qbm_simulation_results}
\begin{figure}[!htb]
    \begin{center}
        \includegraphics[width=1\linewidth]{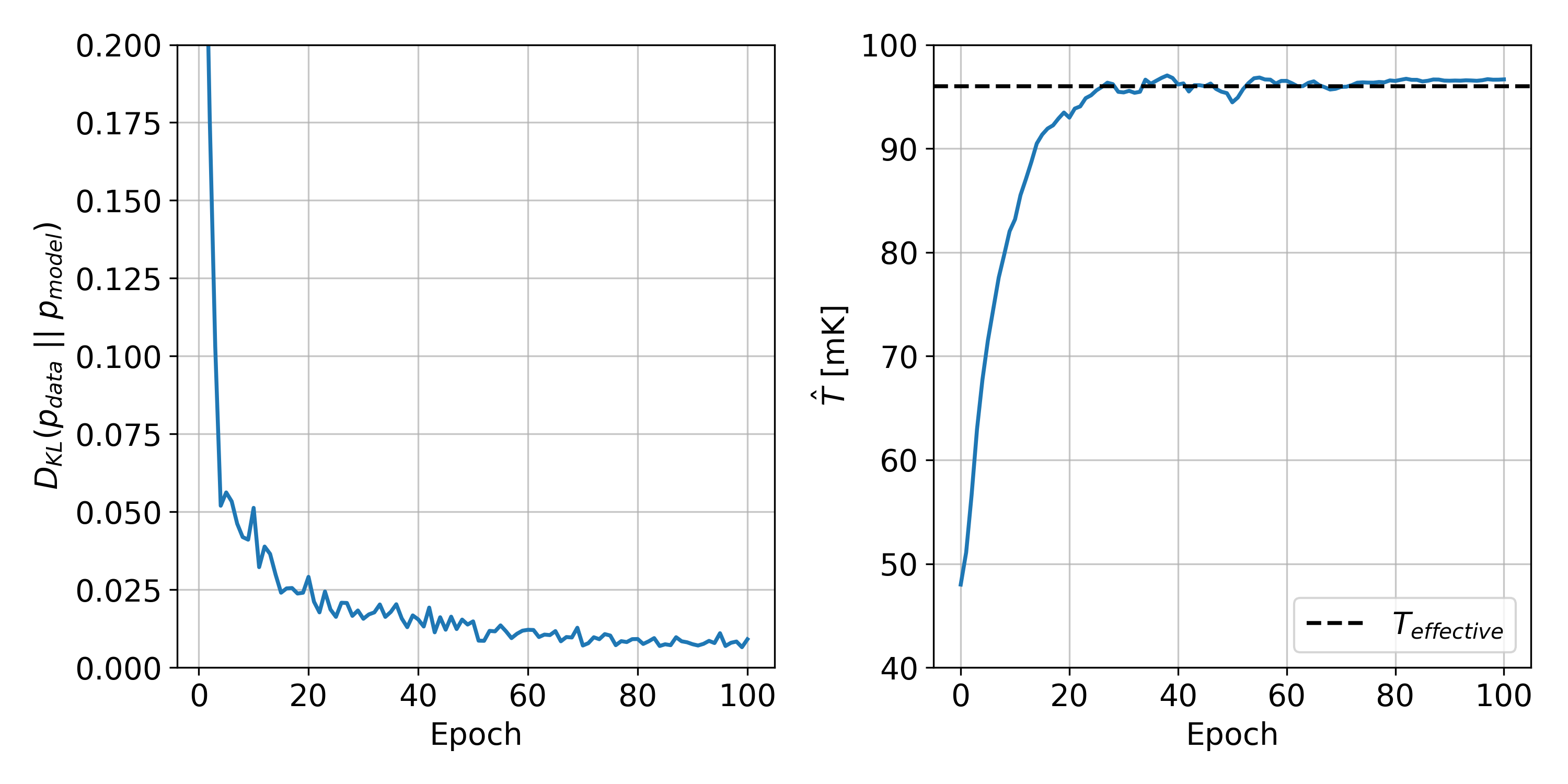}
    \end{center}
    \caption{
        Training results of the 12-qubit model trained using the simulation.
        On the left is the KL divergence \( \DKL{\pdata}{\pmodel} \) plotted against the epochs; each data point was generated using \( 10^4 \) samples at the end of every epoch.
        On the right is the learned temperature estimator \( \hat{T} \) plotted against the epochs, as well as the effective temperature that the simulation was configured to generate samples at.
    }
    \label{fig:train_results_exact}
\end{figure}

\cref{fig:train_results_exact} shows the results of training the simulation-based model on the aforementioned data set.
We use the KL divergence \( \DKL{\pdata}{\pmodel} \) as a way to track the progress of the training and get a read on how well the model learns the data set distribution, because minimizing the KL divergence is equivalent to maximizing the log-likelihood~\cite{murphy_2012}.
The KL divergence is computed at the end of every epoch using a sample set of size \( 10^4 \).
In the left plot of \cref{fig:train_results_exact}, we observe a clear trend of the KL divergence being minimized.
The learning curve reaches an optimal value after about 80 epochs, then remains steady for the next 20 epochs until the end of training.

We designed the simulation such that we can set the effective \( \beta \) to any value we desire.
To verify that the model can learn an accurate value for the estimator \( \betahat \), we configure the simulation to generate samples at an effective value of \( \beta = 0.5 \ \si{\giga\hertz}^{-1} \ (T \approx 96 \ \si{\milli\kelvin}) \) and initialize the model with a value of \( \betahat = 1 \ \si{\giga\hertz}^{-1} \ (\hat{T} \approx 48 \ \si{\milli\kelvin}) \).
The results in the right plot of~\cref{fig:train_results_exact} confirm that it is able to learn a value of \( \betahat \) close to the actual effective \( \beta \).

Overall, the results show that the model can generate samples similar to the training distribution reasonably well when trained using the simulation, i.e., the best case scenario.
Additionally, we are able to verify that the model can accurately learn an estimate of the effective temperature.
We use the results of this model as a baseline to compare the models trained using the Advantage 4.1 in the next subsection with.

\subsection{D-Wave Advantage 4.1-based Model}
Having successfully trained the 12-qubit BQRBM using samples generated via exact simulation, we move to switching the sample generation part to the Advantage 4.1.
We take the same hyperparameters as the simulation and an anneal schedule using \( \squench = 0.55 \), \( \trelative = 20 \ \si{\micro\second} \), and \( \Deltapause = 0 \ \si{\micro\second} \).
We see in~\cref{fig:dkl_min_heatmap} that \( \squench = 0.55 \) has an optimal temperature of around 90 \si{\milli\kelvin} for \( s^* = 1 \), thus we take \( \betahat = 0.5 \ \si{\giga\hertz}^{-1} \ (\hat{T} \approx 96 \ \si{\milli\kelvin}) \) as our initial guess for the effective \( \beta \), and let the model learn from there.

\subsubsection{Results}\label{sec:qbm_annealer_results}
\begin{figure}[!htb]
    \begin{center}
        \includegraphics[width=1\linewidth]{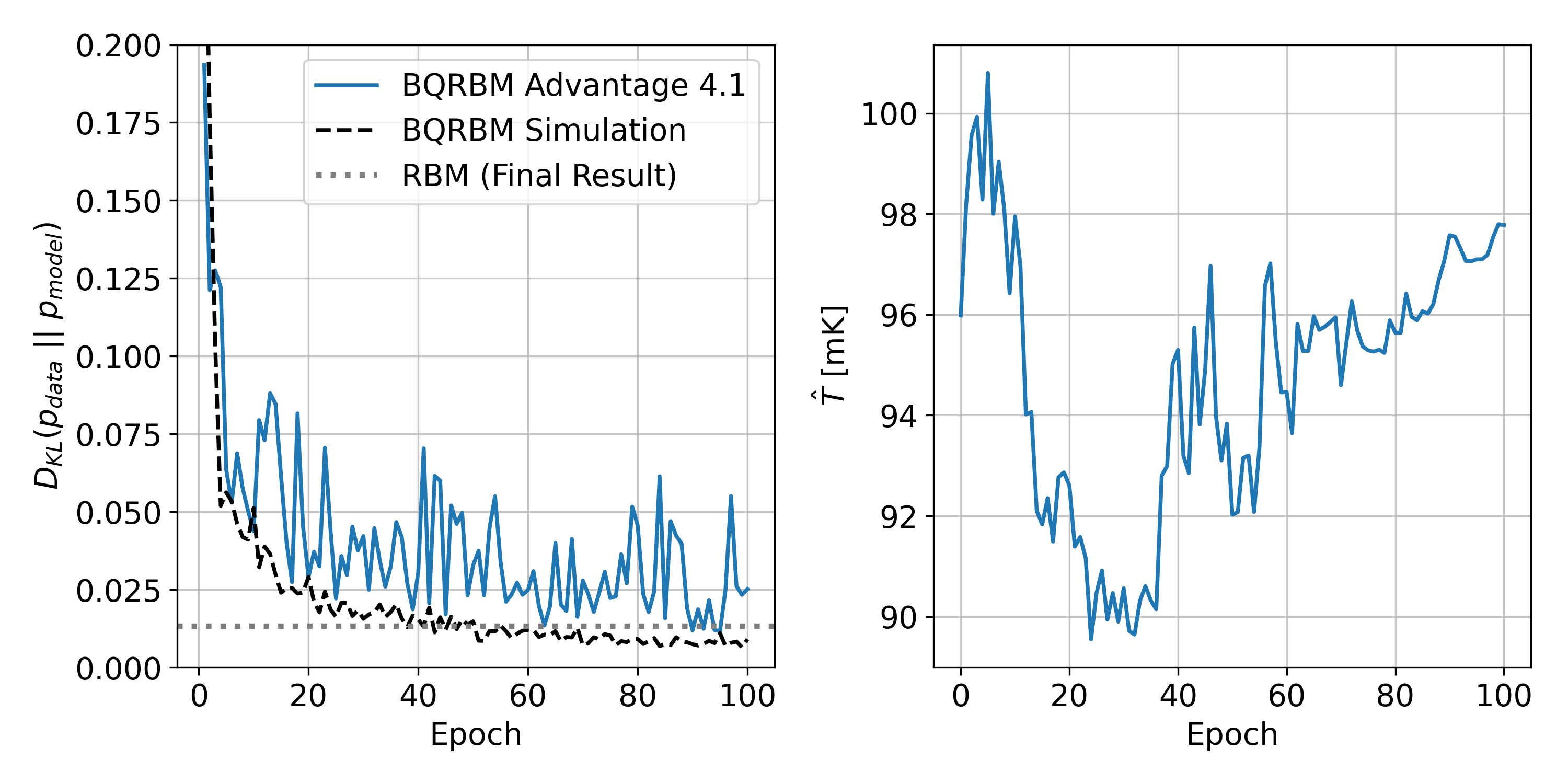}
    \end{center}
    \caption{
        Training results of the 12-qubit model trained using samples generated with the D-Wave Advantage 4.1 compared with that of the simulation and the final results of a classical RBM.
        On the left is the KL divergence \( \DKL{\pdata}{\pmodel} \) plotted against the epochs; each data point was generated using \( 10^4 \) samples at the end of every epoch.
        On the right is the learned temperature estimator \( \hat{T} \) plotted against the epochs.
    }
    \label{fig:train_results_annealer}
\end{figure}

The KL divergences in~\cref{fig:train_results_annealer} and~\cref{tbl:12_qubit_qbm_KL_divergences} show the model trained using the Advantage 4.1 produces samples that resemble the training distribution to some extent, but still underperforms when compared with the simulation and the classically trained RBM.
This is possibly due to the information loss associated with using the D-Wave to approximate the distribution, which likely arises due to noise and errors (see~\cref{sec:challenges}), because after all, real-world systems governed by quantum mechanics are highly sensitive to their environment.

\begin{table}[!htb]
    \centering
    \begin{adjustbox}{max width=\textwidth}
        \begin{tabular}{r r r}
\multicolumn{3}{c}{\(D_{\text{KL}}(p_\text{data} \ || \ p_\text{model})\)} \\
\toprule
\textbf{BQRBM Advantage 4.1} & \textbf{BQRBM Simulation} & \textbf{RBM}\\
\midrule
0.034 $\pm$ 0.014 & 0.008 $\pm$ 0.001 & 0.015 $\pm$ 0.002 \\
\bottomrule
\end{tabular}

    \end{adjustbox}
    \caption{
        KL divergences of the 12-qubit BQRBM models vs.~the classical RBM.
        The values are shown in the format mean \(\pm\) one standard deviation from an ensemble of 100 sample sets consisting of \( 10^4 \) samples each.
}
    \label{tbl:12_qubit_qbm_KL_divergences}
\end{table}

\begin{figure}[!htb]
    \begin{center}
        \includegraphics[width=1\linewidth]{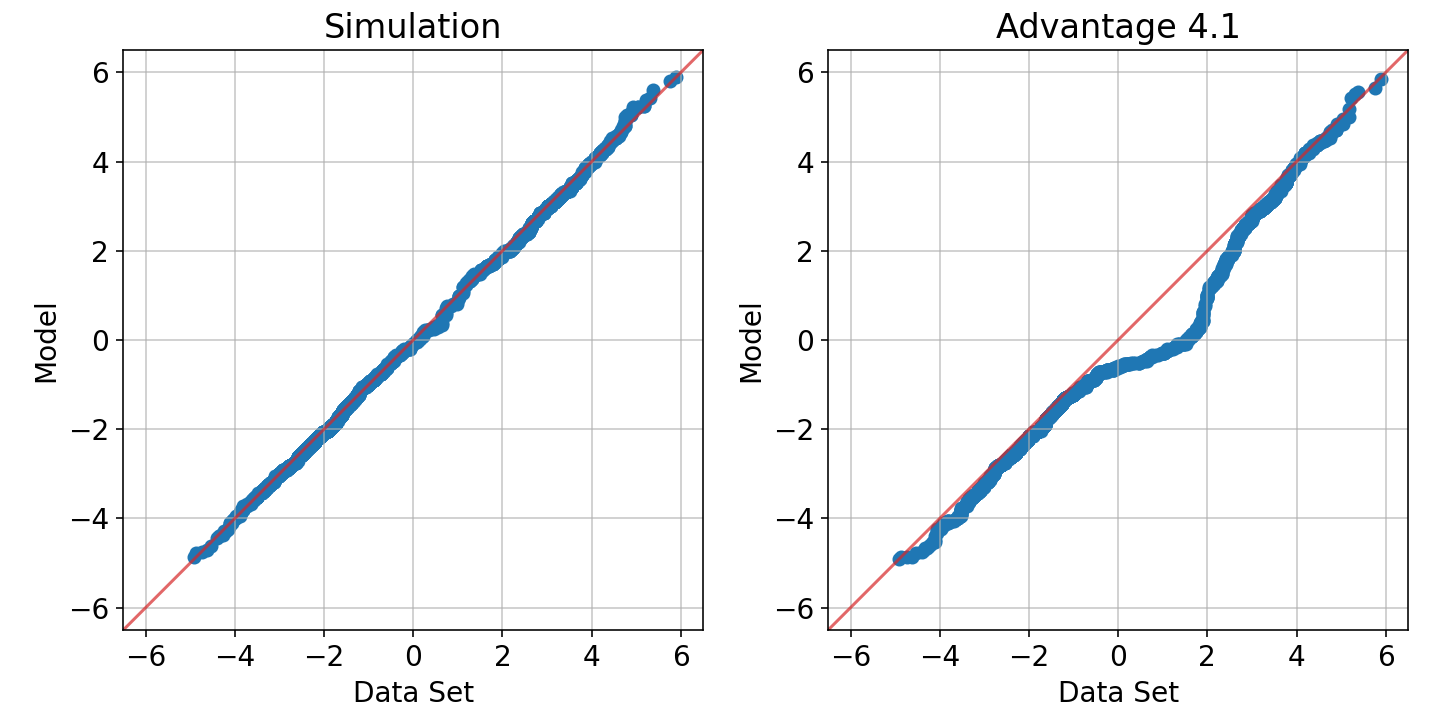}
    \end{center}
    \caption{Log return Q-Q plots of the 12-qubit model trained using the simulation (left) and the D-Wave Advantage 4.1 (right).}
    \label{fig:qq_comparison}
\end{figure}

We notice that the Advantage 4.1-based model struggles most with the trough between the two Gaussian peaks in the training distribution based on the Q-Q plots in~\cref{fig:qq_comparison}.

\begin{figure}[!htb]
    \begin{center}
        \includegraphics[width=0.7\linewidth]{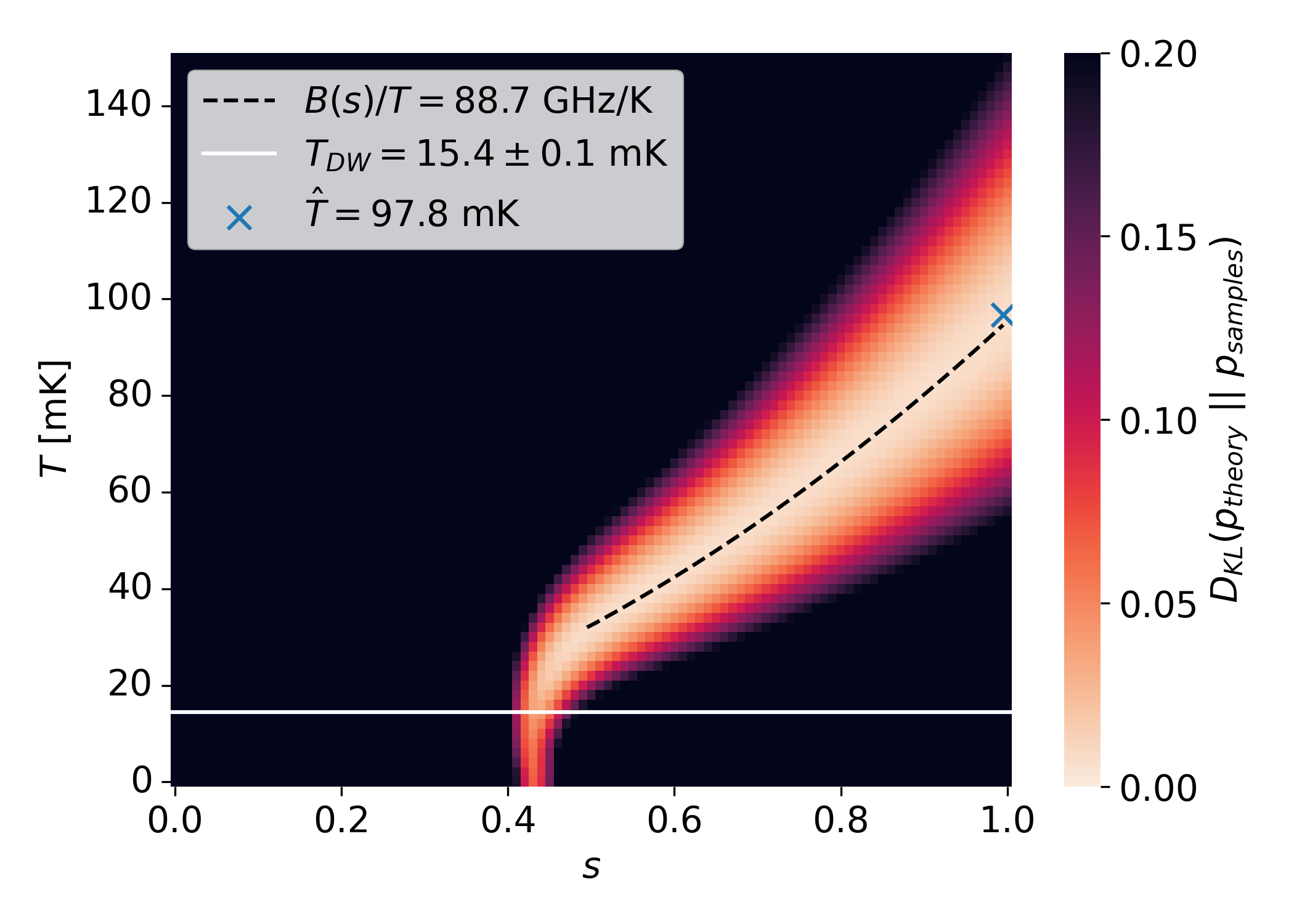}
    \end{center}
    \caption{
        Heatmap of \( \DKL{\ptheory}{\psamples} \) comparing the distributions produced by samples from the Advantage 4.1 to a set of theoretical QBM distributions, using the final \( h_i \) and \( J_{ij} \) values learned by the 12-qubit model trained using the Advantage 4.1.
        The blue cross indicates the learned estimate of the effective temperature.
        The dashed line represents the optimal value of \( B(s) / T = \text{constant} \), computed by taking the value of \( T \) which produces the lowest KL divergence for each \( s \ge 0.5 \).
        Data represents an ensemble average over 10 random gauge sample sets consisting of \( 10^4 \) samples each.
    }
    \label{fig:learned_effective_temperature}
\end{figure}

Although we cannot track the true effective temperature throughout the training process, we are able to see how close the learned effective temperature estimate at the end matches that generated by samples using the final learned weights and biases.
The heatmap shown in \cref{fig:learned_effective_temperature} confirms that the Advantage 4.1-based model's \( \betahat \) value of \( 97.8 \ \si{\milli\kelvin} \) is quite close to the \( 95.6 \ \si{\milli\kelvin} \) computed from the optimal \( B(s) / T \) value.

The results in this section show that a BQRBM can indeed be trained using a D-Wave quantum annealer.
The 12-qubit problem plays a crucial role in our understanding of how one can use a D-Wave quantum annealer to generate Boltzmann distributed samples.
Although the results are not spectacular, and underperform the simulation and classical model, they still show promise.
It will be interesting to rerun this analysis on the next generation of D-Wave quantum annealers to see how much they improve.

\section{The Quantum Market Generator}\label{sec:quantum_market_generator}
With deeper insights into the workings of the BQRBM from the 12-qubit problem, we move to the final stage of training a quantum market generator with 64 visible and 30 hidden units.
All results in this section use the same baseline data set (B) as in~\cref{sec:classical_market_generator}.

\subsection{Setting the Annealer's Hyperparameters}\label{sec:qbm_hyperparameters}
Unlike the 12-qubit problem, the larger problem restricts our ability to perform an in-depth analysis to compare the sample distributions produced by the Advantage 4.1 with that of theory.
Therefore, we have to take a more practical approach when choosing some of the annealer hyperparameters such as the relative chain strength, quench point, and embedding.
To this end, we train a number of models with various settings of these hyperparameters for 20 epochs to get a read on the trend direction, and then choose their values empirically.
For this, we use a mini-batch size of 10, \( s^* = 1 \), constant learning rates of \( \eta = 0.02 \) and \( \eta_{\hat{\beta}} = 0.01 \), and an initial value of \( \betahat = 0.25 \ \si{\giga\hertz}^{-1} \ (\hat{T} \approx 192 \ \si{\milli\kelvin}) \).

The reason we only train for 20 epochs is that the epoch duration is quite high (10-25 minutes) due to latency and load on the annealer, ergo it is not very feasible to train every model for a higher number of epochs.
With an average epoch duration of around 15 minutes, training a model for 20 epochs takes roughly 5 hours, so training for 100 epochs would take around a day.

\subsubsection{Choosing a Relative Chain Strength}\label{sec:qbm_rcs}
The chain strength \( \gamma \) is computed using the relative chain strength \( \rcs \) as
\begin{align}
    \gamma
        &= \rcs \cdot \min\Big\{
            \max\{J_\text{range}\}, \max\big\{\{\abs{h_i}\} \cup \{\abs{J_{ij}}\}\big\}
        \Big\}.
\end{align}

\begin{figure}[!htb]
    \begin{center}
        \includegraphics[width=1\linewidth]{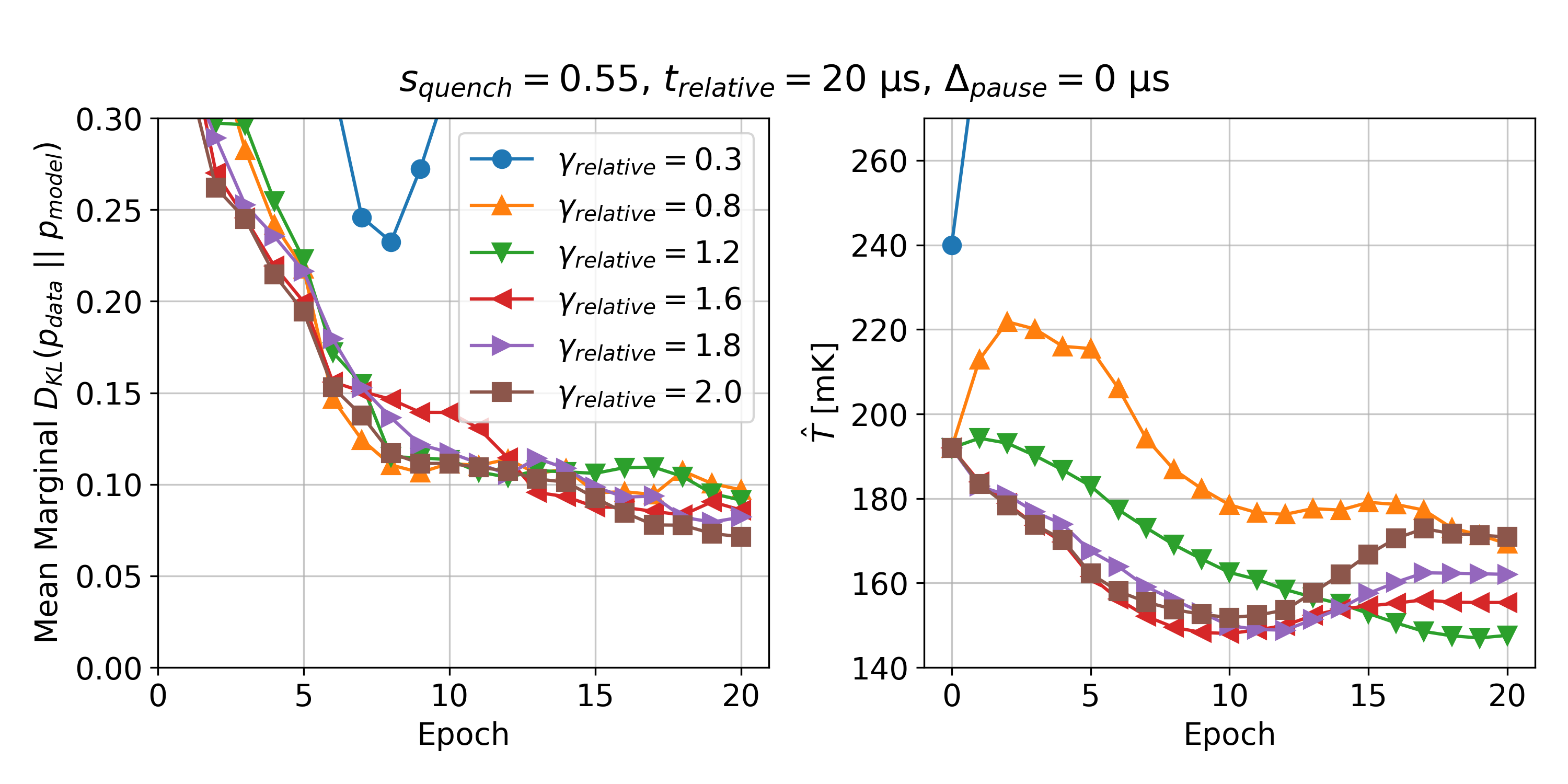}
    \end{center}
    \caption{
        Training results of the relative chain strength \( \rcs \) scan for embedding 1.
        On the left are the mean marginal \( \DKL{\pdata}{\pmodel} \) values, i.e., the average of the KL divergences of the individual currency pairs.
        On the right are the learned estimates of the effective temperature.
        Data plotted on a 5 epoch simple moving average basis to reduce visual noise.
    }
    \label{fig:qbm_log_returns_rcs_comparison}
\end{figure}

We train models using various values of \( \rcs \in [0.3, 2] \), and a pause-and-quench anneal schedule with \( \squench = 0.55 \), \( \trelative = 20 \ \si{\micro\second} \), and \( \Deltapause = 0 \ \si{\micro\second} \).
The results are plotted in~\cref{fig:qbm_log_returns_rcs_comparison} (only a subset depicted).
We find that too low values of \( \rcs \) lead to more chain breaks early on in the training process, which then cause the model to learn a higher temperature, in turn shrinking the allowed range of weights and biases to the point where the model can no longer learn effectively.
Everything indicates that higher relative chain strengths produce better results.
After a number of epochs, we observe \( \max\big\{ \{\abs{h_i}\} \cup \{\abs{J_{ij}}\} \big\} \) grow to a value larger than 0.5, implying that values of \( \rcs \ge 2 \) would not change the results since \( \gamma \) is reaching its limit of \( \max\{J_\text{range}\} = 1 \).
Therefore, we choose a value of \( \rcs = 2 \).

\subsubsection{Choosing an Anneal Schedule}
\begin{figure}[!htb]
    \begin{center}
        \includegraphics[width=1\linewidth]{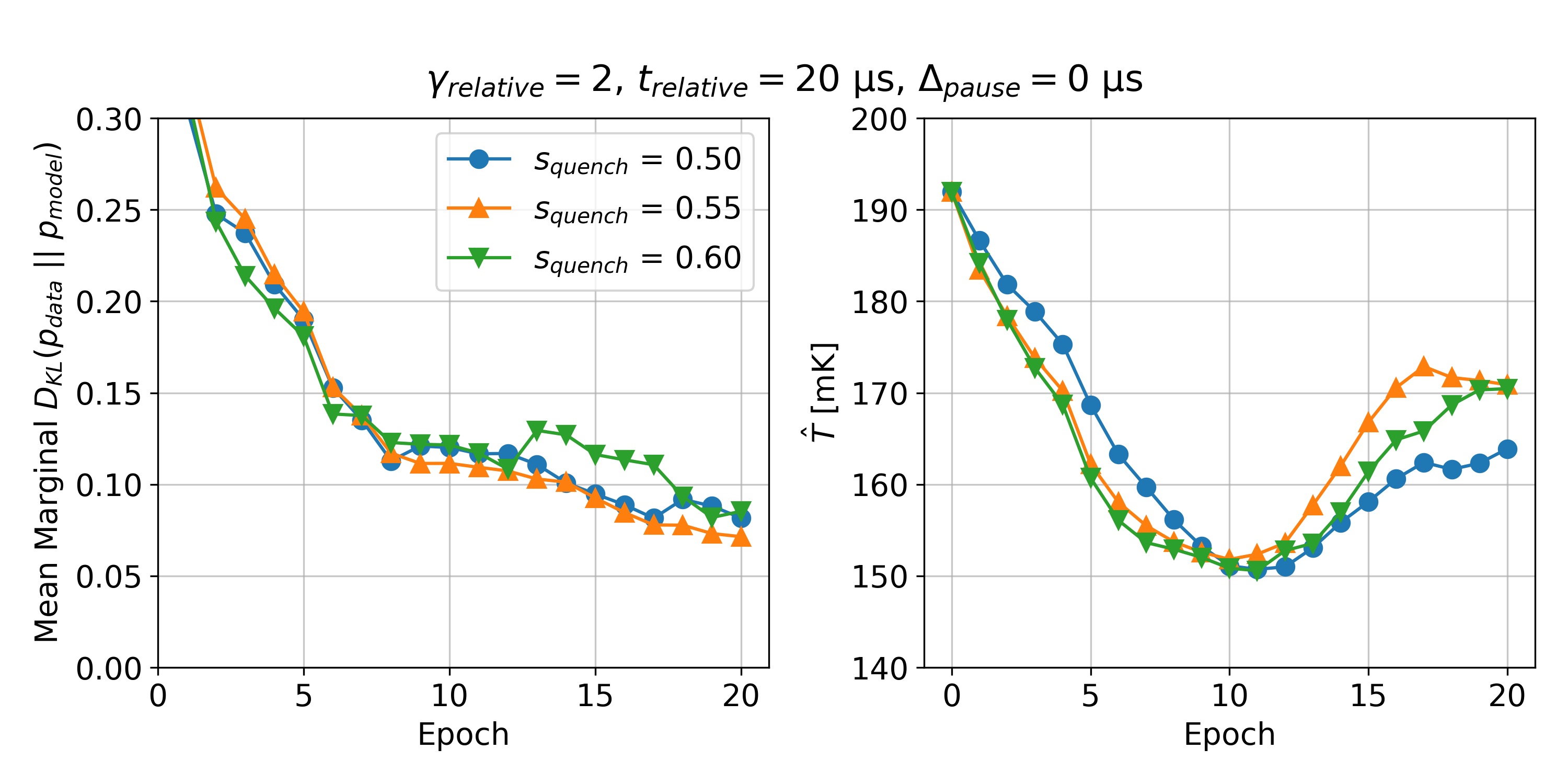}
    \end{center}
    \caption{
        Training results of the \( \squench \) scan for embedding 1.
        On the left are the mean marginal \( \DKL{\pdata}{\pmodel} \) values, i.e., the average of the KL divergences of the individual currency pairs.
        On the right are the learned estimates of the effective temperature.
        Data plotted on a 5 epoch simple moving average basis to reduce visual noise.
    }
    \label{fig:qbm_log_returns_s_quench_comparison}
\end{figure}

We keep \( \trelative = 20 \ \si{\micro\second} \) and \( \Deltapause = 0 \ \si{\micro\second} \) as in the 12-qubit problem, but check to see if a different value of \( \squench \) improves performance.
We try values of \( \squench = 0.5, 0.55, 0.6 \), plotted in~\cref{fig:qbm_log_returns_s_quench_comparison}, and find that \( \squench = 0.55 \) leads to the best KL divergence curve, and thus choose that value going forward.

\subsubsection{Choosing an Embedding}
\begin{figure}[!htb]
    \begin{center}
        \includegraphics[width=1\linewidth]{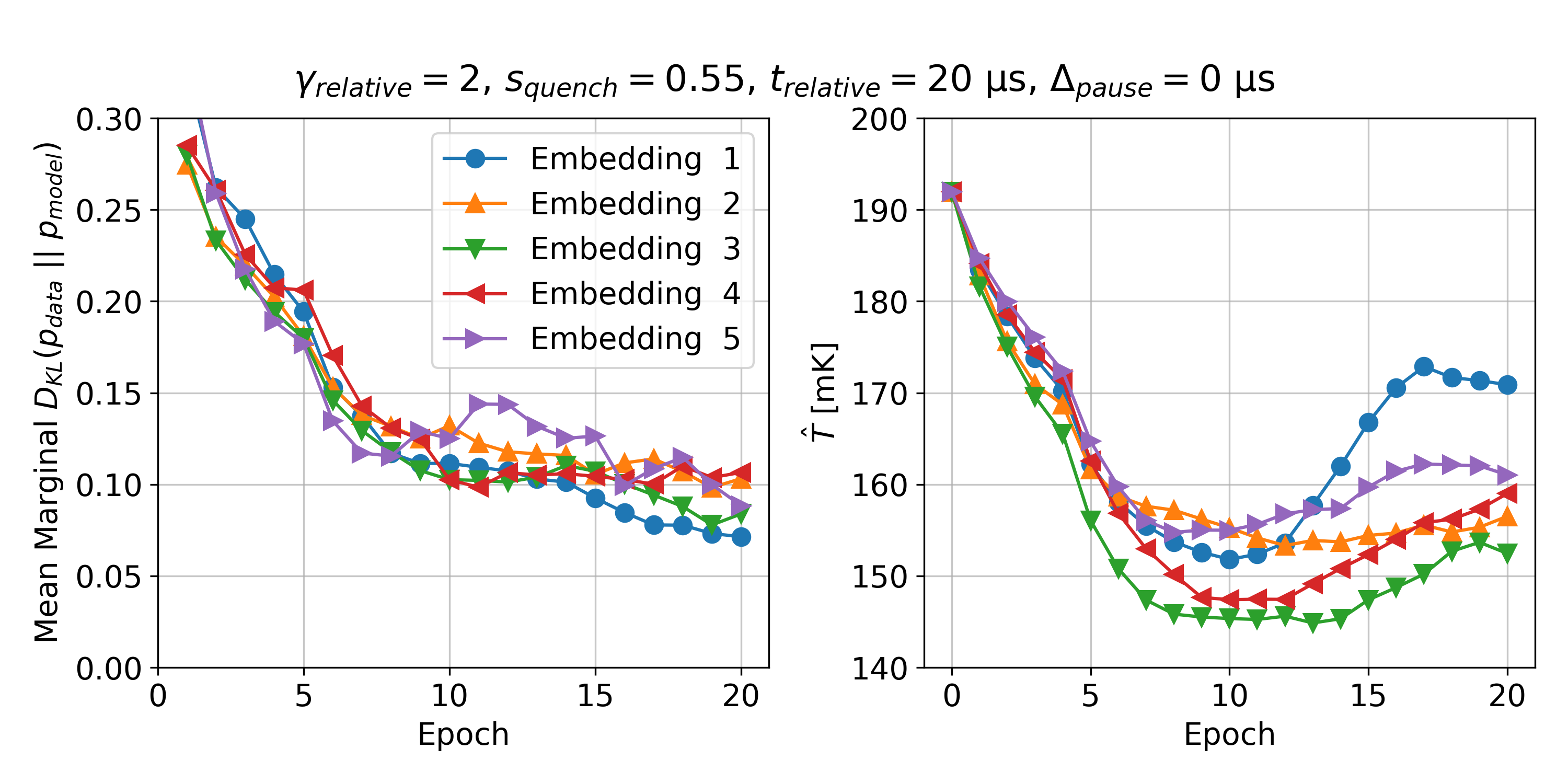}
    \end{center}
    \caption{
        Training results comparing 5 different embeddings.
        On the left are the mean marginal \( \DKL{\pdata}{\pmodel} \) values, i.e., the average of the KL divergences of the individual currency pairs.
        On the right are the learned estimates of the effective temperature.
        Data plotted on a 5 epoch simple moving average basis to reduce visual noise.
    }
    \label{fig:qbm_log_returns_embedding_comparison}
\end{figure}

The final annealer hyperparameter we seek to tune is the embedding.
We try 5 different heuristically generated embeddings each composed of around 400 physical qubits and maximum chain lengths of 7.
The comparison plotted in \cref{fig:qbm_log_returns_embedding_comparison} indicates to us that embedding 1 is likely a good choice to continue with.

\subsection{Results}\label{sec:qbm_log_returns_results}
\begin{figure}[!htb]
    \begin{center}
        \includegraphics[width=1\linewidth]{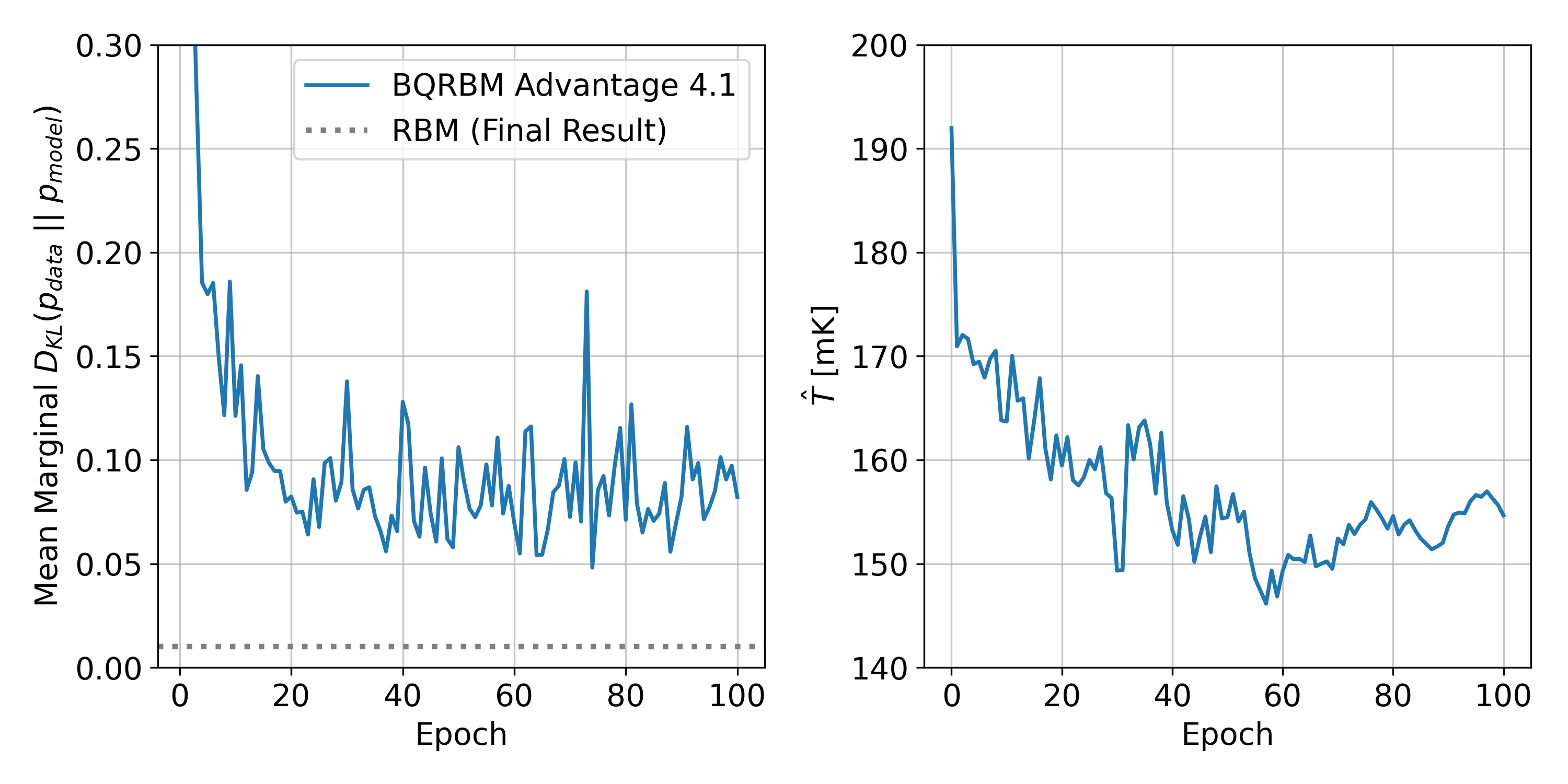}
    \end{center}
    \caption{
        Training results of the BQRBM compared with the final results of the classical RBM.
        On the left is the mean marginal \( \DKL{\pdata}{\pmodel} \) value, i.e., the average of the KL divergences of the individual currency pairs.
        On the right is the learned estimate of the effective temperature.
    }
    \label{fig:qbm_log_returns_full_run}
\end{figure}

With annealer hyperparameters of \( \rcs = 2 \), \( \squench = 0.55 \), and embedding 1, we move to training a full model over 100 epochs.
The training curves are depicted in~\cref{fig:qbm_log_returns_full_run}, where we observe the KL divergence decrease until around epoch 40 where it then oscillates for the remainder of the training process.
Unfortunately, the training results show that the BQRBM model significantly underperforms the classical model.

Poor model performance is further confirmed by the KL divergences in~\cref{tbl:qbm_KL_divergences}.
We see the KL divergences of the BQRBM are about eight times higher than those of the classical RBM.
\begin{table}[!htb]
    \centering
    \begin{adjustbox}{max width=\textwidth}
        \begin{tabular}{l r r}
\multicolumn{3}{c}{\(D_{\text{KL}}(p_\text{data} \ || \ p_\text{model})\)} \\
\toprule
Currency Pair & \textbf{BQRBM} & \textbf{RBM}\\
\midrule
EURUSD & 0.086 $\pm$ 0.044 & 0.010 $\pm$ 0.001 \\
GBPUSD & 0.062 $\pm$ 0.037 & 0.007 $\pm$ 0.001 \\
USDCAD & 0.064 $\pm$ 0.028 & 0.017 $\pm$ 0.002 \\
USDJPY & 0.103 $\pm$ 0.037 & 0.008 $\pm$ 0.001 \\
\midrule
Mean & 0.079 $\pm$ 0.037 & 0.010 $\pm$ 0.001 \\
\bottomrule
\end{tabular}

    \end{adjustbox}
    \caption{
        KL divergences of the BQRBM model vs.~the classical RBM.
        The values are shown in the format mean \(\pm\) one standard deviation from an ensemble of 100 sample sets consisting of \( 10^4 \) samples each.
}
    \label{tbl:qbm_KL_divergences}
\end{table}

\begin{figure}[!htb]
    \begin{center}
        \includegraphics[width=0.7\linewidth]{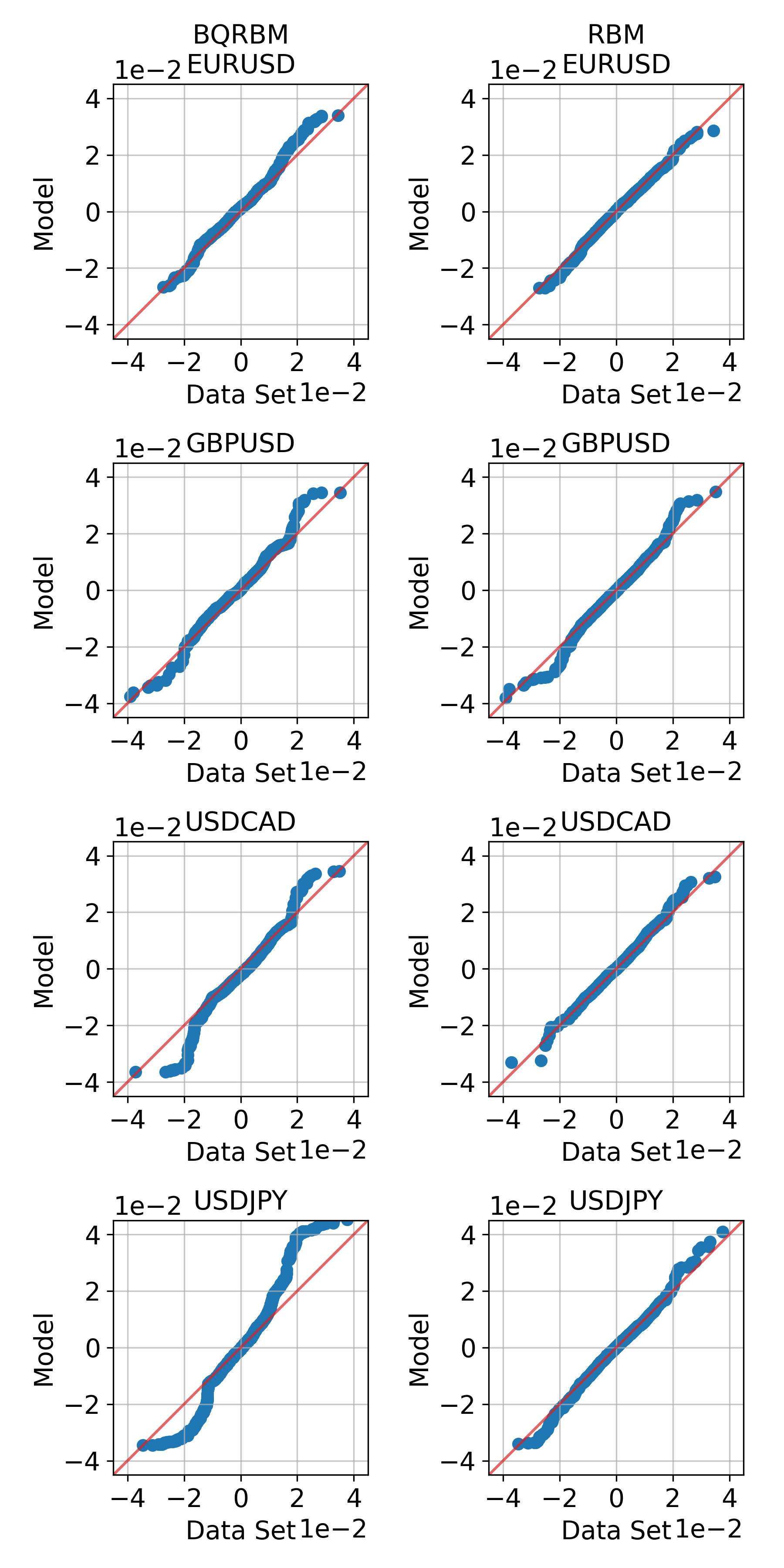}
    \end{center}
    \caption{Log return Q-Q plots of the BQRBM and classical RBM models for each currency pair. Note that these plots only use the same number of samples as the size of the training data set (5165), and thus are not entirely representative of the models' performances.}
    \label{fig:qbm_log_returns_qq}
\end{figure}

The Q-Q plots in~\cref{fig:qbm_log_returns_qq} point out that the BQRBM model struggles the most with the USDJPY and USDCAD marginals.

\cref{tbl:qbm_correlation_coefficients} show that the BQRBM is able to reproduce the structure of the correlation coefficients, albeit to a lesser extent than the classical RBM.
\begin{table}[!htb]
    \centering
    \begin{adjustbox}{max width=\textwidth}
        \begin{tabular}{l r r r r r r}\multicolumn{7}{c}{\textbf{Correlation Coefficients}} \\
\toprule
& \multicolumn{3}{c}{\textbf{Data Set}} & \multicolumn{3}{c}{\textbf{BQRBM}} \\\cmidrule(lr){2-4}
\cmidrule(lr){5-7}
Currency Pairs & Pearson & Spearman & Kendall & Pearson & Spearman & Kendall \\
\midrule
EURUSD/GBPUSD & 0.62 & 0.62 & 0.44 & 0.37 $\pm$ 0.04 & 0.44 $\pm$ 0.05 & 0.30 $\pm$ 0.04 \\
EURUSD/USDCAD & -0.44 & -0.41 & -0.29 & -0.25 $\pm$ 0.03 & -0.30 $\pm$ 0.04 & -0.20 $\pm$ 0.03 \\
EURUSD/USDJPY & -0.26 & -0.30 & -0.21 & -0.12 $\pm$ 0.03 & -0.16 $\pm$ 0.04 & -0.11 $\pm$ 0.02 \\
GBPUSD/USDCAD & -0.42 & -0.37 & -0.26 & -0.24 $\pm$ 0.02 & -0.28 $\pm$ 0.03 & -0.19 $\pm$ 0.02 \\
GBPUSD/USDJPY & -0.14 & -0.21 & -0.15 & -0.12 $\pm$ 0.02 & -0.15 $\pm$ 0.03 & -0.10 $\pm$ 0.02 \\
USDCAD/USDJPY & 0.00 & 0.06 & 0.04 & 0.05 $\pm$ 0.02 & 0.06 $\pm$ 0.02 & 0.04 $\pm$ 0.01 \\
\midrule
& \multicolumn{3}{c}{\textbf{RBM}} & \\\cmidrule(lr){2-4}
Currency Pairs & Pearson & Spearman & Kendall & & & \\
\midrule
EURUSD/GBPUSD & 0.48 $\pm$ 0.01 & 0.53 $\pm$ 0.01 & 0.38 $\pm$ 0.01 &  &  &  \\
EURUSD/USDCAD & -0.33 $\pm$ 0.01 & -0.34 $\pm$ 0.01 & -0.24 $\pm$ 0.01 &  &  &  \\
EURUSD/USDJPY & -0.21 $\pm$ 0.01 & -0.25 $\pm$ 0.01 & -0.17 $\pm$ 0.01 &  &  &  \\
GBPUSD/USDCAD & -0.31 $\pm$ 0.01 & -0.33 $\pm$ 0.01 & -0.22 $\pm$ 0.01 &  &  &  \\
GBPUSD/USDJPY & -0.15 $\pm$ 0.01 & -0.18 $\pm$ 0.01 & -0.13 $\pm$ 0.01 &  &  &  \\
USDCAD/USDJPY & 0.06 $\pm$ 0.01 & 0.07 $\pm$ 0.01 & 0.05 $\pm$ 0.01 &  &  &  \\
\bottomrule
\end{tabular}

    \end{adjustbox}
    \caption{Correlation coefficients of the data set vs.~samples generated by the BQRBM and classical RBM models. The BQRBM and RBM values are shown in the format mean \(\pm\) one standard deviation from an ensemble of 100 sample sets consisting of \( 10^4 \) samples each.}
    \label{tbl:qbm_correlation_coefficients}
\end{table}

Interestingly, the BQRBM is able to reproduce the volatilities for the most part except for the USDJPY, as seen in~\cref{tbl:qbm_volatilities}.
\begin{table}[!htb]
    \centering
    \begin{adjustbox}{max width=\textwidth}
        \begin{tabular}{l r r r}
\multicolumn{4}{c}{\textbf{Historical Volatilities}} \\
\toprule
Currency Pair & \textbf{Data Set} & \textbf{BQRBM} & \textbf{RBM}\\\midrule
EURUSD & 9.78\% & 10.17\% $\pm$ 0.79\% & 9.98\% $\pm$ 0.11\% \\
GBPUSD & 8.98\% & 8.86\% $\pm$ 0.47\% & 9.34\% $\pm$ 0.11\% \\
USDCAD & 8.56\% & 8.44\% $\pm$ 0.39\% & 8.98\% $\pm$ 0.13\% \\
USDJPY & 10.02\% & 11.58\% $\pm$ 1.44\% & 10.26\% $\pm$ 0.13\% \\
\bottomrule
\end{tabular}

    \end{adjustbox}
    \caption{
        Historical volatilities of the data set vs.~samples generated by the BQRBM and classical RBM models.
        The BQRBM and RBM values are shown in the format mean \(\pm\) one standard deviation from an ensemble of 100 sample sets consisting of \( 10^4 \) samples each.
    }
    \label{tbl:qbm_volatilities}
\end{table}

\begin{figure}[!htb]
    \begin{center}
        \includegraphics[width=1\linewidth]{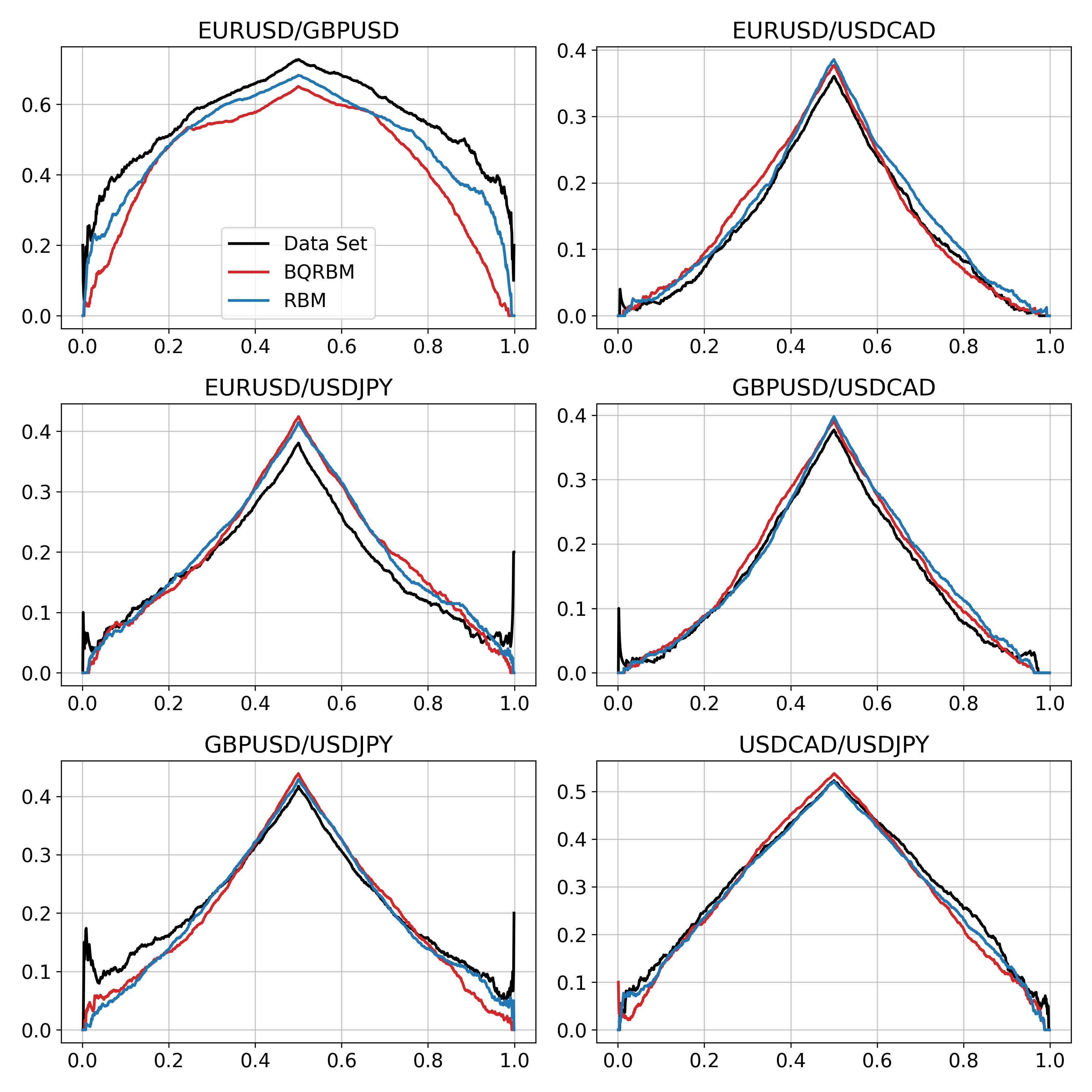}
    \end{center}
    \caption{Tail concentration functions of the data set vs.~samples generated by the BQRBM and classical RBM models.}
    \label{fig:qbm_log_returns_tail_concentrations}
\end{figure}

\cref{tbl:qbm_tails} shows the quantum model struggles more on the tails, particularly with the USDJPY, as well as EURUSD.
This is further confirmed by the tail concentration functions in~\cref{fig:qbm_log_returns_tail_concentrations}.
\begin{table}[!htb]
    \centering
    \begin{adjustbox}{max width=\textwidth}
        \begin{tabular}{l r r r}
\multicolumn{4}{c}{\textbf{Lower Tails (1st Percentile)}} \\
\toprule
Currency Pair & \textbf{Data Set} & \textbf{BQRBM} & \textbf{RBM} \\\midrule
EURUSD & -1.64\% & -2.02\% $\pm$ 0.24\% & -1.80\% $\pm$ 0.04\% \\
GBPUSD & -1.47\% & -1.45\% $\pm$ 0.12\% & -1.59\% $\pm$ 0.04\% \\
USDCAD & -1.40\% & -1.43\% $\pm$ 0.17\% & -1.54\% $\pm$ 0.05\% \\
USDJPY & -1.70\% & -2.32\% $\pm$ 0.38\% & -2.03\% $\pm$ 0.07\% \\
\bottomrule \\
\multicolumn{4}{c}{\textbf{Upper Tails (99th Percentile)}} \\
\toprule
Currency Pair & \textbf{Data Set} & \textbf{BQRBM} & \textbf{RBM} \\\midrule
EURUSD & 1.62\% & 1.70\% $\pm$ 0.23\% & 1.59\% $\pm$ 0.04\% \\
GBPUSD & 1.42\% & 1.49\% $\pm$ 0.07\% & 1.45\% $\pm$ 0.04\% \\
USDCAD & 1.51\% & 1.50\% $\pm$ 0.12\% & 1.61\% $\pm$ 0.04\% \\
USDJPY & 1.59\% & 2.13\% $\pm$ 0.50\% & 1.56\% $\pm$ 0.04\% \\
\bottomrule
\end{tabular}

    \end{adjustbox}
    \caption{
        Lower and upper tails, i.e., 1st and 99th percentiles, of the data set vs.~samples generated by the BQRBM and classical RBM models.
        The BQRBM and RBM values are shown in the format mean \(\pm\) one standard deviation from an ensemble of 100 sample sets consisting of \( 10^4 \) samples each.
    }
    \label{tbl:qbm_tails}
\end{table}

\subsection{Comparison to Gate-Based Models}
In the paper \textit{Quantum Versus Classical Generative Modelling in Finance} by Coyle et al.~\cite{coyle_2020}, they use a 32-qubit gate-based quantum computer to train both classical RBMs and quantum circuit Born machines (QCBMs) on a similar forex log returns data set.
Their work shows that the QCBM produces better results than the RBM, but they mostly focus on smaller models using 4, 6, 8, and 12 qubits due to the limited number of qubits available.
Therefore, it is not exactly a fair comparison against the results of our 94-qubit model here since the bits of precision are significantly higher, but the BQRBM trained in this section appears to perform better than their QCBM when comparing the Q-Q plots.

\subsection{Summary}
Overall, the BQRBM model trained using the Advantage 4.1 produces lackluster results when compared with the classical RBM, and does not motivate training additional models on the transformed and volatility indicator enhanced data sets.
This could possibly be due to hyperparameters, as our grid search of the space was limited by the training time requirements, but it is not immediately clear if there is a better way to do this.
This could also be due to the fact that the model is so large that it uses a significant portion of the QPU and requires chains with lengths of up to 7, resulting in added complexity.

In conclusion, it does not appear that the Advantage 4.1-trained BQRBM can produce results good enough to replace the classical RBM.
It will be interesting to see how much this improves with future generations of annealers, although it will likely take serious advances in the technology to outperform the classical RBM.

\clearpage
\section{Challenges of Using a D-Wave Annealer to Train QBMs}\label{sec:challenges}
Using a D-Wave quantum annealer to train quantum Boltzmann machines is a difficult task, and there are many challenges which need to be overcome in order to do so.
In this section we touch on some of these difficulties and discuss some possible methods to mitigate them.
Around the time this thesis was started, Pochart et al. released a paper~\cite{pochart_2021} in which they discuss challenges associated with using a D-Wave annealer to sample Boltzmann random variables.

\subsection{Choosing an Embedding}
Mapping the logical qubits to physical qubits is nontrivial.
D-Wave provides a heuristic method to find embeddings, but in practice it cannot be guaranteed that the returned embedding is optimal.
As we saw first hand, different embeddings can produce different results.
Therefore, it is recommended to generate multiple embeddings and compare them against each other, and choose the one that performs the best.
Additionally, it is worth noting that an optimal embedding on one QPU might not be optimal on another of the same generation.

\subsubsection{Chain Strength}
Depending on how large the problem is, one will likely need to use an embedding that is not direct, i.e., one that requires chains of physical qubits to represent single logical qubits due to limited connectivity.
This brings about an additional hyperparameter that needs to be tuned.
Rather than setting the chain strength directly though, it is recommended to use the relative chain strength as mentioned in~\cref{sec:qbm_rcs}.
It is best to do a comparison in the beginning to get an idea of what a good relative chain strength might be, as it is problem dependent.

\subsection{Sampling the Proper Distribution}
The most important thing when using a quantum annealer to train a Boltzmann machine is making sure the annealer is sampling from the proper distribution.
In the case of the quantum Boltzmann machine we need samples generated according to \( \rho = \frac{1}{Z} e^{-\beta H(s^*)} \).
For smaller problems it is easy to compare results obtained from the annealer with exact computed distributions (as in the 12-qubit problem), but it is not as simple for larger problems.
For larger problems there is the possibility to use advanced methods, such as they did in~\cite{marshall_2019} with the use of an entropic sampling technique~\cite{barash_2019} based on population annealing to estimate degeneracies, and in turn use those to compute classical Boltzmann distributions to compare with, but that might not always be practical.
Alternatively, one can try a hyperparameter grid search as in~\cref{sec:qbm_hyperparameters}

\subsubsection{Effective Temperature}
One of the most important hyperparameters is that of the effective inverse temperature \( \beta \).
In practice, we divide our weights and biases by a factor of \( -\betahat B(s^*) \) (as per~\cref{eq:qbm_scaling}) in order to cancel out the effective temperature so that we can sample the problem we wish to, thus it is crucial for proper parameter scaling.
For the case of \( s^* = 1 \) we have the ability to treat the effective temperature as a learnable parameter (as in~\cref{sec:learning_beta}), for which we use \( \betahat \) as an estimator of.
This is not so straightforward for \( s^* < 1 \) though, because of the initial Hamiltonian and the D-Wave's inability to measure the qubits in the \( x \)-direction.

\subsubsection{Anneal Schedules}
The ability to configure the anneal schedule as allowed by the D-Wave annealer means that there are a number of different ways one can tweak the annealing process such that the results returned minimize the KL divergence between the theoretical distribution one wishes to approximate and the samples returned by the annealer.
In an ideal world, the way to get the desired distribution is to anneal slowly at first, then quench the system at the point \( s^* \) and measure the qubits~\cite{amin_2018}.
Unfortunately, the research conducted in this thesis seems to indicate that the current generation of D-Wave annealers cannot quench fast enough to prevent any nontrivial dynamics occurring after \( s^* \), and all sample sets we collected are more similar to classical Boltzmann distributions than quantum ones.
With that said, the annealer can still be used to assist in the training of a classical Boltzmann machine.

\subsection{QPU Limitations and Imperfections}
The properties of the QPU itself must also be taken into account.
There is no doubt that D-Wave is a top-notch manufacturer of quantum annealers, but even with all of their expertise the QPUs are still subject to imperfections and errors.
It is possible for some areas of the chip to perform better than others, or for some of the qubits to have readout biases (although biases can be mitigated by using gauge transformations as detailed in~\cref{sec:gauge}).

\subsubsection{Maximum Sample Set Size}
One of the main limitations of the D-Wave annealer for this purpose is that of the maximum sample set size.
When sampling, the D-Wave one can only generate sample sets with a maximum size of \( 10^4 \) samples, which is adequate for the intended purpose of optimization, but can fall short when one wants to use it as a sampler for a QBM.
It is natural to think that one could just combine the results from multiple sample sets, but this is not necessarily the case.
Due to the spin-bath polarization effect (see error sources below), one cannot combine sample sets because of the possibility of previous samples affecting future ones~\cite{pochart_2021}.

\subsubsection{Time Requirements}
Another issue is the time requirements.
Most models detailed here require little QPU access time to train, around 5-10 minutes, but this can add up and get expensive if one is doing a hyperparameter grid search.
Additionally, there is the total training time, which as we saw can be quite substantial due the latency to the cloud platform combined with the load queue of the solver.
To illustrate this point, training the simulation-based 12-qubit model takes roughly 2 seconds per epoch (the majority of which is spent computing the density matrix), whereas the Advantage 4.1-based model takes 2-5 minutes per epoch.

\subsubsection{Error Sources}
There are a number of sources from which errors can arise on a D-Wave quantum annealer.
D-Wave does an excellent job at detailing these errors in their documentation~\cite{dwave_ice_errors,dwave_other_errors}, so we will only briefly touch on them here with high-level information obtained from the aforementioned references.
\begin{itemize}
    \item \textbf{Integrated Control Errors (ICE)} are errors due to the accuracy at which the \( h_i \) and \( J_{ij} \) values can be implemented.
        In mathematical terms this is because the problem the QPU solves is closer to
        \begin{align}
            H_{\text{Ising}}^\delta = \sum_{i=1}^{n} (h_i + \delta_{h_i}) \sigma_i^z + \sum_{i=1}^{n}\sum_{j=i+1}^n (J_{ij} + \delta_{J_{ij}}) \sigma_i^z \sigma_j^z,
        \end{align}
        for some small \( \delta_{h_i} \) and \( \delta_{J_{ij}} \).
    \item \textbf{Temperature} errors arise due to fluctuations in the physical temperature of the device, which can change depending on how frequently the QPU is programmed.
    \item \textbf{High-Energy Photon Flux} errors can occur in the presence of photons with energies higher than that expected at the effective temperature dependent equilibrium, which can lead to higher energy solutions. These photons originate from cryogenic filtering at higher temperature phases.
    \item \textbf{Readout Fidelity} errors can occur when the bit string returned by the annealer differs from that arrived at by the QPU by one or more bit flips.
        For reference, D-Wave annealers have a readout fidelity of >99\%.
    \item \textbf{Programming Errors} can occur when the problem implemented by the QPU suffers from programming issues resulting in the implemented problem's low-energy subspace not having an overlap with that of the desired problem.
    \item \textbf{Spin-Bath Polarization Effect} errors can arise when the current flowing through the qubits during the annealing process causes the spins to obtain a polarization which can bias the measurements.
\end{itemize}

\chapter{Conclusion}
\label{ch:conclusion}
\section{Summary}
We started with an analysis of the forex log returns data set in~\cref{ch:data_analysis}, analyzing the data from a number of aspects to get an understanding of the intricacies.
After that, we moved to training classical RBM models in~\cref{ch:rbm}, where we were able to produce good results similar to those in~\cite{kondratyev_2019}.
The outlier power transformation detailed in~\cref{sec:outlier_transform} shows much promise, as models trained on the transformed data sets perform noticeably better than those trained on the base data sets.

After establishing a classical baseline to compare our quantum models with, we studied a small 12-qubit problem in~\cref{sec:qbm_12_qubit_problem}, through which we gained a deeper understanding of how to sample quantum Boltzmann random variables using a D-Wave quantum annealer, specifically the Advantage 4.1.
There, we were able to match sample distributions returned by the annealer to theoretical distributions from the family of distributions corresponding to the density operator \( \rho(s,T) = \frac{1}{Z}e^{-\beta H(s)} \).
Our findings indicate that, with the anneal schedules and parameters used here, the Advantage 4.1 is not able to sample from just any quantum Boltzmann distribution, rather only those that are classical Boltzmann-like in nature.
To be more specific, the samples we obtained from the annealer resemble a subset of the family of distributions that satisfies \( B(s) / T = \text{constant} \), as indicated by the streak patterns observed in~\cref{fig:dkl_min_heatmap}.
This occurs when the distribution is similar to one late in the anneal process, i.e., when \( e^{-\beta H(s^*)} \approx e^{-\beta B(s^*) H_\text{final}} \).
This is likely due to the annealer not being able to quench the system fast enough, allowing for nontrivial dynamics to occur, as the shortest allowed quench durations are still quite long relative to the qubit oscillation frequency in terms of gigahertz.
How closely the annealer can approximate a desired classical Boltzmann distribution was found to be dependent on both the embedding and the anneal schedule, thus it is highly recommended to tune these accordingly.

With the information that we can only reliably sample classical Boltzmann distributions, we moved to training a bound-based quantum restricted Boltzmann machine (BQRBM) with a freeze-out point of \( s^* = 1 \), essentially reducing the problem to a classical RBM trained using quantum assistance.
The difficulty of choosing the effective temperature was in this case easily circumvented by treating \( \beta \) as a learnable parameter as described in~\cref{sec:learning_beta} and verified in~\cref{sec:qbm_simulation_results}.
We trained BQRBM models using both a simulation and the Advantage 4.1 annealer, allowing us to compare exactly how close the Advantage 4.1-trained model is to the theory.
Additionally, we trained a classical RBM to use as a reference point.
In short, the BQRBM model trained using the Advantage 4.1 underperforms both the classical RBM and the simulation, as seen in~\cref{sec:qbm_annealer_results}.
The simulation-based model shows promise though, outperforming the classical RBM, offering hope for future annealer-trained models if annealers can further reduce the information loss associated with sampling (quantum) Boltzmann distributions.

Finally, we used the knowledge gained about how to train a small BQRBM and applied it to training a larger one in~\cref{sec:quantum_market_generator} using the log returns data set, mapping 94 logical qubits to 398 physical qubits with chain lengths of up to 7.
This model proved to be more challenging to train because setting the annealer hyperparameters (chain strength, anneal schedule, and embedding) cannot be done as in the 12-qubit problem due to the fact that we cannot simulate such a large system.
In practice, we had to choose these values by doing a limited hyperparameter scan, which was difficult due to increased training times that averaged around 15 minutes per epoch.
Longer epoch times originated from a combination of solver load and latency from Europe to the North American West Coast.
This meant that training a model for 100 epochs would have taken around a day, and if we wanted to fully train the models for all hyperparameters in our scan it would have taken weeks.

The results in~\cref{sec:qbm_log_returns_results} show that the BQRBM was able to learn to produce synthetic data similar to the log returns data set distribution to some extent, but drastically underperforms the classical RBM.
This could likely have been improved with a more exhaustive hyperparameter scan, but that was not necessarily feasible given the time requirements, and it is unclear if the results would have been significantly better given that even the 12-qubit BQRBM trained on the annealer underperforms the classical RBM.

In this thesis we laid out a framework with which one can train quantum Boltzmann machines using both simulations and D-Wave quantum annealers.
As part of this thesis, the Python package \texttt{qbm}~\cite{qbm} was developed to make it easier to train and study QBMs.
This package is open source and available to the public to encourage further study of QBMs.

Overall, this thesis furthered not only our understanding of QBMs, but that of D-Wave annealer sampling in general.
We hope that this work will be useful for future research and development.

\section{Future Directions}
Throughout this thesis we came across several directions which we would have liked to explore more in depth but did not have the time to.

It would be interesting to investigate if adding technical indicators to the log returns data set could increase model performance.
Given that the log returns data set used here only takes into account the currency pairs' behavior over one day (excluding the volatility indicators), technical indicators calculated using data over a historical window could enrich the data set with vital information to help the model better learn the complexities of the distribution.

The discretization procedure for converting continuous data into bit vectors could probably be further improved.
As we saw in~\cref{sec:classical_market_generator}, the models that used the outlier power-transformed data sets generate samples with lower KL divergences and better reproduced the correlations between the currency pairs.

Of most interest is simulating the time-dependent Schr\"odinger equation of the D-Wave annealer to determine how fast the system needs to quench in order to freeze out the dynamics.
This would give a good indication of how much quantum annealers need to improve in order to be able to sample from arbitrary quantum Boltzmann distributions.

Studying additional anneal schedule formats would also be a very interesting direction.
Reverse annealing was tested to a small extent here only to see if it produced drastically different results than forward annealing, but was left out of the final research because the results did not show any significant improvements and led to added complexity due to the need to choose what state the system was initialized in and if the system was reinitialized to the same state after each measurement or not.

\appendix
\begin{appendices}
\label{ch:appendix}
\chapter{Definitions and Methodologies}
\section{Correlation Coefficients}\label{app:correlation_coefficients}
The Pearson correlation coefficient is defined as
\begin{align}
    \rho_{X,Y} = \frac{\text{cov}{(X,Y)}}{\sigma_X \sigma_Y} \in [-1, 1],
\end{align}
and measures the linear correlation between the random variables \( X \) and \( Y \).
Therefore, it must be noted that this does not capture nonlinear relations, and should not be relied upon to tell the full story.
Additionally, this measure is quite sensitive to outliers.

The Spearman rank correlation coefficient is defined as
\begin{align}
    r_s = \rho_{R(X),R(Y)} = \frac{\text{cov}{\big(R(X),R(Y)\big)}}{\sigma_{R(X)} \sigma_{R(Y)}} \in [-1, 1],
\end{align}
and is the Pearson correlation coefficient of the rank of the random variables \( X \) and \( Y \).
The main difference to the Pearson correlation coefficient is that the Spearman measures the monotonic relationship, regardless of linearity.
The Spearman correlation coefficient is also less sensitive to outliers than the Pearson.

The Kendall rank correlation coefficient is defined as
\begin{align}
    \tau = \frac{2}{n(n-1)} \sum_{i=1}^{n}\sum_{j=i+1}^{n} \text{sign}(x_i - x_j) \text{sign}(y_i - y_j) \in [-1, 1],
\end{align}
where \( (x_1, y_1), \dots, (x_n, y_n) \) are pairs of observations of the random variables \( X \) and \( Y \).

It is important to keep in mind how one interprets the correlation coefficients.
The sign of the correlation coefficient determines whether the variables are negatively or positively correlated, and the magnitude determines how strong the correlation effects are.
As a loose guide, correlation coefficient values of 0.1, 0.3, and 0.5 can be termed small, medium, and large, respectively~\cite{research_design_and_statistical_analysis}.
In general, one must be careful when interpreting the correlation coefficients; it is important to understand what the values mean, and what they do not.
Section 3.4.2 "Interpreting the Correlation Coefficient" of~\cite{research_design_and_statistical_analysis} offers further insight and points out some pitfalls to watch out for.

In this thesis the correlation coefficients are computed using the respective functions from the SciPy Python package~\cite{python_scipy}.

\section{Annualized Volatility}\label{app:annualized_volatility}
In finance, the annualized volatility of a time series vector \( \vec{x} \) is computed as
\begin{align}
    \text{vol}(\vec{x}) = \sqrt{252} \cdot \text{std}(\vec{x}),
\end{align}
where the factor of \( \sqrt{252} \) comes from the square root of the number of trading days in a year, i.e., it's the annualization factor.

\section{Learning Rate Decay Schedule}\label{app:lr_exp_decay}
The learning rate at epoch \( t \) is given by
\begin{align}
    \eta^{(t)}
        &= \eta^{(0)} \cdot \min\bigg\{1, 2^{-\frac{t - t_\text{decay}}{T_\text{decay}}}\bigg\},
\end{align}
where \( \eta^{(0)} \) is the initial learning rate, \( t_\text{decay} \) is the epoch at which the decay begins, and \( T_\text{decay} \) is the decay period.

\section{Autocorrelation Analysis}\label{app:autocorrelation_analysis}
When studying results from an MCMC-based model it is important to be aware that sequentially generated samples are not always statistically independent, that is, there is some thermalization threshold that corresponds to the minimum number of sampling steps between samples to consider them as statistically independent.

For a time series \( x_1, \dots, x_n \), the lag-\( k \) autocorrelation function is defined as~\cite{time_series_analysis}
\begin{align}
    \rho_k
        &= \frac{\text{cov}(x_t, x_{t+k})}{\sigma_{x}^2}.
\end{align}
The autocorrelation function is essentially the Pearson correlation coefficient, except instead of comparing two different variables it compares the same variable at different times.
In this thesis we use the statsmodels Python package~\cite{python_statsmodels} to compute the autocorrelation function as there are some caveats when computing it in practice with large chains, e.g., there are some tricks such as using a Fourier transformation to make the computations more efficient.

The integrated autocorrelation time is a reasonable estimate of how many steps in between samples we should have before we can consider them to be (to a degree) statistically independent.
In this thesis we use the emcee Python package~\cite{python_emcee} to estimate the integrated autocorrelation time, which follows the approach laid out by Goodman and Weare in~\cite{goodman_weare_2010}.

\section{Kullback-Leibler Divergence}\label{app:kl_divergence}
The Kullback-Leibler divergence~\cite{kullback_1951} is a measure of how much the probability distribution \( q \) differs from the reference probability distribution \( p \).
It is defined as
\begin{align}
    \DKL{p}{q}
        &= \sum_{x\in\mathcal{X}} p(x) \log\frac{p(x)}{q(x)},
\end{align}
where \( \mathcal{X} \) is the probability space.
It can be interpreted as the amount of information loss associated with using \( q \) to approximate \( p \).
We also note that the KL divergence is a distance, but not a metric (rather a divergence), because of the asymmetry that \( \DKL{p}{q} \ne \DKL{q}{p} \).

\subsection{Kullback-Leibler Divergence in Practice}\label{app:kl_divergence_in_practice}
Due to the limited maximum sample size of \( 10^4 \) when using a D-Wave annealer and the inability to concatenate sample sets due to spin-bath polarization effects~\cite{pochart_2021}, it makes computation of the KL divergence quite difficult because we cannot get a proper read on the probability distribution when the number of possible states is high.
Even for the small 12-qubit problem there are still \( 2^{12} = 4096 \) possible states, thus \( 10^4 \) samples are not entirely representative of the true distribution.
This problem is only exacerbated when working with larger system sizes.

Therefore, in this thesis we take a histogram-based approach to approximate the KL divergence.
All KL divergences are computed using 32 bins since this is close to the number of bins computed using the Freedman-Diaconis rule on some of the sample sets for the 12-qubit problem.

When computing the \( q \) distribution from a sample set of limited size, it is often the case that some probabilities come out to zero, which in turn leads to issues computing the KL divergence due to zeros in the denominator of the argument of the log.
Luckily, there is a way around this if we know the true probability of measuring such a state to be nonzero.
Due to the quantum nature of this problem and the fact that no state has a truly zero probability (although some infinitesimally small), we can take such an approach.

The method we use to mitigate this problem is called smoothing~\cite{han_kl_divergence}, in which we add some small probability \( \epsilon \) to the \( q \) distribution probabilities that are observed to be zero, then take the sum of the added probabilities and evenly subtract it from the nonzero probabilities in order to ensure the distribution remains normalized.
For example, if \( \{q_1 = 1/3, q_2 = 2/3, q_3 = 0, q_4 = 0\} \), then the corresponding smoothed distribution is \( \{q_1 = 1/3 - \epsilon, q_2 = 2/3 - \epsilon, q_3 = \epsilon, q_4 = \epsilon\} \).

Furthermore, we call it \textit{relative} smoothing when the smoothed probabilities are taken to be relative to the reference distribution \( p \).
This is useful when it is difficult to choose a constant value of \( \epsilon \), e.g., when the reference distribution probabilities vary widely and can coincide with \( \epsilon \).
For example, if \( \{q_1 = 1/3, q_2 = 2/3, q_3 = 0, q_4 = 0\} \), then the corresponding relative smoothed distribution is \( \{q_1 = 1/3 - \epsilon (p_3 + p_4)/2, q_2 = 2/3 - \epsilon (p_3 + p_4)/2, q_3 = \epsilon p_3, q_4 = \epsilon p_4\} \).

We take a value of \( \epsilon = 10^{-6} \) when computing \( \DKL{\pdata}{\pmodel} \) because it is small enough that it will not coincide with any of the \( \pdata \) values since the data sets contain only a few thousand samples, thus the smallest value of \( \pdata \) is roughly on the order of \( 10^{-4} \).
When computing \( \DKL{\ptheory}{\psamples} \) though, we opt to use relative smoothing with a value of \( \epsilon = 10^{-6} \) since sometimes some probabilities of \( \ptheory \) can be close to \( \epsilon \) and give a false sense of agreement with the smoothed \( \psamples \).

\section{Tail Concentration Functions}\label{app:tail_concentration_functions}
The lower tail concentration function is defined as~\cite{venter_2002}
\begin{align}
\begin{split}
    L(z)
        &= \frac{p(U_1 \le z, U_2 \le z)}{z} \\
        &= \frac{C(z,z)}{z},
\end{split}
\end{align}
and the upper as
\begin{align}
\begin{split}
    R(z)
        &= \frac{p(U_1 > z, U_2 > z)}{1-z} \\
        &= \frac{1 - 2z + C(z,z)}{1-z},
\end{split}
\end{align}
where \( U_1 \) and \( U_2 \) are uniform random variables on the interval \( [0, 1] \), and \( C(u_1, u_2) \) is the copula of \( (U_1, U_2) \).

In practice, we compute \( U_1 \) and \( U_2 \) as the normalized rank of the observations of the random variables \( X \) and \( Y \), respectively.
The way to interpret the concentration functions is that they represent the probability that \( X \) and \( Y \) simultaneously take on extreme values.
When plotted, the lower tail concentration function is used for \( 0 \le z \le 0.5 \) and the upper for \( 0.5 < z \le 1 \).

A nice explanation with animations can be found at~\cite{charpentier_2012}.

\section{Exact Computation of \( \rho \)}\label{app:exact_rho_computation}
For the density matrix
\begin{align}
    \rho = \frac{1}{Z} e^{-\beta H},
\end{align}
we can compute this as
\begin{align}
    \rho
        &= \frac{1}{\tr(A)} S A S^{-1},
\end{align}
where
\begin{align}
    A = \text{diag}\Big(e^{-\beta(\lambda_1 - \min_i\{\lambda_i\})}, \dots, e^{-\beta(\lambda_{2^n} - \min_i\{\lambda_i\})}\Big),
\end{align}
where \( \{\lambda_i\} \) are the eigenvalues of \( H \), and \( S \) is the matrix of eigenvectors that transforms \( H \) to and from its eigenspace.
We subtract \( \min_i\{\lambda_i\} \) from the eigenvalues in practice to avoid computing the exponential of a large number which can lead to divergence in floating point calculations.

\section{Constants}\label{app:constants}
The values of \( A(s) \) and \( B(s) \) are in terms of \si{\giga\hertz}, and consequently so is the Hamiltonian.
The density matrix is of the form \( \rho = e^{-\beta H(s)} \), where \( \beta = 1 / kT \), so in order to obtain the (effective) temperature we need the argument of the exponential to be dimensionless, i.e., \( k \) must be in terms of \( \si{\giga\hertz} \cdot \si{\kelvin}^{-1} \).
This is achieved by taking a value of
\begin{align}
\begin{split}
    k
        &= \frac{k_B}{h} \\
        &\approx \frac{1.380649 \cdot 10^{-23} \ \si{\joule} \cdot \si{\kelvin}^{-1}}{6.62607015 \cdot 10^{-34} \ \si{\joule} \cdot \si{\hertz}^{-1}} \\
        &\approx 2.083661912 \cdot 10^{10} \ \si{\hertz} \cdot \si{\kelvin}^{-1} \\
        &= 20.83661912 \ \si{\giga\hertz} \cdot \si{\kelvin}^{-1}.
\end{split}
\end{align}

\chapter{Restricted Boltzmann Machine}
\section{Conditional Probabilities}\label{app:conditional_probabilities_derivation}
This derivation follows along the lines of that found on p. 658-659 of~\cite{goodfellow_deep_learning}.
We start by noting~\cref{eq:rbm_joint_probability}
\begin{align}
    p(\vec{v,h}) = \frac{1}{Z} e^{-E(\vec{v},\vec{h})}.
\end{align}
From this we can derive the conditional probability using~\cref{eq:rbm_energy,eq:rbm_partition_function}
\begin{align}
\begin{split}
    p(\vec{h} | \vec{v})
        &= \frac{p(\vec{v},\vec{h})}{p(\vec{v})} \\
        &= \frac{1}{p(\vec{v})} \frac{1}{Z} \exp( \vec{a}\T\vec{v} + \vec{b}\T\vec{h} + \vec{v}\T\mat{W}\vec{h} ) \\
        &= \frac{1}{Z'} \exp\bigg( \sum_{j=1}^{n_h} b_j h_j + \sum_{j=1}^{n_h} (\vec{v}\T\mat{W})_j h_j \bigg) \\
        &= \frac{1}{Z'} \prod_{j=1}^{n_h} \exp\big( b_j h_j + (\vec{v}\T\mat{W})_j h_j \big),
\end{split}
\end{align}
with
\begin{align}
    Z' = \sum_\vec{h} \exp( \vec{b}\T\vec{h} + \vec{v}\T\mat{W}\vec{h} ),
\end{align}
where \( \sum_{\vec{h}} \) denotes the sum over all possible configurations of \( \vec{h} \).
This leads us to
\begin{align}
\begin{split}
    p(h_j = 1 | \vec{v})
        &= \frac{\tilde{p}(h_j = 1 | \vec{v})}{\tilde{p}(h_j = 0 | \vec{v}) + \tilde{p}(h_j = 1 | \vec{v})} \\
        &= \frac{\exp\big( b_j + (\vec{v}\T\mat{W})_j \big)}{1 + \exp\big( b_j + (\vec{v}\T\mat{W})_j \big)} \\
        &= \sigma\big( b_j + (\vec{v}\T\mat{W})_j \big).
\end{split}
\end{align}
Finally, we have
\begin{align}
    p(\vec{h} | \vec{v}) = \prod_{j=1}^{n_h} \sigma\big( (2\vec{h} - 1) \odot (\vec{b} + \mat{W}\T\vec{v}) \big)_j.
\end{align}
Analogously for \( p(\vec{v} | \vec{h}) \) one finds
\begin{align}
    p(\vec{v} | \vec{h}) = \prod_{i=1}^{n_v} \sigma\big( (2\vec{v} - 1) \odot (\vec{a} + \mat{W}\vec{h}) \big)_i.
\end{align}

\section{Log-Likelihood Derivative}\label{app:rbm_log_likelihood_derivation}
For the data set distribution \( p_\text{data} \) and parameters \( \theta = (\mat{W}, \vec{a}, \vec{b}) \) the log-likelihood is given by
\begin{align}
\begin{split}
    \ell(\theta)
        &= \sum_{\vec{v}} p_{\text{data}}(\vec{v}) \log p(\vec{v}) \\
        &= \sum_{\vec{v}} p_{\text{data}}(\vec{v}) \log \sum_\vec{h} p(\vec{v},\vec{h}) \\
        &= \sum_{\vec{v}} p_{\text{data}}(\vec{v}) \log \bigg(\frac{1}{Z} \sum_\vec{h} e^{-E(\vec{v},\vec{h})}\bigg) \\
        &= \sum_{\vec{v}} p_{\text{data}}(\vec{v}) \log \sum_\vec{h} e^{-E(\vec{v},\vec{h})} - \log \sum_{\vec{v},\vec{h}} e^{-E(\vec{v},\vec{h})}.
\end{split}
\end{align}
Taking the partial derivative we find
\begin{align}
\begin{split}
    \partial_{\theta} \ell(\theta)
        &= \sum_{\vec{v}} p_{\text{data}}(\vec{v}) \frac{\sum_\vec{h} e^{-E(\vec{v},\vec{h})} \partial_{\theta}\big( -E(\vec{v},\vec{h}) \big) }{\sum_\vec{h} e^{-E(\vec{v},\vec{h})}}
            - \frac{\sum_{\vec{v},\vec{h}} e^{-E(\vec{v},\vec{h})} \partial_{\theta}\big( -E(\vec{v},\vec{h}) \big) }{\sum_{\vec{v},\vec{h}} e^{-E(\vec{v},\vec{h})}} \\
        &= \sum_{\vec{v}} p_{\text{data}}(\vec{v}) \Big\langle \partial_{\theta}\big( -E(\vec{v},\vec{h}) \big) \Big\rangle_{p(\vec{h}|\vec{v})}
        - \Big\langle \partial_{\theta}\big( -E(\vec{v},\vec{h}) \big) \Big\rangle_{p(\vec{v},\vec{h})}.
\end{split}
\end{align}
This gives us
\begin{align}
\begin{split}
    \partial_{w_{ij}} \ell(\theta)
        &= \langle v_i h_j \rangle_{\text{data}} - \langle v_i h_j \rangle_{\text{model}}, \\
    \partial_{a_i} \ell(\theta)
        &= \langle v_i \rangle_{\text{data}} - \langle v_i \rangle_{\text{model}}, \\
    \partial_{b_j} \ell(\theta)
        &= \langle h_j \rangle_{\text{data}} - \langle h_j \rangle_{\text{model}},
\end{split}
\end{align}
where \( \langle \ \cdot \ \rangle_{\text{data}} \) denotes the expectation value with respect to the data set distribution, and \( \langle \ \cdot \ \rangle_{\text{model}} \) denotes the expectation value with respect to the model distribution.

\chapter{Quantum Boltzmann Machine}
\section{Log-Likelihood Derivative}\label{app:qbm_log_likelihood_derivation}
This derivation follows along the lines of that laid out in~\cite{amin_2018}.
We start with the log-likelihood
\begin{align}
\begin{split}
    \ell(\theta)
        &= \sum_{\vec{v}} p_{\text{data}}(\vec{v}) \log p(\vec{v}) \\
        &= \sum_{\vec{v}} p_{\text{data}}(\vec{v}) \log \frac{\tr(\Lambda_\vec{v} e^{-H})}{\tr(e^{-H})} \\
        &= \sum_{\vec{v}} p_{\text{data}}(\vec{v}) \Big[ \log\tr(\Lambda_\vec{v} e^{-H}) - \log\tr(e^{-H}) \Big],
\end{split}
\end{align}
where \( \sum_{\vec{v}} \) denotes the sum over all possible configurations of \( \vec{v} \).
Taking the partial derivative yields
\begin{align}
    \label{eq:qbm_log_likelihood_derivative}
    \partial_\theta \ell(\theta)
        &= \sum_{\vec{v}} p_{\text{data}}(\vec{v}) \bigg[ \frac{\tr(\Lambda_\vec{v} \partial_\theta e^{-H})}{\tr(\Lambda_\vec{v} e^{-H})} - \frac{\tr(\partial_\theta e^{-H})}{\tr(e^{-H})} \bigg].
\end{align}
Due to the noncommutativity of \( H \) and \( \partial_\theta H \), we need to use the trick laid out in~\cite{amin_2018} where we take \( e^{-H} = (e^{-\delta\tau H})^n \) with \( \delta\tau \equiv 1 / n \), which allows one to write
\begin{align}
    \partial_\theta e^{-H}
        &= -\sum_{m=1}^{n} e^{-m\delta\tau H} \delta\tau \partial_\theta He^{-(n-m)\delta\tau H} + \mathcal{O}(\delta\tau^2).
\end{align}
Taking the limit as \( n \rightarrow \infty \) of both sides gives
\begin{align}
\begin{split}
    \partial_\theta e^{-H}
        &= \lim_{n\rightarrow\infty} -\sum_{m=1}^{n} e^{-m\delta\tau H} \delta\tau \partial_\theta He^{-(n-m)\delta\tau H} + \mathcal{O}(\delta\tau^2) \\
        &= -\int_{0}^{1} d\tau e^{-\tau H} \partial_\theta H e^{(\tau-1)H}.
\end{split}
\end{align}
From here one can take the trace of both sides to arrive at
\begin{align}
\begin{split}
    \tr(\partial_\theta e^{-H})
        &= -\tr\bigg( \int_{0}^{1} d\tau e^{-\tau H} \partial_\theta H e^{(\tau-1)H} \bigg) \\
        &= -\int_{0}^{1} d\tau \tr\big(e^{-\tau H} \partial_\theta H e^{(\tau-1)H} \big) \\
        &= -\int_{0}^{1} d\tau \tr\big(e^{(\tau-1)H} e^{-\tau H} \partial_\theta H \big) \\
        &= -\int_{0}^{1} d\tau \tr\big(e^{-H} \partial_\theta H \big) \\
        &= -\tr\big(e^{-H} \partial_\theta H \big),
\end{split}
\end{align}
which gives
\begin{align}
\begin{split}
    \frac{\tr(\partial_\theta e^{-H})}{\tr(e^{-H})}
        &= -\frac{\tr(e^{-H} \partial_\theta H)}{\tr(e^{-H})} \\
        &= -\tr(\rho \partial_\theta H) \\
        &= -\langle \partial_\theta H \rangle.
\end{split}
\end{align}
Unfortunately, due to the additional factor of \( \Lambda_\vec{v} \) in the first term of \cref{eq:qbm_log_likelihood_derivative}, one arrives at
\begin{align}
\begin{split}
    \tr(\Lambda_\vec{v} \partial_\theta e^{-H})
        &= -\tr\bigg( \int_{0}^{1} d\tau \Lambda_\vec{v} e^{-\tau H} \partial_\theta H e^{(\tau-1)H} \bigg) \\
        &= -\int_{0}^{1} d\tau \tr\big(\Lambda_\vec{v} e^{-\tau H} \partial_\theta H e^{(\tau-1)H} \big),
\end{split}
\end{align}
which is nontrivial to compute in practice.

\section{Log-Likelihood Lower Bound}\label{app:qbm_log_likelihood_lower_bound}
This derivation follows along the lines of that laid out in~\cite{amin_2018}.
The Golden-Thompson inequality that \( \tr(e^{A}e^{B}) \ge \tr(e^{A+B}) \) allows one to write (for small \( \epsilon > 0 \))
\begin{align}
    \tr(e^{-H} e^{\log(\Lambda_\vec{v}+\epsilon)}) \ge \tr(e^{-H+\log(\Lambda_\vec{v}+\epsilon)}).
\end{align}
Taking the limit \( \epsilon \rightarrow 0 \) yields
\begin{align}
    \tr(\Lambda_\vec{v}e^{-H}) \ge \tr(e^{-H_\vec{v}}),
\end{align}
where
\begin{align}
    H_\vec{v} &= \braket{\vec{v} | H | \vec{v}}
\end{align}
is the \textit{clamped} Hamiltonian.
This is called clamped because the visible qubits are held to the classical state of the visible vector \( \vec{v} \) due to an infinite energy penalty imposed by the \( \log(\Lambda_\vec{v} + \epsilon) \) term.
Using this we can write the inequality
\begin{align}
\begin{split}
    p(\vec{v})
        &= \frac{\tr(\Lambda_\vec{v} e^{-H})}{\tr(e^{-H})} \\
        &\ge \frac{\tr(e^{-H_\vec{v}})}{\tr(e^{-H})},
\end{split}
\end{align}
which in turn allows for the log-likelihood to be bounded as
\begin{align}
    \ell(\theta) \ge \tilde{\ell}(\theta),
\end{align}
where
\begin{align}
    \tilde{\ell}(\theta)
        &= \sum_\vec{v} p_\text{data}(\vec{v}) \log\frac{\tr(e^{-H_\vec{v}})}{\tr(e^{-H})}.
\end{align}
\section{Log-Likelihood Lower Bound Derivative}\label{app:qbm_log_likelihood_lower_bound_derivative}
This derivation follows along the lines of that laid out in~\cite{amin_2018}.
Taking the partial derivative of the log-likelihood lower bound yields
\begin{align}
\begin{split}
    \label{eq:qbm_log_likelihood_derivative_lower_bound}
    \partial_\theta \tilde{\ell}(\theta)
        &= \sum_{\vec{v}} p_{\text{data}}(\vec{v}) \bigg[ \frac{\tr(\partial_\theta e^{-H_\vec{v}})}{\tr(e^{-H_\vec{v}})} - \frac{\tr(\partial_\theta e^{-H})}{\tr(e^{-H})} \bigg] \\
        &= \sum_{\vec{v}} p_{\text{data}}(\vec{v}) \bigg[ \frac{\tr(-e^{-H_\vec{v}} \partial_\theta H_\vec{v})}{\tr(e^{-H_\vec{v}})} - \frac{\tr(-e^{-H} \partial_\theta H)}{\tr(e^{-H})} \bigg] \\
        &= \sum_{\vec{v}} p_{\text{data}}(\vec{v}) [ \tr(-\rho_\vec{v} \partial_\theta H_\vec{v}) - \tr(-\rho \partial_\theta H) ] \\
        &= \sum_{\vec{v}} p_{\text{data}}(\vec{v}) [ \langle -\partial_\theta H_\vec{v} \rangle_\vec{v} - \langle -\partial_\theta H \rangle ] \\
        &= \overline{\langle -\partial_\theta H_\vec{v} \rangle_\vec{v}} - \langle -\partial_\theta H \rangle.
\end{split}
\end{align}
Plugging in our parameters we get
\begin{align}
\begin{split}
    \label{eq:qbm_log_likelihood_partials}
    \partial_{w_{ij}} \tilde{\ell}(\theta)
        &= \overline{\langle \sigma_i^z \sigma_j^z \rangle_\vec{v}} - \langle \sigma_i^z \sigma_j^z \rangle \\
        &= \langle \sigma_i^z \sigma_j^z \rangle_\text{data} - \langle \sigma_i^z \sigma_j^z \rangle_\text{model}, \\
    \partial_{b_i} \tilde{\ell}(\theta)
        &= \overline{\langle \sigma_i^z \rangle_\vec{v}} - \langle \sigma_i^z \rangle \\
        &= \langle \sigma_i^z \rangle_\text{data} - \langle \sigma_i^z \rangle_\text{model},
\end{split}
\end{align}
where \( \langle \ \cdot \ \rangle_{\text{data}} \) denotes the expectation value with respect to the data set distribution, and \( \langle \ \cdot \ \rangle_{\text{model}} \) denotes the expectation value with respect to the model distribution.

When restrictions are imposed on connections within the hidden layer, the clamped Hamiltonian reduces to
\begin{align}
    H_\vec{v}
        &= -\sum_{i=1}^{n} \big(\Gamma_i \sigma_i^x + b_i'(\vec{v}) \sigma_i^z\big),
\end{align}
where \( b_i'(\vec{v}) = b_i + (\mat{W}\T\vec{v})_i \).
This allows one to rewrite the clamped density matrix as
\begin{align}
\begin{split}
    \rho_\vec{v}
        &= \frac{1}{Z_\vec{v}} \exp\bigg( \sum_{i=1}^{n} \big(\Gamma_i \sigma_i^x + h_i'(\vec{v}) \sigma_i^z\big) \bigg) \\
        &= \frac{1}{Z_\vec{v}} \prod_{i=1}^{n} \exp \big(\Gamma_i \sigma_i^x + b_i'(\vec{v}) \sigma_i^z\big) \\
        &= \prod_{i=1}^{n} \rho_\vec{v}^{(i)}.
\end{split}
\end{align}
With this we can compute the expectation values as
\begin{align}
\begin{split}
    \langle \sigma_i^z \rangle_\vec{v}
        &= \tr(\rho_\vec{v}^{(i)}\sigma_i^z) \\
        &= \frac{\tr\bigg[ \exp \big(\Gamma_i \sigma_i^x + b_i'(\vec{v}) \sigma_i^z\big) \sigma_i^z \bigg]}{\tr\bigg[ \exp \big(\Gamma_i \sigma_i^x + b_i'(\vec{v}) \sigma_i^z\big) \bigg]} \\
        &= \frac{b_i'(\vec{v})}{D_i(\vec{v})} \tanh\big(D_i(\vec{v})\big),
\end{split}
\end{align}
where \( D_i(\vec{v}) = \sqrt{\Gamma_i^2 + b_i'(\vec{v})^2} \).

The last equality above is obtained by using that for traceless \( A \) with \( \det A < 0 \) we can write
\begin{align}
    \exp(A) = \cosh\Big(\sqrt{\abs{\det A}}\Big) I + \frac{1}{\sqrt{\abs{\det A}}}\sinh\Big(\sqrt{\abs{\det A}}\Big) A.
\end{align}
This is obtained by using Cayley-Hamilton theorem along with the series expansion of the matrix exponential and grouping the terms.

\section{Effective \( \beta \) as a Learnable Parameter}\label{app:learning_beta}
This derivation follows along the lines of that laid out in~\cite{xu_2021}.
Suppose the D-Wave annealer samples according to a classical Boltzmann distribution \( p_\text{DW} \) of energies \( E_\text{DW} = \beta E \), i.e.,
\begin{align}
\begin{split}
    p_\text{DW}
        &= \frac{1}{Z_\text{DW}} e^{-E_\text{DW}} \\
        &= \frac{1}{Z_\text{DW}} e^{-\beta E}.
\end{split}
\end{align}
Then we can take the partial derivative of the corresponding negative log-likelihood
\begin{align}
    -\partial_{\beta} \log p_\text{DW}
        &= E - \langle E \rangle,
\end{align}
and after averaging over all configurations we get
\begin{align}
    \Delta\beta
        &= \langle E \rangle_\text{data} - \langle E \rangle_\text{model},
\end{align}
which we can use to treat the effective inverse temperature as a learnable parameter.

\end{appendices}

\printbibliography

\end{document}